\documentclass[useAMS,usenatbib,usegraphicx]{mn2e}
\usepackage{pdflscape}
\usepackage[export]{adjustbox}
\usepackage{subfig}
\usepackage{breqn}
\usepackage{hyperref}
\usepackage{caption}
\usepackage{enumerate}

\hypersetup{colorlinks=true, breaklinks=true, citecolor=blue, linkcolor=blue, menucolor=blue, urlcolor=blue}

\newcommand{\mnuc}{$M_{K,nuclear}$\ }
\newcommand{\mbulge}{$M_{K,bulge}$\ }

\title[The host galaxies of AGN with powerful relativistic jets]{The host galaxies of active galactic nuclei with powerful relativistic jets}

\author[A. Olgu\'{\i}n-Iglesias et al.]{A. Olgu\'{\i}n-Iglesias$^{1,2}$, J. Le\'on-Tavares$^{1,3}$, J. K. Kotilainen$^{2,4}$, V. Chavushyan$^{1}$
\newauthor
M. Tornikoski$^{5}$, E. Valtaoja$^{4}$, C. A\~norve$^{6}$, J. Valdes$^{1}$, L. Carrasco$^{1}$\\ 
$^{1}$Instituto Nacional de Astrof\'{\i}sica \'Optica y Electr\'onica (INAOE), Apartado Postal 51 y 216, 72000 Puebla, M\'exico\\
$^{2}$Finnish Centre for  Astronomy with ESO (FINCA), University of Turku, V\"ais\"al\"antie 20, FI-21500  Piikki\"o, Finland\\
$^{3}$Sterrenkundig Observatorium, Universiteit Gent, Krijgslaan 281-S9, B-9000 Gent, Belgium\\
$^{4}$Tuorla Observatory, Department  of Physics and Astronomy, University of Turku, 20100 Turku, Finland\\
$^{5}$Aalto University Mets\"ahovi Radio Observatory,  Mets\"ahovintie 114, FIN-02540 Kylm\"al\"a, Finland\\
$^{6}$Facultad de Ciencias de la Tierra y del Espacio de la Universidad Aut\'onoma de Sinaloa, C.P. 80010, Culiac\'an Sinaloa, M\'exico}

\begin{document}

%\date{Accepted 1988 December 15. Received 1988 December 14; in original form 1988 October 11}

%\pagerange{\pageref{firstpage}--\pageref{lastpage}} \pubyear{2002}
 
\maketitle

\label{firstpage}
\begin{abstract}
We present deep Near-infrared (NIR) images of a sample of 19 intermediate-redshift ($0.3<z<1.0$)  radio-loud active galactic nuclei  (AGN) with powerful  relativistic jets  ($L_{1.4GHz} >10^{27}$ WHz$^{-1}$),  previously classified as flat-spectrum radio quasars. We also compile host galaxy and nuclear magnitudes for blazars from literature.  The combined sample (this work and compilation) contains 100 radio-loud AGN with host galaxy detections and a broad range of radio luminosities $L_{1.4GHz} \sim 10^{23.7} - 10^{28.3}$~WHz$^{-1}$, allowing us to divide our sample into high-luminosity blazars (HLBs) and low-luminosity blazars (LLBs). The host galaxies of our sample are bright and seem to follow the $\mu_{e}$-$R_{eff}$ relation for ellipticals and bulges. The two populations of blazars show different behaviours in the \mnuc - \mbulge plane, where a statistically significant correlation is observed for HLBs. Although it may be affected by selection effects, this correlation suggests a close coupling between the accretion mode of the central supermassive black hole and its host galaxy, that could be interpreted in terms of AGN feedback. Our findings are consistent with semi--analytical models where low--luminosity AGN emit the bulk of their energy in the form of radio jets, producing a strong feedback mechanism, and high--luminosity AGN are affected by galaxy mergers and interactions, which provide a common supply of cold gas to feed both nuclear activity and star formation episodes.\smallskip


\end{abstract}
\section{Introduction}\label{sec:intro}

Tight empirical relations between the black hole mass and properties of its host galaxy bulge \citep[e.g.][]{magorrian_1998,gebhardt_2000,ferrarese_merrit_2000,tremaine_2002,gultekin_2009} suggest a synergic connection between the growth of the black hole and the evolution of its host galaxy. The energy  (radiative and mechanical) deposited  by the AGN on its environment is thought to fundamentally influence its host galaxy -- for a recent review on AGN feedback see \citet{fabian_2012} and \citet{heckman_2014}. \smallskip

However, details on the mechanisms allowing nuclear activity to play a significant role on the formation and evolution of its host galaxy, remain elusive. Bearing this in mind, one may envision a close coupling between the relativistic jet launched by some black holes and their host galaxies, the so called radio-mode AGN feedback \citep{croton_2006,bower_2006}. So the question arises: Is there any relation between the host galaxy (its black hole) and the jet it launches? Are there any consequences on the host galaxy evolution because launching a powerful jet?\smallskip

Since radio jets are believed to be efficient to distribute (and affect) matter and energy from nuclear \citep[e.g][]{leontavares_2013} to galactic (kpc) scales \citep[e.g.][]{emonts_2005,nesvadba_2008,morganti_2013,tadhunter_2014}, the more powerful the jet, the larger the chance we might have to uncover imprints of the jet on its host galaxy. Then, an effective way to try to address the above question is by a thorough analysis of galaxies hosting AGN with relativistic jets covering a wide range of power (BL Lac objects and flat spectrum radio quasars [FSRQ], grouped together as blazars). However, host galaxies of blazars are usually outshone  by the highly-beamed synchrotron emission from the jet which makes a considerable challenge to detect and resolve the host galaxy in these type of objects. \smallskip

Despite these issues, significant effort to study the host galaxies of this type of AGN have been conducted  \citep[e.g.][]{stickel_1991,kotilainen_1998bllacs,falomo_1999,falomo_2000,scarpa_2000,urry_2000,nilsson_2003,cheung_2003,kotilainen_2005,Hyvonen_2007,leontavares_2011_mbh}, albeit concentrating on low-luminosity and nearby sources (mostly BL Lacs), for a recent compilation of results on BL Lacs host galaxies see \citet{falomo_2014}. \smallskip

In this work, we report the properties of  galaxies  hosting  high--luminosity FSRQ being successfully resolved with our deep NIR imagery. We compare the host galaxy properties in our sample with those reported in the literature for blazar sources. Our sample is described in \S2 and the observational data is presented in \S3. In \S 4 we present the structural analysis of the host galaxy  and these results are discussed in section \S 5. Our results are summarized in \S 6.  Throughout the manuscript we adopt cosmological parameters of $\Omega_{m}=0.3$, $\Omega_{\Lambda} = 0.7$ and a Hubble constant of   $H_{0} = 70$ Mpc$^{-1}$ km s$^{-1}$.

%%%%%%%%%%%%%%%%%%%%%%%%%%%%%%%%%%%%%%%%%%%%
%%%%%%%%%%%%%%%%%%%%%%%%%%%%%%%%%%%%%%%%%%%%
%%%%%%%%%%%%%%%%%%%%%%%%%%%%%%%%%%%%%%%%%%%%
%%%%%%%%%%%%%%%%%%%%%%%%%%%%%%%%%%%%%%%%%%%%
%%%%%%%%%%%%%%%%%%%%%%%%%%%%%%%%%%%%%%%%%%%5

\begin{figure}
\begin{tabular}{l}
	\includegraphics*[width=0.49\textwidth]{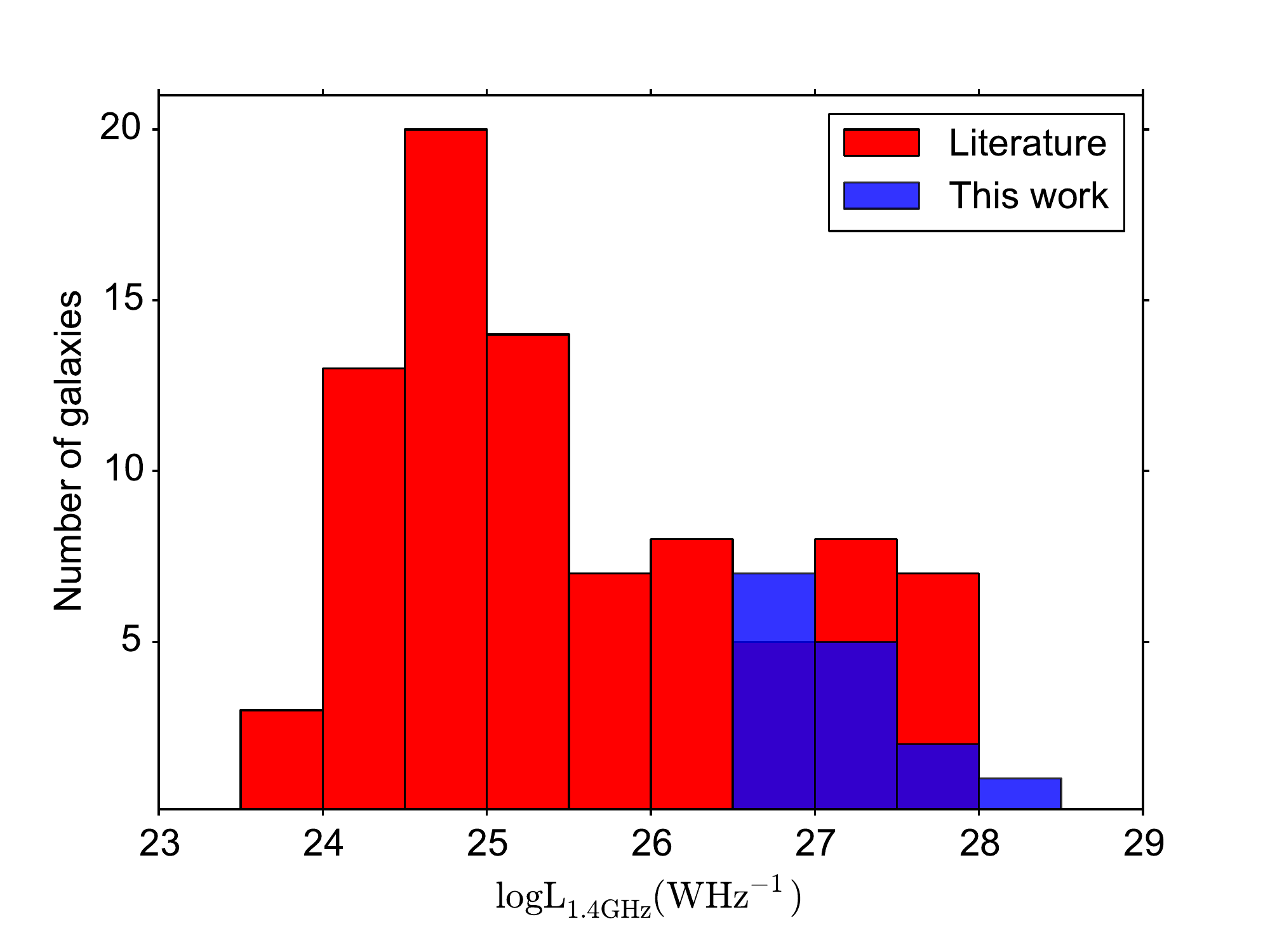}\\
	\includegraphics*[width=0.49\textwidth]{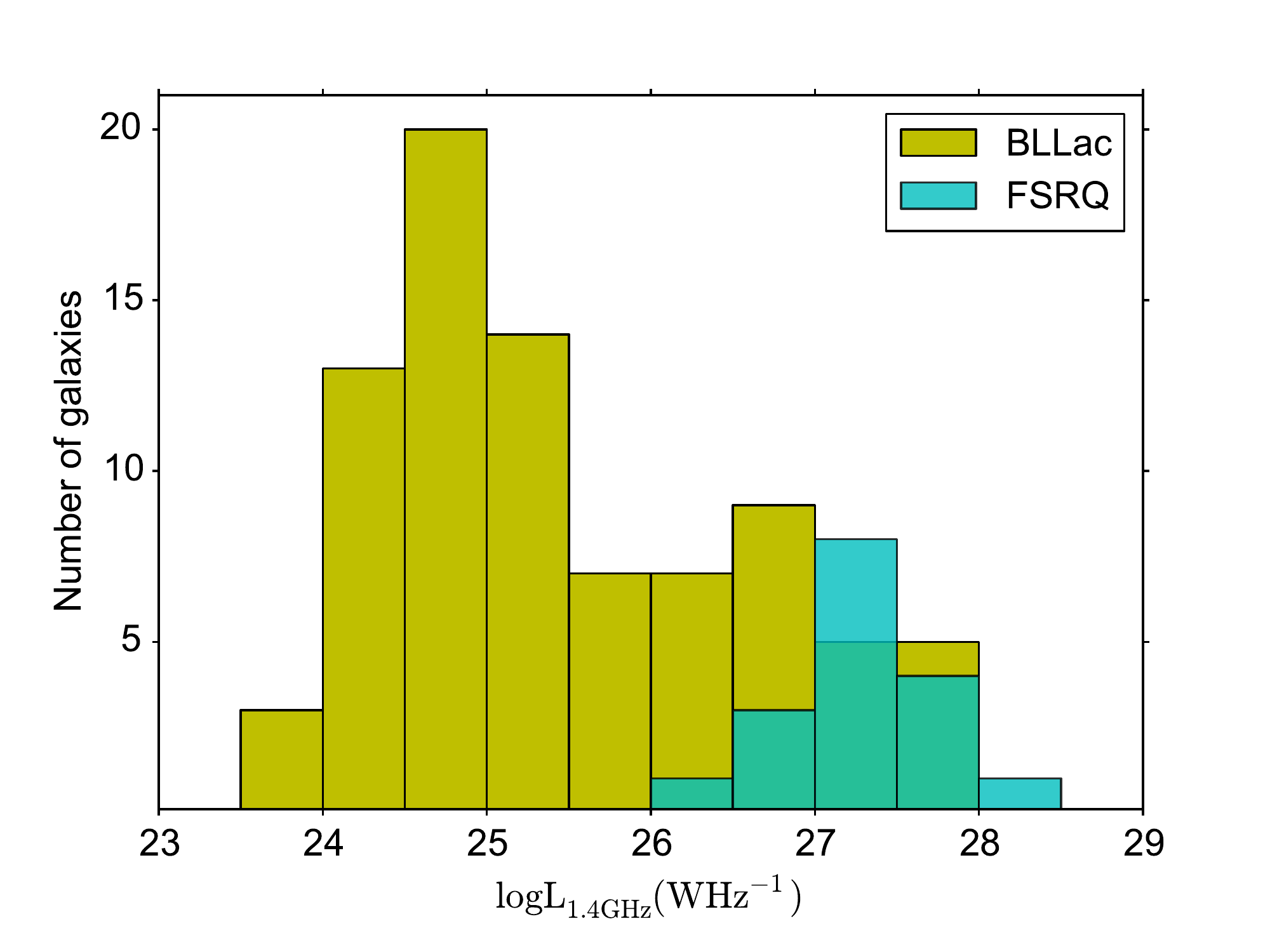} \\
	\includegraphics*[width=0.49\textwidth]{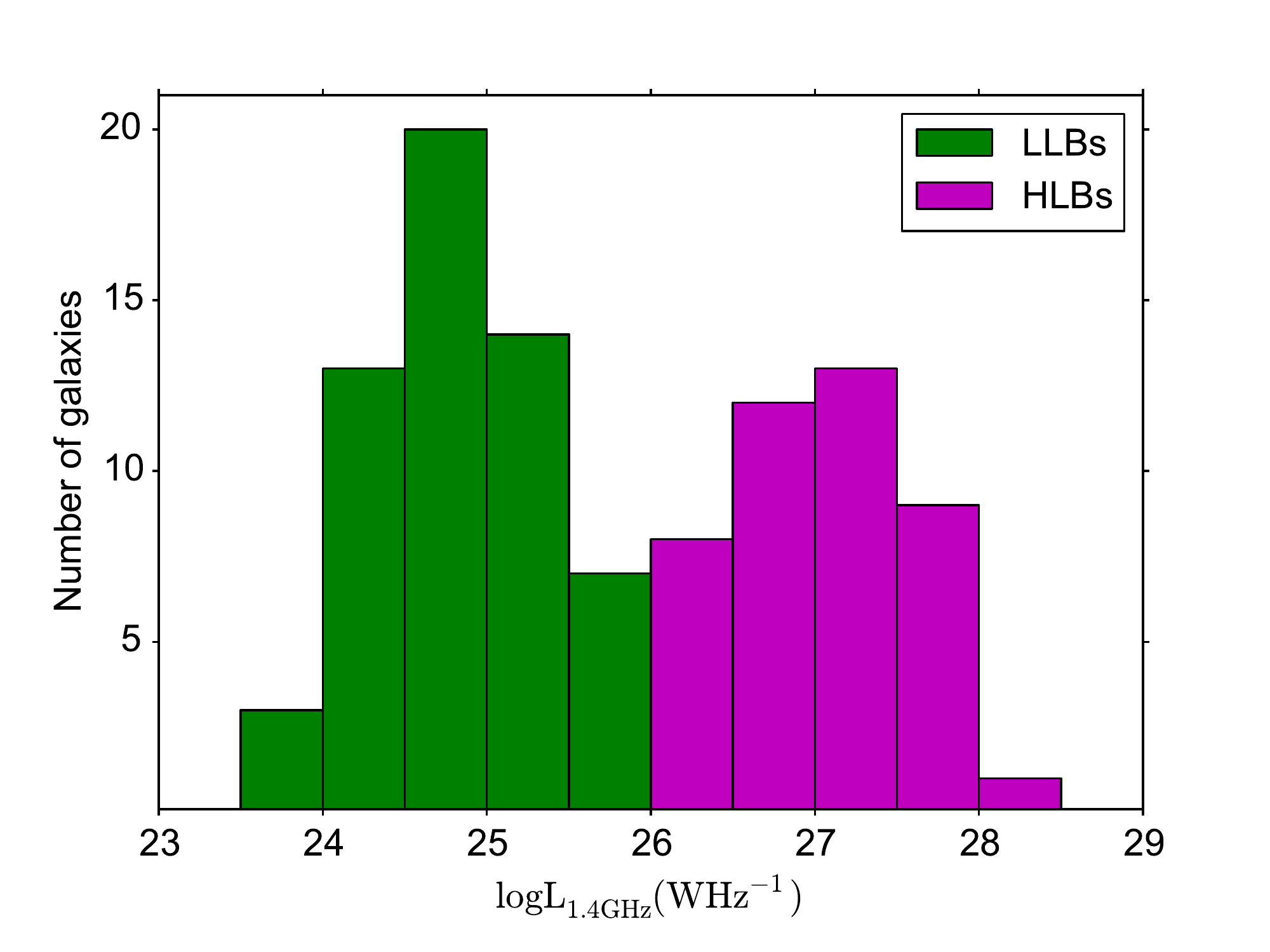} \\
   \end{tabular}
  \caption{Distribution of the 1.4GHz luminosities of the combined sample. Top panel shows the distribution of radio luminosity for sources compiled from literature (red) and the sample analysed in this work (blue). In the middle panel, the combined sample is divided according to the classical blazar classification scheme. BL Lac type objects (yellow) and FSRQ (cian) populate the low-end and high-end luminosities of the distribution, respectively. In the bottom panel, the sample is divided according to the adopted classification (LLBs and HLBs, green and magenta fills, respectively). Given the bimodality shown, we use $L_{1.4GHz}$ $=10^{26}$ $W$ $Hz^{-1}$ as the dividing value.}
\label{fig:sample}
\end{figure}

\section{Sample}\label{sec:sample}
%%%%%%%%%%%%%%%%%%
%%%%%%%%%%%%%%%%%%
\begin{table*}
 \begin{minipage}{150mm}
  \caption{Main properties of the galaxies analysed in this work and observations log. All the targets were observed with the NOTcam on the Nordic Optical Telescope (NOT). All galaxies are classified as FSRQs/HLBs.}
\label{table:sample}
  \begin{tabular}{@{}llcccccccl@{}}
  \hline
Source	&	Other name	&	z	&	RA	&	DEC	&	UT Date &	Filter 	&	Seeing	&	Exposure time \\	
		&	 			&		&		&		&			&			&	(")		&	(s)     		 \\
\multicolumn{1}{|c|}{(1)}&	\multicolumn{1}{|c|}{(2)} 	& (3)	 &	(4)	& (5)		&	(6)	&	(7) &	(8)	&	\multicolumn{1}{|c|}{(9)}	  		 \\	
\hline
0003$-$066		&	NRAO5				&	0.347	&	00:06:13.8	&	$-$06:23:35.3		&	15-Aug-11	&	J	&		0.75	 	&	2250	\\
0048$-$097		&	PKS~0048$-$09		&	0.634	&	00:50:41.3	&	$-$09:29:05.2		&	15-Sep-11	&	J	&	 	0.86	  	&	1450	\\
0059$+$581		&	TXS~0059				&	0.644	&	01:02:45.7	&	$+$58:24:11.1		&	15-Aug-11	&	J	&	  	0.61	  	&	4050	\\
0133$+$476		&	DA55					&	0.859	&	01:36:58.6 	&	$+$47:51:29.0	 	&	14-Sep-11	&	H	&	  	0.68	  	&	3325	\\
0202$+$149		&	4C~15.05				&	0.405	&	02:04:50.4	&	$+$15:14:11.0		&	15-Sep-11	&	J	&	  	0.63	  	&	3150	\\
0306$+$102		&	PKS~0306$-$102		&	0.863	&	03:09:03.6	&	$+$10:29:16.3		&	15-Sep-11	&	H	&	  	0.86	  	&	4050	\\
1150$+$497		&	4C$-$49.22			&	0.334	&	11:53:24.4	&	$+$49:31:08.8		&	9-May-11		&	J	&	  	0.63	  	&	900		\\
1156$+$295		&	4C~29.45				&	0.724	&	11:59:31.8	&	$+$29:14:43.8		&	9-May-11		&	J	&	  	0.54	  	&	3600	\\
1219$+$044		&	PKS~1219$-$04		&	0.966	&	12:22:22.5	&	$+$04:13:15.7		&	11-Jun-11	&	H	&	  	0.73	  	&	3650	\\
1308$+$326		&	AUCVN				&	0.998	&	13:10:28.6	&	$+$32:20:43.7		&	9-May-11		&	H	&	  	1.10	  	&	4650	\\
1510$-$089		&	PKS~1510$-$08		&	0.360	&	15:12:50.5	&	$-$09:05:59.8		&	11-Jun-11	&	J	&	  	0.91	  	&	1800	\\
1546$+$027		&	PKS~1546$-$027		&	0.414	&	15:49:29.4	&	$+$02:37:01.1		&	11-Jun-11	&	J	&	  	0.85	  	&	1800	\\
1641$+$399		&	3C~345				&	0.593	&	16:42:58.8	&	$+$39:48:36.9		&	9-May-11		&	J	&	  	0.61	  	&	1900	\\
1642$+$690		&	4C$-$69.21			&	0.751	&	16:42:07.8	&	$+$68:56:39.7		&	12-Jun-11	&	J	&	  	0.59	  	&	3600	\\
1828$+$487		&	3C~380				&	0.692	&	18:29:31.7	&	$+$48:44:46.1		&	12-Jun-11	&	J	&	  	0.83	  	&	3600	\\
1849$+$670		&	4C$+$66.20			&	0.657	&	18:49:16.1 	&	$+$67:05:42.0		&	14-Aug-11	&	J	&	  	0.89	  	&	3600	\\
1928$+$738		&	4C~73.18				&	0.302	&	19:27:48.4	&	$+$73:58:01.5		&	14-Aug-11	&	J	&	  	0.79	  	&	1800	\\
2216$-$038		&	4C$-$03.79			&	0.901	&	22:18:52.0	&	$-$03:35:36.8		&	14-Sep-11	&	H	&	  	0.86	  	&	3600	\\
2234$+$282		&	B2~2234+28A 			&	0.795	&	22:36:22.4	&	$+$28:28:57.4		&	14-Aug-11	&	H	&	  	0.61	  	&	3600	\\
\hline
\end{tabular}\\
Columns: (1) and (2) give the designation and name of the source;
(3) the redshift of the object; (4) and (5) the J2000 right ascension and declination of the source; 
(6) the observation date;
(7) the filter used for the observation: J=NOTcam standard J filter (1.165$\mu$m-1.328$\mu$m), H=NOTcam standard H filter (1.484$\mu$m-1.780$\mu$m);
(8) the seeing during the observation, and
(9) the total exposure time.
\end{minipage}
\end{table*}
%%%%%%%%%%%%%%%%%%
%%%%%%%%%%%%%%%%%%
The sample of sources analyzed in this work is a sub--sample of variable radio-loud AGN monitored at 7mm (S$_{7mm}~>1~Jy$) with the Aalto University Mets\"ahovi Radio Observatory in Finland \footnote{\href{http://metsahovi.aalto.fi/en/}{http://metsahovi.aalto.fi/en/}}, since the last 30 years \citep{terasranta_1992,terasranta_1998,leontavares_2011,nieppola_2011}. According to the AGN unification scheme \citep{antonucci_1993,urry_padovani_1995}, FSRQ and BL Lacs are those AGN whose relativistic jets point towards the Earth. Jet viewing angles have been estimated in \citet{hovatta_2009} for most sources in our sample. The distribution of jet-viewing angle comprised in our sample is  narrow  and skewed towards small angles (i.e. $< 10$ deg). Then, for the purpose of this investigation, the distribution of jet viewing angles of our sample can be adopted as statistically indistinguishable. Therefore, hereafter we shall assume that the range of observed radio luminosities in the sample is merely associated to the intrinsic power (and speed) of the jet.\smallskip

The sources considered in this work have also been regularly observed with the Very Long Baseline Array (VLBA) within the MOJAVE programme \citep{mojave}. Thus, having well sampled millimeter light curves (via monitoring with the Mets\"ahovi Radio Observatory) and information about the evolution of the inner parsec scale jet structure (via VLBA observations), allow us to explore a possible connection between the host galaxy (its black hole) and intrinsic properties of the jet (e.g. Doppler factor, intrinsic speed) -- this connection will be explored in a companion manuscript.\smallskip

We select intermediate redshift ($0.3 < z <1.0$) sources so that $J-$ and $H-$band observations could cover the  typical rest-frame wavelength of $\sim~1\mu m$, thus allowing us to probe the old stellar population of galactic bulges. The sample of sources for which host galaxy imaging has been attempted in this work is listed on Table~\ref{table:sample}, comprising 19 FSRQs.\smallskip

Additionally to these sources (hereafter, this work), we performed a large compilation of blazars (78 BL Lacs and 7 FSRQ; hereafter the compiled sample or literature sample) with host galaxy detection \citep{kotilainen_1998,kotilainen_1998bllacs,falomo_1999,falomo_2000,scarpa_2000,urry_2000,nilsson_2003,heidt_2004,cheung_2003,odowd_2005,kotilainen_2005, nilsson_2009}, redshifts $0.0 < z < 1.3$ and 1.4~GHz flux density measurements reported in the literature.\smallskip 

To allow for comparison, all magnitudes in our and compiled samples were transformed to $K$--band. We use our adopted cosmology ({section \ref{sec:intro}}) together with rest-frame colours of giant ellipticals, more specifically: $R-K=2.7$ and $I-K = 2.0$ from \citet{kotilainen_1998} and $H-K=0.22$ reported in  \citet{recillas_1990}), and the following nuclear colours: $R-K=2.95$ \citep{kotilainen_2005}, $H-K = 1.10$ \citep{kotilainen_1998} and  J-K=1.44 \citep{cheung_2003}. The galaxies in the sample of \citet{heidt_2004} and \citet{nilsson_2009} were studied in I band. By interpolating between R and J band fluxes, we derived a $I-K=2.34$ colour to account for the nuclear part of the galaxies.\smallskip

Absolute magnitudes derived from Hubble Space Telescope (HST) and ground-based images of BL Lac objects are in very good agreement with each other \citep[as shown by][]{falomo_1999}. Moreover, previous studies have not found effective radius dependence with wavelength, although some trend has been noticed by \citet{Hyvonen_2007}.\smallskip

Top panel of Figure~\ref{fig:sample} shows the distribution of radio luminosity at 1.4~GHz  (L$_{1.4GHz}$) for our sample, along with the compiled sample.
As it can be seen from the top and middle panels of Figure~\ref{fig:sample}, previous host galaxies studies have focused on sources with low-luminosity AGN, mostly BL Lacs. At first glance, a bimodal distribution can be gleaned from inspection of Figure~\ref{fig:sample}. The bimodality of $L_{1.4GHz}$ reflects the distribution of blazar types (see middle panel of Figure \ref{fig:sample}), most BL Lacs populating the low-end luminosity of the distribution and all FSRQs populating the high-end luminosity of the distribution. We have not changed the galaxy classification from their original studies, although we have checked (using more recent spectra) that most BL Lacs populating the high-end luminosity of the distribution ($L_{1.4GHz} \geq 10^{26}$ W Hz$^{-1}$) are misclassified FSRQs.\smallskip

The division of blazar sources between BL Lacs and FSRQs based solely on the exhibition of broad emission lines is far from being accurate. As an example, we invoke the source 2201$+$044 previously classified as BL Lac \citep[e.g.][]{sbarufatti_2005}, however, prominent broad emission lines can be easily identified on its optical spectrum \citep{sbarufatti_2006}. A recent study by \citet{giommi_2012, giommi_2013} discusses more in detail the deficiency of the BL Lac/FSRQ classification. They propose an alternative blazar classification scheme considering two physically different AGN classes: low-excitation radio galaxies (LERGs) and high-excitation radio galaxies (HERGs). In LERGs, the black hole accretes material via a geometrically thick advection--dominated accretion flow (ADAF), with low accretion rates and radiative efficiency. On the other hand, in HERGs, the black hole accretes material very efficiently via an optically thick and geometrically thin accretion disc, whose UV emission is capable of ionizing the broad-- and narrow--emission lines.

Considering the above, we avoid the BL Lac/FSRQ classification and instead we assume that intrinsic differences in blazars properties are due to differences in their accretion modes and that these differences are reflected in their radiative efficiencies and in turn, radio luminosity L$_{1.4GHz}$, since the luminosity of the accretion disc and jet radio power are known to be correlated \citep{giommi_2012}. Specifically, black holes accreting through ADAFs should show low radiative efficiencies, and in turn low L$_{1.4GHz}$, whereas black holes accreting through accretion discs, should show high radiative efficiencies and therefore, high L$_{1.4GHz}$. In order to set a dividing L$_{1.4GHz}$ value, we assume that the bimodality presented in the L$_{1.4GHz}$ distribution of Figure \ref{fig:sample} is largely caused by the dominance of either of the two accretion modes. Hence, we call low--luminosity blazars (LLBs), those sources with $L_{1.4GHz} < 10^{26}$ WHz$^{-1}$ and high--luminosity blazars (HLBs), those that show $L_{1.4GHz} \geq 10^{26}$ WHz$^{-1}$. In this way, the LLBs sources comprises 57\% of our sample and the HLBs, 43\%.\smallskip

In a recent study, \citet[][see also, \citealt{buttiglione_2010}]{best_2012} find that LERGs and HERGs appear to switch in dominance at $L_{1.4GHz}\sim10^{26}WHz^{-1}$. However, both populations are found across a wide range of radio luminosities ($\sim10^{22} WHz^{-1} < L_{1.4GHz} < \sim10^{27} WHz^{-1}$). Moreover, when the jet viewing angle is closely aligned to the line of sight, $L_{1.4GHz}$ strongly depends on the unknown intrinsic speed and power of the jet. Hence, because of relativistic beaming, the fraction of jets from LERGs that might take values $L_{1.4GHz}\geq10^{26}WHz^{-1}$ is even larger in samples constituted entirely by blazars, like ours.

Based on the foregoing, we can not categorically link LERGs to LLBs or HERGs to HLBs, specially for sources populating the low luminosity end of HLBs, where powerful and highly beamed LERGs are more likely to be located. However, we assume that the differences in radio luminosities are due to a predominant accretion mode in each sample

\section{Observations}

Observations were made with the Nordic Optical Telescope (NOT)\footnote{\href{http://www.not.iac.es/}{http://www.not.iac.es/}} at La Roque de los Muchachos, La Palma, Canarias, Spain. They were conducted between 09 May and 15 September 2011 using the near-infrared Camera (NOTCam)\footnote{\href{http://www.not.iac.es/instruments/notcam/}{http://www.not.iac.es/instruments/notcam/}} on the NOT. NOTcam field of view is $4' \times  4'$ with a pixel scale of $0.234 "$/pixel designed to be used in the range from 0.8 to 2.5$\mu$m in the bands J, H and K.  \smallskip

Observing the targets in the red part of the rest-frame spectrum is important since the stars in elliptical galaxies (expected hosts of blazars) are  mostly population II (red, old and low mass). Therefore, to target the rest-frame R- or I-band emission from the host galaxies, we observed the sources in the J (1.250$\mu$m) and H (1.626$\mu$m) filters, for redshifts $0.3<z<0.8$ and $0.8<z<1.0$, respectively.\smallskip

A dithering pattern was used for the observations to allow an accurate sky subtraction. The dithering step was 40 arcsec, with 50 seconds exposures for each step. Table \ref{table:sample} shows the journal of the observations.\smallskip

Data reduction for the images was performed using the NOTCam quicklook package in IRAF \footnote{\href{http://www.not.iac.es/instruments/notcam/guide/observe.html\#reductions}{http://www.not.iac.es/instruments/notcam/guide/\\observe.html\#reductions}}. We corrected beforehand the distortion of the camera used (WF-camera) which is significant, especially at the corners. It is corrected using distortion models for each band constructed with high quality data of a rich field of stars ($\sim$300 2MASS sources). Next, a masterflat was created. Two pairs of skyflats were observed each night for a better estimation of normalized median combined masterflat. Skyflats were interpolated over bad pixels, using a bad pixel mask and corrected for the dc-gradient in differential images. After that, the dithered images are combined to get a sky template, which is subtracted from each image. Finally, using field stars as reference points, the images were aligned and combined to obtain a co--added image which are used for our analysis. \smallskip

\section{Host galaxy images}
\subsection{Photometric Decomposition}

We analyse quantitatively the structure of the host galaxies in our sample by modeling their surface brightness (see Figure \ref{figureA1} in Appendix) following the methodology presented in \citet{leontavares_2014} where the 2D image decomposition code GALFIT \citep{peng_2011} was used. GALFIT uses a least-squares technique to minimize the $\chi^2$ from the residual image (the product of subtracting the model from the observed galaxy). Models are composed of analytical functions that, in turn, are composed of different parameters which are free to vary until the $\chi^2$ is reduced.\smallskip

The first component used to model the galaxies in the sample is the sky. The modeling of the sky is performed using a flat plane with the ability to tilt in the $x$ and $y$ directions. With the aim of obtaining more precise sky estimations, this component is modeled in regions of the image with the minimum amount of objects. \smallskip

\begin{figure*}
\includegraphics[width=1.19\textwidth,center]{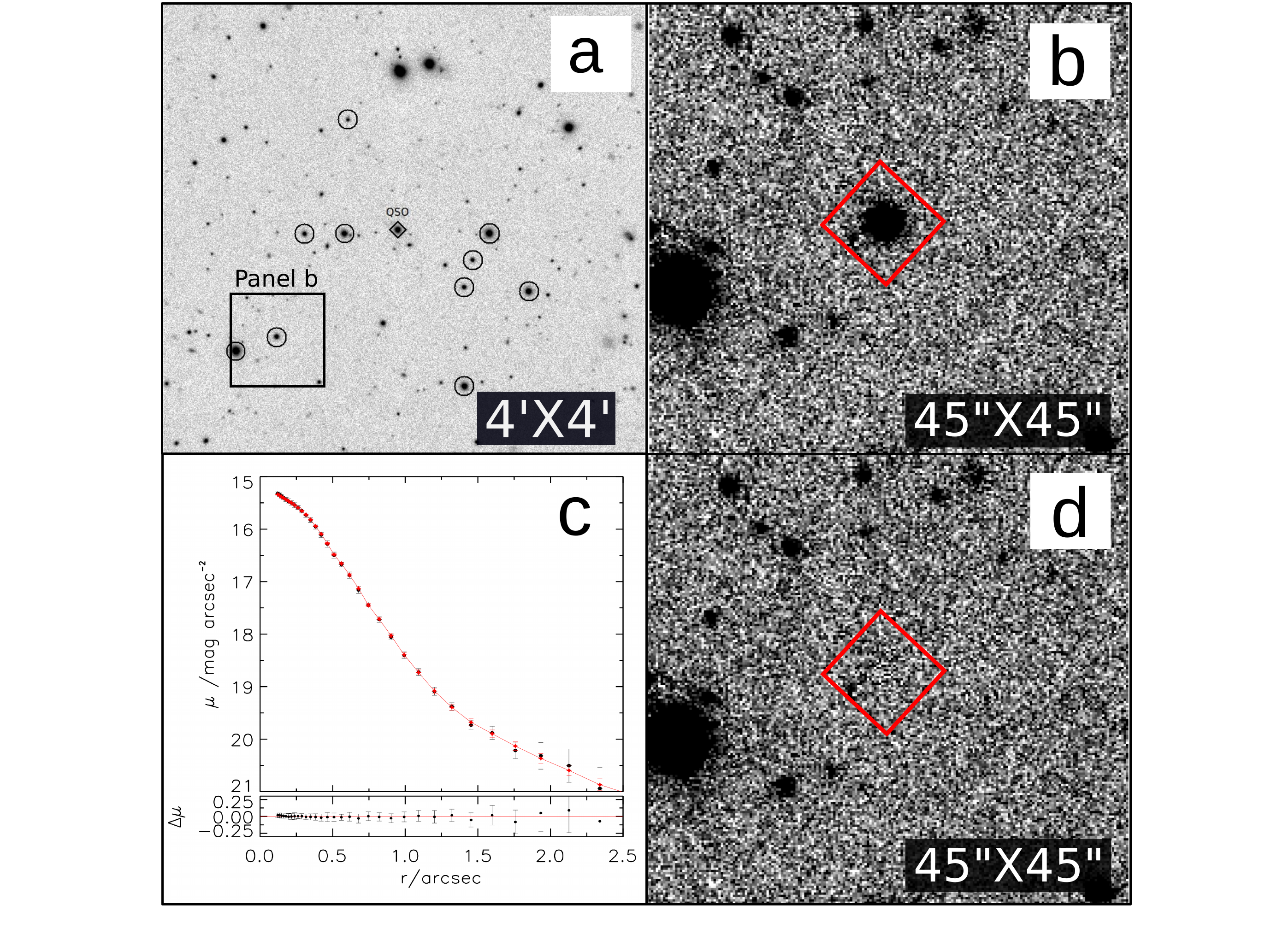}
\caption{Example of PSF model procedure. In panel (a) we show the target (1828+487, rhombus) and the selected stars to make the PSF model (circles). In Panel (b) we show and example of a random star (red rhombus) from the field used to test the PSF model. The location of panel (b) is indicated in panel (a). In panel (c) we show the azimuthally averaged radial surface brightness profiles of the random star (black data points) and the PSF model (red solid line). (d) Subtracted PSF model residuals.} \label{fig:psf_test}
\end{figure*}

The next component, is the point spread function (PSF). The PSF modeling is performed by simultaneously fitting the highest number of stars in the field as possible. We use a variable number of Gaussian and exponential functions convolved in a 100 pixels $\times$ 100 pixels ($23"\times23"$) box. We select non--saturated stars (although of any magnitude), with no close companions (closer than $\sim 2.5"$) and preferentially close to the source. If inside the convolution box there is any other object, it is masked out using the SExtractor segmentation image \citep{bertin_1996}. We finally test our PSF by fitting random stars in the field to ensure our model is suitable for our analysis. Figure \ref{fig:psf_test} shows an example of the PSF modeling procedure.\smallskip

Since the AGN emission in our images is unresolved, we use the PSF model to represent it. In addition to the AGN emission, the PSF is needed to model the central region of the host galaxies. A poor PSF model might lead to over/under estimations of the host galaxy parameters so, it has to be stressed that the PSF modeling is the key step towards an appropriate fit.\smallskip

The final component is the host galaxy. As radio loud AGN are expected to be hosted by early type galaxies, \citep{kotilainen_1998,falomo_2000,scarpa_2000, nilsson_2003}, we use the S\'ersic profile, described as:

%%%%%%%%%%%%%%%%%%%%%%%%%%%%%%%%%%%%%%%%%%%%%%%%%%%%%%%%%%%
%%%%%%%%%%%%%%%%%%%%%%%%%%%%%%%%%%%%%%%%%%%%%%%%%%%%%%%%%%%

	\begin{center}
	\begin{equation}
	I(R)=I_e exp\left[ -\kappa\left( \left(\frac{R}{R_e} \right)^{1/n} -1\right)\right]
	\label{sersic}
	\end{equation}
	 \end{center}

%%%%%%%%%%%%%%%%%%%%%%%%%%%%%%%%%%%%%%%%%%%%%%%%%%%%%%%%%%%
%%%%%%%%%%%%%%%%%%%%%%%%%%%%%%%%%%%%%%%%%%%%%%%%%%%%%%%%%%%
\noindent	 	 
where $I(R)$ is the surface brightness at the radius R, and $\kappa$ is a parameter coupled to the S\'ersic index $n$ in such way that $I_e$ is the surface brightness at the effective radius $R_e$ (radius where the galaxy contains half of the light).  See \citet{graham_2005} for a formal definition of the parameters involved in the S\'ersic profile.\smallskip

Even though we expected the host galaxies to be better described by a S\'ersic profile, we did not discard the possibility of the existence of disc components in their sub-structures, so we explored the exponential profile alone and together with the S\'ersic profile. However, in accordance with previous works \citep{kotilainen_1998,kotilainen_1998bllacs,falomo_1999,falomo_2000,scarpa_2000,urry_2000,nilsson_2003,heidt_2004,cheung_2003,odowd_2005,kotilainen_2005, nilsson_2009}, the latter did not show better results.\smallskip

Once we define the analytical functions needed for the fit, we proceed to run GALFIT. We use the initial guesses from the catalogue of objects derived by SExtractor for each image, and a S\'ersic index $n=4$ (de Vaucouleurs profile, which describes how the surface brightness in elliptical galaxies varies as a function of the distance from the center). As with the PSF, we masked out the objects inside the convolution box by implementing the SExtractor segmentation image. If the targets have very close companions, we fit them simultaneously with the galaxy (e.g. 2234$+$282) since their light distributions merge and a simple mask is not enough to remove them from the fit.\smallskip

We fit the sky background first and left fixed during the model fitting, in this way, the total number of parameters is reduced when the other components are being computed. Once the background is fitted, we run GALFIT again to fit the nuclear part, and once again to fit the host galaxy. The parameters obtained from these fits are used as initial guesses for the final run, where we fit all the components together to obtain the final model.\smallskip

%%%%%%%%%%%%%%%%%%%%%%%%%%%%%%%%%%%%%%%%%%%%%%%%%%%%%%%
%%%%%%%%%%%%%%%%%%%%%%%%%%%%%%%%%%%%%%%%%%%%%%%%%%%%%%%
%%%%%%%%%%%%%%%%%%%%%%%%%%%%%%%%%%%%%%%%%%%%%%%%%%%%%%%%
%%%%%%%%%%%%%%%%%%%%%%%%%%%%%%%%%%%%%%%%%%%%%%%%%%%%%%%%
%%%%%%%%%%%%%%%%%%%%%%%%%%%%%%%%%%%%%%%%%%%%%%%%%%%%%%%%
%%%%%%%%%%%%%%%%%%%%%%%%%%%%%%%%%%%%%%%%%%%%%%%%%%%%%%%%

\subsection{Uncertainties}
Uncertainties of parameters derived from GALFIT, are typically $\sim0.01$ mag and $\sim0.05$ arcsec. However, the actual values are larger and are originated in the uncertainties of the appropriate functional form of galaxy components.\smallskip

In order to estimate uncertainties, we follow the method in \cite{greene_2008}. We first identified model parameters and assumptions that could contribute most significantly to errors.\smallskip

Uncertainties in the PSF model due to temporal and spatial variations would affect the galaxy structural parameters and magnitudes. The sky background should just slightly influence the model magnitudes, since the galaxies in our sample are observed in NIR bands and thus the sky counts must be zero (however, we take it into account since it may vary up to 0.5 mag).\smallskip

To account for uncertainties due to PSF, we used several PSF models assuming that the differences in the fits would represent the uncertainties due to our PSF model imperfection. To account for uncertainties due to sky, we run several sky fits in separated regions of 300 pixels $\times$ 300 pixels ($70"\times70"$). Using the different PSF models and sky values, we perform several GALFIT runs with these variations. The fits obtained were used to make a statistic where the best-fit value is the mean and the errors are $\pm1\sigma$ for each of the parameters of the galaxy model.

\begin{landscape}
 \begin{table}
 \centering
 \begin{minipage}{220mm}
  \caption{Host galaxies parameters derived from 2D analysis. All K--corrections were performed using the \href{http://kcor.sai.msu.ru/}{K-corrections calculator}\\  \citep{chilingarian_2010, chilingarian_2012}}
\label{table:galfit}
  \begin{tabular}{@{}ccccccccccc@{}}
  \hline
Name	 &	$m_{nuclear}$		&	$M_{nuclear}$\footnote{Corrected for extinction only. Nuclei are assumed to have flat power spectra and therefore have negligible K--corrections.}		&	$m_{host}$		&	$M_{host}$\footnote{Corrected for exttinction and K--correction.} &  Re	&	Re	 & $\mu_{e}$\footnote{Corrected for Galactic extinction, K--correction and cosmological dimming.}	 & n  					&    $\chi^2_{best fit}$& 	$\chi^2_{bestfit}$/$\chi^2_{PSF}$  \\ 
&	&	&  &	&  (arcsec)  &	(kpc)	 & ($mag/arcsec^{-2}$)	 &   		&     	&	\\
(1)	 & (2)	&	(3)	& (4)	&   (5)	& (6)	& (7) & (8)	 &  (9) &  (10)     &(11)		\\
\hline
0003$+$066	&	17.29	$\pm$	0.16	&	$-$24.03	$\pm$	0.42	&	16.68	$\pm$	0.21	&	-24.64	$\pm$	0.42	&	1.99	$\pm$	0.67	&	9.77	$\pm$	3.29	&	19.50	$\pm$	0.51	&	3.25	$\pm$	0.42	&	1.157	&	0.42	\\
0048$-$097	&	14.91	$\pm$	0.22	&	$-$27.97	$\pm$	0.45	&	17.34	$\pm$	0.11	&	-25.54	$\pm$	0.27	&	2.54	$\pm$	0.22	&	17.60	$\pm$	1.52	&	20.40	$\pm$	0.47	&	3.00	$\pm$	0.23	&	1.223	&	0.48	\\
0059$+$581	&	18.30	$\pm$	0.16	&	$-$24.62	$\pm$	0.35	&	$>$18.28			&	$>-$24.64			&	--			&	--			&	--			&	--			&	1.300	&	0.99	\\
0133$+$476	&	13.64	$\pm$	0.12	&	$-$30.05	$\pm$	0.13	&	$>$15.20			&	$>-$28.49			&	--			&	--			&	--			&	--			&	1.176	&	0.99	\\
0202$+$149	&	18.73	$\pm$	0.11	&	$-$22.98	$\pm$	0.29	&	18.71	$\pm$	0.23	&	-23.00	$\pm$	0.42	&	1.50	$\pm$	0.14	&	8.12	$\pm$	0.76	&	20.65	$\pm$	0.52	&	4.00	$\pm$	0.36	&	1.147	&	0.50	\\
0306$+$102	&	15.75	$\pm$	0.20	&	$-$27.95	$\pm$	0.17	&	16.93	$\pm$	0.21	&	-26.77	$\pm$	0.22	&	1.75	$\pm$	0.13	&	13.46	$\pm$	1.00	&	19.90	$\pm$	0.53	&	3.30	$\pm$	0.31	&	1.140	&	0.55	\\
1150$+$497	&	14.60	$\pm$	0.11	&	$-$26.62	$\pm$	0.45	&	16.07	$\pm$	0.05	&	-25.15	$\pm$	0.46	&	0.90	$\pm$	0.29	&	4.33	$\pm$	1.40	&	15.60	$\pm$	0.42	&	4.00	$\pm$	0.50	&	1.455	&	0.56	\\
1156$+$295	&	14.29	$\pm$	0.10	&	$-$28.94	$\pm$	0.32	&	17.04	$\pm$	0.07	&	-26.90	$\pm$	0.17	&	1.17	$\pm$	0.45	&	8.47	$\pm$	3.26	&	19.20	$\pm$	0.41	&	3.50	$\pm$	0.25	&	1.309	&	0.53	\\
1219$+$044	&	16.23	$\pm$	0.22	&	$-$27.77	$\pm$	0.24	&	$>$16.95			&	$>-$27.05			&	--			&	--			&	--			&	--			&	1.157	&	0.98	\\
1308$+$326	&	13.61	$\pm$	0.12	&	$-$30.48	$\pm$	0.22	&	$>$14.70			&	$>-$29.39			&	--			&	--			&	--			&	--			&	1.150	&	0.99	\\
1510$-$089	&	15.19	$\pm$	0.24	&	$-$26.22	$\pm$	0.25	&	16.32	$\pm$	0.24	&	-25.09	$\pm$	0.33	&	1.99	$\pm$	0.30	&	10.03	$\pm$	1.51	&	17.90	$\pm$	0.51	&	4.00	$\pm$	0.47	&	1.157	&	0.54	\\
1546$+$027	&	14.56	$\pm$	0.19	&	$-$27.21	$\pm$	0.28	&	16.45	$\pm$	0.18	&	-25.32	$\pm$	0.26	&	1.20	$\pm$	0.63	&	6.58	$\pm$	3.45	&	18.10	$\pm$	0.45	&	3.00	$\pm$	0.15	&	1.128	&	0.58	\\
1641$+$399	&	15.76	$\pm$	0.24	&	$-$26.94	$\pm$	0.30	&	16.04	$\pm$	0.16	&	-26.66	$\pm$	0.35	&	1.37	$\pm$	0.18	&	9.10	$\pm$	1.20	&	18.05	$\pm$	0.41	&	3.82	$\pm$	0.33	&	1.080	&	0.54	\\
1642$+$690	&	18.15	$\pm$	0.16	&	$-$25.18	$\pm$	0.35	&	17.61	$\pm$	0.21	&	-25.72	$\pm$	0.36	&	2.00	$\pm$	0.69	&	14.68	$\pm$	5.06	&	21.60	$\pm$	0.42	&	4.00	$\pm$	0.26	&	1.147	&	0.44	\\
1828$+$487	&	16.03	$\pm$	0.22	&	$-$27.08	$\pm$	0.19	&	16.91	$\pm$	0.06	&	-26.20	$\pm$	0.38	&	0.63	$\pm$	0.38	&	4.51	$\pm$	2.72	&	17.20	$\pm$	0.47	&	3.10	$\pm$	0.50	&	1.111	&	0.52	\\
1849$+$670	&	15.66	$\pm$	0.15	&	$-$27.31	$\pm$	0.29	&	16.57	$\pm$	0.18	&	-26.40	$\pm$	0.37	&	2.57	$\pm$	0.13	&	17.89	$\pm$	0.90	&	20.75	$\pm$	0.53	&	3.50	$\pm$	0.50	&	1.120	&	0.58	\\
1928$+$738	&	14.75	$\pm$	0.10	&	$-$26.22	$\pm$	0.18	&	16.38	$\pm$	0.16	&	-24.59	$\pm$	0.48	&	3.51	$\pm$	0.53	&	15.71	$\pm$	2.37	&	20.20	$\pm$	0.46	&	3.40	$\pm$	0.44	&	1.195	&	0.51	\\
2216$-$038	&	14.48	$\pm$	0.20	&	$-$29.34	$\pm$	0.28	&	17.52	$\pm$	0.06	&	-26.30	$\pm$	0.42	&	2.62	$\pm$	0.21	&	20.42	$\pm$	1.64	&	20.00	$\pm$	0.46	&	3.25	$\pm$	0.49	&	1.137	&	0.52	\\
2234$+$282	&	16.90	$\pm$	0.23	&	$-$26.58	$\pm$	0.21	&	18.18	$\pm$	0.17	&	-25.30	$\pm$	0.41	&	1.19	$\pm$	0.68	&	8.91	$\pm$	5.09	&	19.50	$\pm$	0.55	&	3.00	$\pm$	0.46	&	1.159	&	0.56	\\
\hline
\end{tabular}

Column (1) gives the galaxy name;
(2) and (3) the apparent and absolute nuclear magnitude for the best-fit model in the observed band;
(4) and (5) the apparent and absolute host galaxy magnitude for the best-fit model in the observed band. When the host galaxy is not detected, we determine an upper limit by simulating a host galaxy. We assume a de Vaucouleurs profile and an effective radius equals to a typical value for HLBs ($R_e=10kpc$). We increase the simulated host galaxy luminosity until it becomes detectable within the associated errors of the luminosity profile;
(6) and (7) the bulge model effective radius in arcsec and in kpc, respectively;
(8) the bulge model surface brightness at the effective radius;
(9) the bulge model S\'ersic index;
(10) the reduced chi squared for the best-fit model, and
(11) the ratio between best-fit (S\'ersic + PSF) $\chi^2$ and PSF-fit $\chi^2$.
\end{minipage}
\end{table}
\end{landscape}

%%%%%%%%%%%%%%%%%%%%%%%%%%%%%%%%%%%%%%%%%%%%%%%%%%%%%%%%%%%%%%%
%%%%%%%%%%%%%%%%%%%%%%%%%%%%%%%%%%%%%%%%%%%%%%%%%%%%%%%%%%%%%%%
%%%%%%%%%%%%%%%%%%%%%%%%%%%%%%%%%%%%%%%%%%%%%%%%%%%%%%%%%%%%%%%

\section{Results and discussion}
\subsection{The host galaxies}\label{sec:host_galaxies}

%%%%%%%%%%%%%%%%%%
%%%%%%%%%%%%%%%%%%
%%%%%%%%%%%%%%%%%%
%%%%%%%%%%%%%%%%%%

\begin{figure}
\includegraphics[width=.52\textwidth]{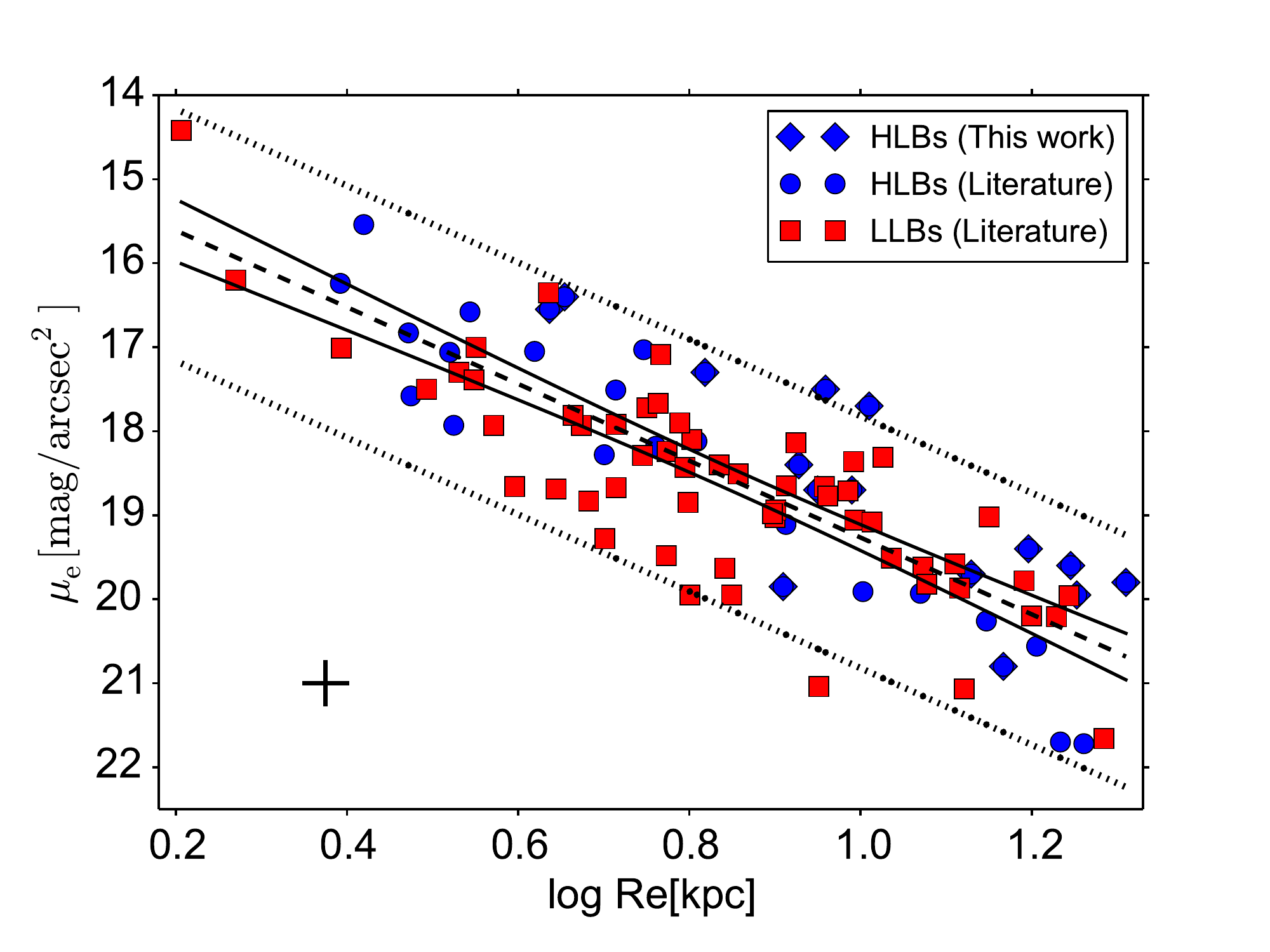}
\caption{The Kormendy relation (symbols are explained in the figure). The effective radius (log $R_{e}$) is plotted versus the surface brightness at that radius ($\mu_{e}$). A statistically significant correlation is found between these parameters (Kendall rank correlation, $\tau=0.63,$ $p=2.2\times10^{-16}$). We show the 95\% prediction bands (dotted lines) and the 95\% confidence intervals (solid lines). A typical error bar is shown in the lower left corner. The best linear fit (segmented line) relation obtained is $\mu_{e}=(4.57\pm0.35)logR_{e}+(14.69\pm0.31)$.} 
\label{fig:kormendy}
\end{figure}
%%%%%%%%%%%%%%%%%%
%%%%%%%%%%%%%%%%%%
%%%%%%%%%%%%%%%%%%
%%%%%%%%%%%%%%%%%%
%%%%%%%%%%%%%%%%%%%%%%%%%%%%%%%%%%%%%%%%%%%%
%%%%%%%%%%%%%%%%%%%%%%%%%%%%%%%%%%%%%%%%%%%%
%%%%%%%%%%%%%%%%%%%%%%%%%%%%%%%%%%%%%%%%%%%%

%%%%%%%%%%%%%%%%%%%%%
%%%%%%%%%%%%%%%%%%%%%

\begin{table*}
\centering
 \begin{minipage}{155mm}
\caption{Comparison of average host galaxy properties between different sub-samples of this work.}
\begin{tabular}{l l l l l l}
\hline
\multicolumn{1}{|c|}{Sub--sample}\footnote{Bulge and nuclear colour transformations are shown in section \ref{sec:sample}} & 
\multicolumn{1}{|c|}{$M_{K,bulge}$} & 
\multicolumn{1}{|c|}{$M_{K,nuclear}$} &
\multicolumn{1}{|c|}{$R_{eff}$}&
\multicolumn{1}{|c|}{$\mu_K$}&
\multicolumn{1}{|c|}{$logL_{1.4GHz}$}\\
\multicolumn{1}{|c|}{} & 
\multicolumn{1}{|c|}{} & 
\multicolumn{1}{|c|}{} &
\multicolumn{1}{|c|}{($kpc$)}&
\multicolumn{1}{|c|}{($mag/arcsec^2$)}&
\multicolumn{1}{|c|}{($WHz^{-1}$)}\\
\multicolumn{1}{|c|}{(1)} & 
\multicolumn{1}{|c|}{(2)} & 
\multicolumn{1}{|c|}{(3)} &
\multicolumn{1}{|c|}{(4)}&
\multicolumn{1}{|c|}{(5)}&
\multicolumn{1}{|c|}{(6)}\\
\hline
\multicolumn{6}{|c|}{HLBs}\\
\hline
All							& $-26.40\pm1.21$ 			&		$-28.20\pm1.94$ 			&	$9.34\pm5.52$     &$18.46\pm1.65$ &$27.05\pm0.53$  \\
This work 						& $ -26.20\pm 0.90$ 			&       $ -28.07\pm 1.59$   		&	$ 11.30\pm4.84$   &$18.59\pm1.49$&$27.14\pm0.45$	\\
Literature 					& $-26.52\pm1.33$ 			&       $-28.27\pm2.10$			&	$8.29\pm5.59$	  &$18.36\pm1.75$  &	$27.03\pm0.54$\\
NIR (This work $\&$ Literature) 	&$ -26.84\pm1.09$ 			&       $ -28.95\pm1.63$ 		&	$8.96\pm5.19$	  &$18.31\pm1.37$& $27.18\pm0.41$\\
NIR (Literature)  			& $-27.49\pm 0.85$			&       $ -29.83\pm1.10$ 		&	$ 6.63 \pm 4.42$	  &$17.82\pm0.95$	& $27.23\pm0.36$ \\
Optical (Literature) 		& $-25.40\pm 0.80$			& 	    $ -26.45 \pm 1.40$		&	$ 10.21\pm 6.16$	  &$18.79\pm2.10$	&$26.81\pm0.62$\\
\hline
\multicolumn{6}{|c|}{LLBs (all from literature)}\\
\hline
All						&		$-25.56\pm0.58$ 		  &		$-25.45\pm1.23$ 			&		$7.90\pm4.10$      &$18.65\pm1.24$ &  $24.81\pm0.52$  \\
Optical					&		$-25.48\pm0.53$ 		  &		$-25.46\pm1.23$ 			&		$8.47\pm4.06$      &$18.80\pm1.22$ & $24.83\pm0.51$  \\
NIR						&		$-26.19\pm0.57$ 		  &		$-25.32\pm1.25$ 			&		$3.83\pm1.31$      &$17.55\pm0.78$ &  $24.62\pm0.52$  \\
\hline
\end{tabular}
\label{table:comparison}

Column (1) gives the sub--sample analysed; 
(2) the average absolute $K-$ band bulge magnitude; 
(3) the average absolute $K-$ band nuclear magnitude; 
(4) the average effective radius in kpc;
(5) the surface brightness at the effective radius;
(6) the average 1.4GHz luminosity of the sub--sample.
\end{minipage}
\end{table*}

We successfully detect the host galaxy in 79\% of our sample, being all best fitted by a single component model; a bulge, represented by the S\'ersic profile. For the 4 unresolved sources, we estimated upper limits following the method described in \citet{kotilainen_2007}, assuming an elliptical galaxy ($n=4$) with a typical size of $R_{eff}= 10$ kpc. The best fit 2D surface brightness decomposition is shown for each target in the Appendix (Figure \ref{figureA1}), and the best fit model parameters are summarized in Table \ref{table:galfit}.\smallskip

We estimate an average S\'ersic index $<n>=3.47\pm0.38$ for the host galaxies in our sample, with average and median $K$--band host galaxy magnitudes $<M(K)_{bulge}> _{This work}= -26.20~\pm~0.90$ and $<M(K)_{bulge}> _{This work}= -26.34~\pm~0.90$, respectively. Our imagery analysis is in consistency with previous studies \cite[][and references therein]{falomo_2014}, where it is suggested that the most common type of galaxy hosting radio--loud AGN are bright, giant ellipticals.\smallskip

It is known that elliptical galaxies and bulges follow a tight inverse relation between $\mu_{e}$ and $R_{e}$ known as the Kormendy relation \citep{kormendy_1977}. This relation is explored in Figure~\ref{fig:kormendy} for the combined sample of radio-loud AGN host galaxies. As it can be seen, the overall sample follows the Kormendy relation (Kendall rank correlation, $\tau=0.63,$ $p=2.2\times10^{-16}$) and the best-fit linear relation obtained is $\mu_{e}=(4.57\pm0.35)logR_{e}+(14.69\pm0.31)$, see dashed-line in Figure~\ref{fig:kormendy}. \smallskip 

The later expression is consistent with those derived in $K$-band for BL Lacs \citep{cheung_2003,kotilainen_2005} and inactive elliptical galaxies \citep[e.g.][]{pahre_1995}. Figure~\ref{fig:kormendy}, thus suggests that blazar host galaxies resemble those of inactive elliptical galaxies, in terms of dynamics and structural parameters. This comes in-line with previous findings, see \cite[][and references therein]{falomo_2014}.\smallskip

%%%%%%%%%%%%%%%%%%%%%%%%%%%%%%%%%%%%%%%%%%%%
%%%%%%%%%%%%%%%%%%%%%%%%%%%%%%%%%%%%%%%%%%%5

\begin{figure}
\begin{tabular}{l}
	\includegraphics*[width=0.49\textwidth]{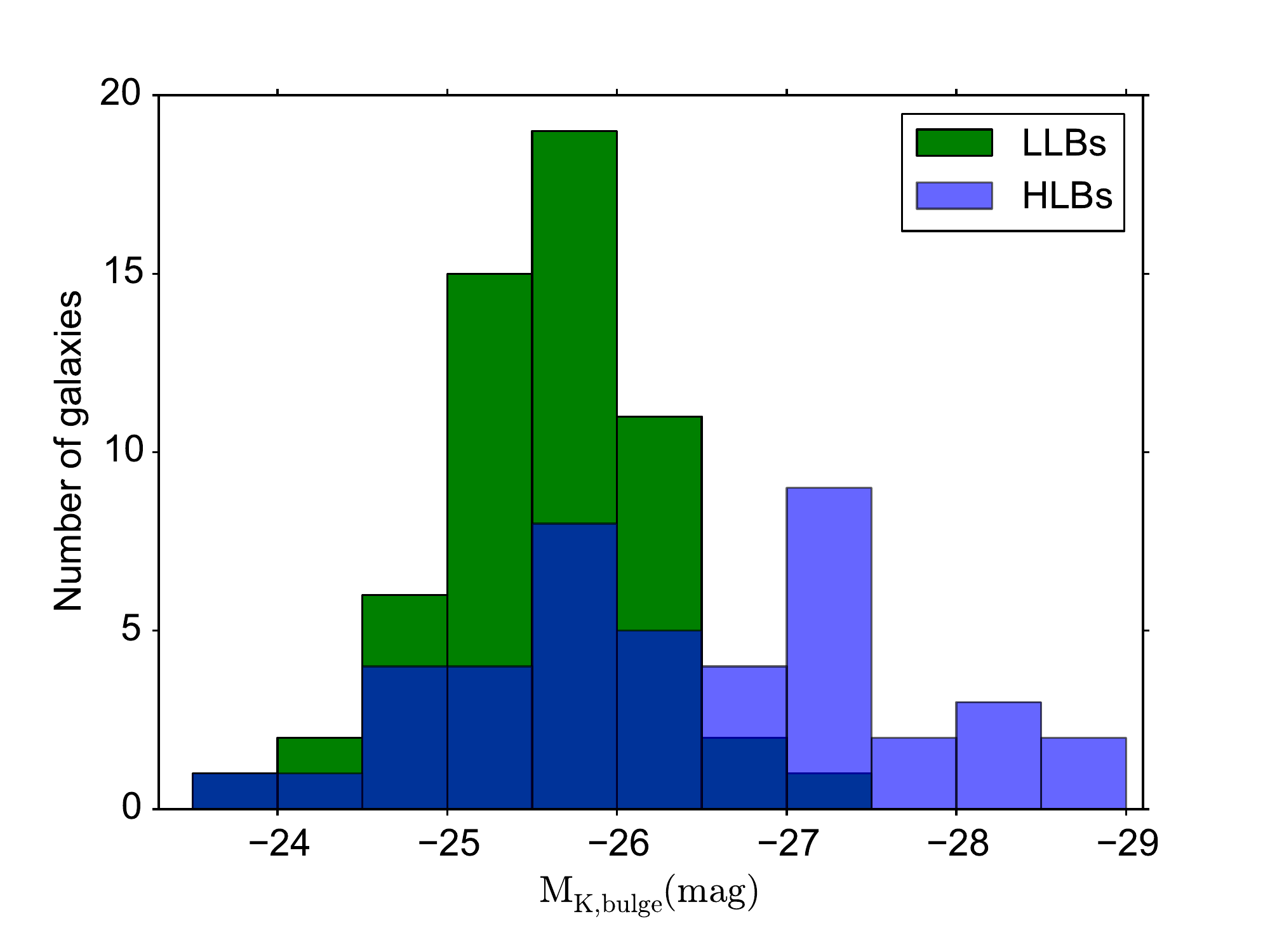}\\
	\includegraphics*[width=0.49\textwidth]{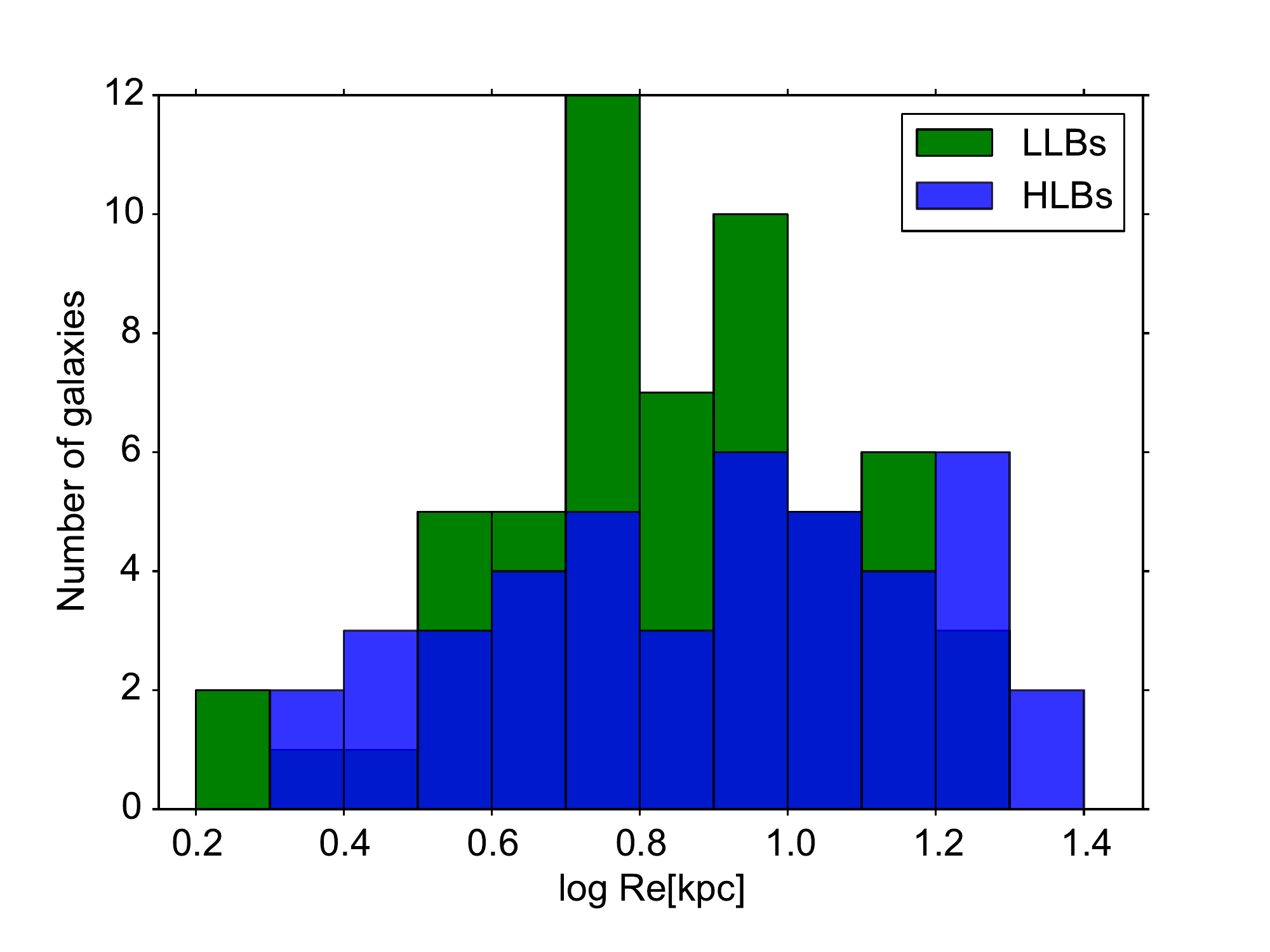} \\
   \end{tabular}
  \caption{Distributions of the K-band bulge magnitudes (upper panel) and bulge effective radius ($R_{e}$) (lower panel) for the LLBs and HLBs. A K-S test shows that the distribution of bulge magnitude of the LLBs and HLBs are different, whereas the effective radii are drawn from the same parent population.}
\label{fig:Mhost_re}
\end{figure}
%%%%%%%%%%%%%%%%%%%%%%%%%%%%%%%%%%%%%%%%%%%%%%%%%%%%%%%%%%%%%%
%%%%%%%%%%%%%%%%%%%%%%%%%%%%%%%%%%%%%%%%%%%%%%%%%%%%%%%%%%%%%%%%

In Figure~\ref{fig:Mhost_re} we investigate the difference between galaxies hosting LLBs and HLBs. The bulge magnitude distribution (top panel) for LLBs appears to be narrow, with an average and median $K$--band magnitude $<M(K)_{bulge}> = -25.56~\pm~0.58$ and $<M(K)_{bulge}>= -25.57~\pm~0.58$, respectively. On the other hand, HLBs span a wide range of bulge magnitudes, with an average and median of $<M(K)_{bulge}> = -26.40~\pm~1.21$ and $<M(K)_{bulge}> = -26.12~\pm~1.21$, respectively. Although the  distributions  of bulge magnitudes for LLBs and HLBs, seem to overlap over a of host galaxy magnitudes ($\sim~3~mag$), the two distributions however, are  significantly different ($P = 1\times10^{-4}$, obtained by a  Kolmogorov-Smirnov test). Thus, suggesting an intrinsic relation between the bulge magnitude and the central engine mode of the AGN nuclei.\smallskip

Since our division (LLBs and HLBs) is associated to the power of the AGN, then the latter result can be interpreted also as a close connection between the bulge magnitude and the power of the jet  that is  digging its way out of the host galaxy. We discuss this finding with more detail in Section~\ref{sec:agn}.\smallskip

The distributions of effective radius for bulges hosting LLBs and HLBs are displayed in lower panel of Figure~\ref{fig:Mhost_re}. While the average effective radius for LLBs is $<R_e>= 7.90~\pm~4.10$ and the median is $<R_e>= 6.83~\pm~4.10$, the average effective radius for HLBs is $<R_e>= 9.34~\pm~5.52$ and the median is $<R_e>= 8.18~\pm~5.52$. It appears that both distributions are indistinguishable in terms of bulge size. This perception is confirmed by a Kolmogorov-Smirnov test  ($P=0.41$). Thus, suggesting that there is no difference in bulge size between LLBs and HLBs. Nevertheless, some caution might be exercised when interpreting the above result. This obeys  to the fact that estimations of effective radius have been compiled from literature and those imaging studies have been conducted at different band filters.  Whereas previous studies  have not found an effective radius of BL Lacs bulges dependence with wavelength, some trend has been noticed by \citet{Hyvonen_2007} in the sense that effective radius increases towards shorter wavelengths.\smallskip

A summary of the host galaxy properties for the different sub--samples in this work is shown in table \ref{table:comparison}. The combined sample (this work and compiled sample) of blazar host galaxies contains 57 LLBs and 43 HLBs (previously classified as 78 BL Lacs and 22 FSRQ, of which 15 new host galaxy detections were obtained in this work, all of them classified as FSRQ) and a broad range of radio luminosities  $\log L_{1.4GHz} = 23.7 - 28.3$ W~Hz$^{-1}$ (represented in column 6), allowing for an investigation on whether host galaxy properties are connected with the properties of the radio jet.

%%%%%%%%%%%%%%%%%%%%%%%%%%%%%5%%%%%%%%%%%%%%%%%%%%%%%%%%%%%%%%%%%%%
%%%%%%%%%%%%%%%%%%%%%%%%%%%%%5%%%%%%%%%%%%%%%%%%%%%%%%%%%%%%%%%%%%%
%%%%%%%%%%%%%%%%%%%%%%%%%%%%%%%%%%%%%%%%%%%%%%%%%%%%%%%%%%%%%%%%%%%%%%%%5
%%%%%%%%%%%%%%%%%%%%%%%%%%%%%%%%%%%%%%%%%%%%%%%%%%%%%%%%%%%%%%%%%%%%%%%%5
\begin{table}
\begin{center}
\begin{tabular}{|*{3}{c|}}
\hline
\multicolumn{3}{|c|}{For HLBs (43 elements)}\\
\hline
&$\tau$ & $p$\\
\hline
\multicolumn{1}{|l|}{$z-M_{K,bulge}$}			& $0.40$ 	& $9.2\times10^{-5}$\\
\multicolumn{1}{|l|}{$z-M_{K,nuclear}$}			& $0.37$	& $3.0\times10^{-4}$\\
\multicolumn{1}{|l|}{$M_{K,bulge}-M_{K,nuclear}$}		&$0.53$		& $6.6\times10^{-7}$  \\
\multicolumn{1}{|l|}{$m_{K,bulge}-m_{K,nuclear}$}		& $0.53$	& $8.3\times10^{-7}$  \\
\hline
\multicolumn{3}{|c|}{For HLBs with  $z<0.6$ (20 elements)}\\
\hline
\multicolumn{1}{|l|}{$z-M_{K,bulge}$}	& $0.12$& \multicolumn{1}{|c|}{$0.50$}\\
\multicolumn{1}{|l|}{$z-M_{K,nuclear}$	}	& $0.04$& \multicolumn{1}{|c|}{$0.82$}\\
\multicolumn{1}{|l|}{$M_{K,bulge}-M_{K,nuclear}$}&$0.76$& \multicolumn{1}{|c|}{$5.6\times10^{-6}$}  \\
\multicolumn{1}{|l|}{$m_{K,bulge}-m_{K,nuclear}$}&$0.73$& \multicolumn{1}{|c|}{$1.3\times10^{-5}$} \\
\hline
\end{tabular}
\caption{Correlation tests between $z$, $M_{K,bulge}$ and $M_{K,nuclear}$ for the complete sample of HLBs and for the subsample of HLBs with $z<0.6$, where the $z-M_{K,bulge}$ and $z-M_{K,nuclear}$ correlations disappear, suggesting that the $M_{K,bulge}-M_{K,nuclear}$ correlation is not significantly affected by selection effects. We show the Kendall rank correlation coefficient $\tau$ and the probability that the correlation is given by chance $p$. For correlations between magnitudes we show partial Kendall rank correlation tests in order to remove their common dependence on distance. We consider a correlation statistically significant when $p<0.05$. All magnitudes have been transformed to K--band assuming the colours in section \ref{sec:sample}}
\label{table:correlations}
\end{center}
\end{table}

\subsection{Host-galaxy -AGN connection}\label{sec:agn}
%%%%%%%%%%%%%%%%%%%%%%%%%%%%%%%%%%%%%%%%%%%%%%%%%%%%%%%%%%%%%%%
%%%%%%%%%%%%%%%%%%%%%%%%%%%%%%%%%%%%%%%%%%%%%%%%%%%%%%%%%%%%%%%%

\begin{figure*}
\begin{tabular}{ll}
	\includegraphics[width=1\textwidth,center]{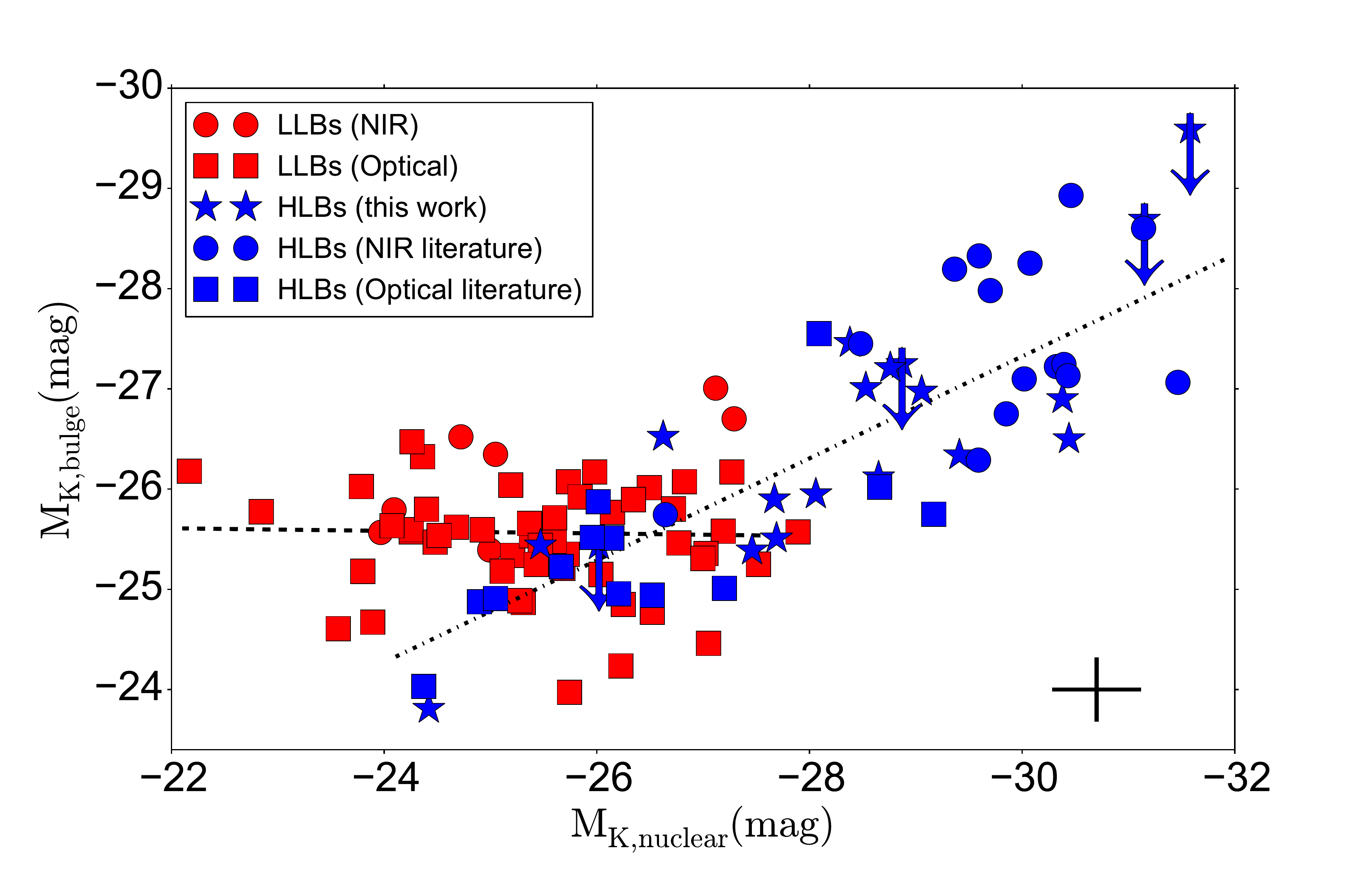}\\
   \end{tabular}
  \caption{Plot of the nuclear K-band magnitude versus the bulge K-band magnitude (symbols are explained in the figure). We show the best linear fits for LLBs and HLBs (dashed and dotdashed lines, respectively). A statistically significant partial correlation ($\tau=0.53$, $p=6.6\times10^{-7}$) is found for HLBs. Upper limits for unresolved galaxies analyzed in this work are shown as down arrows. A typical error bar is shown in the lower right corner.}
\label{fig:host_agn}
\end{figure*}
%%%%%%%%%%%%%%%%%%%%%%%%%%%%%%%%%%%%%%%%%%%%%%%%%%%%%%%%%%%%%%
%%%%%%%%%%%%%%%%%%%%%%%%%%%%%%%%%%%%%%%%%%%%%%%%%%%%%%%%%%%%%%

Figure~\ref{fig:host_agn} displays the $M_{K,bulge}$ - $M_{K,nuclear}$ relation for the galaxies in the combined sample. Bulge and nuclear magnitudes were transformed to K--band assuming the colours presented in section \ref{sec:sample}.\smallskip

For the bulges of LLBs (red symbols in Figure \ref{fig:host_agn}), we observe a narrow range of $M_{K,bulge}$ ($M_{K,bulge}= -25.56\pm0.58$) which is consistent with the standard candle assumption \citep[][and references therein find a typical value $M_{K,bulge}\sim-25.7$]{urry_2000,falomo_2014} made on imaging redshift estimation studies, widely applied to BL Lac objects \citep{romanishin_1987,falomo_1996,falomo_1999,sbarufatti_2005, treves_2007,nilsson_2008,meisner_2010,kotilainen_2011,nilsson_2012,stadnik_2014}.\smallskip

For black holes accreting matter through ADAFs, the bulk of energy is expected to be emitted in the form of radio jets which, even when they are less powerful than jets from black holes accreting matter through accretion discs \citep{Cattaneo_2009}, produce a strong feedback mechanism capable of declining star formation \citep{best_heckman_2012}. Thus, in the $M_{nuclear}$ - $M_{bulge}$ plot, the nuclear luminosity should increase while the bulge luminosity decreases.\smallskip 

The slightly negative trend for LLBs in this work (red symbols of Figure \ref{fig:host_agn}) might hint to the latter picture. However, the slope is almost flat (although negative) and the correlation absent.\smallskip

\begin{figure}
\begin{tabular}{l}
	\includegraphics*[width=0.52 \textwidth,center]{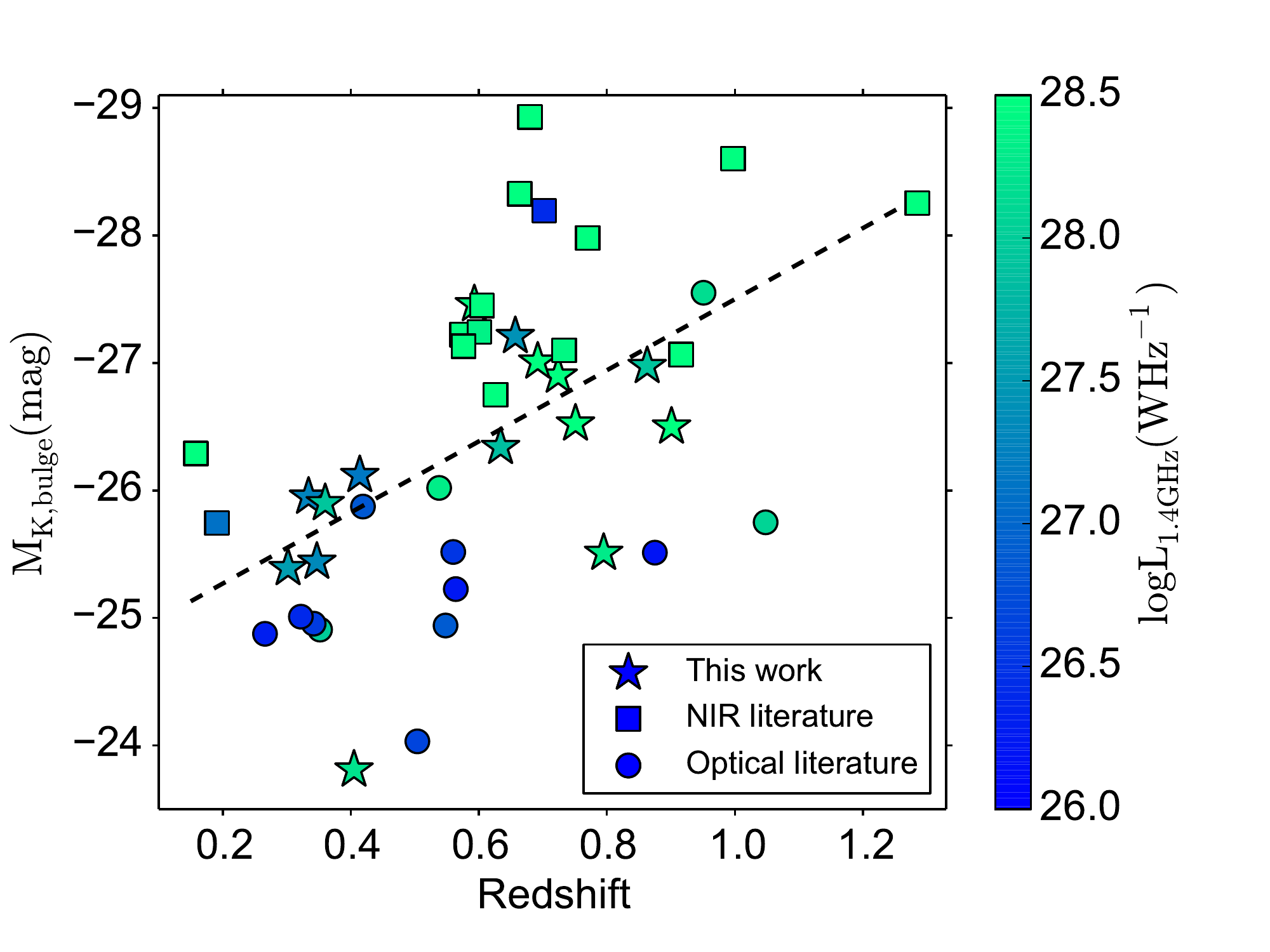}\\
	\includegraphics*[width=0.52\textwidth,center]{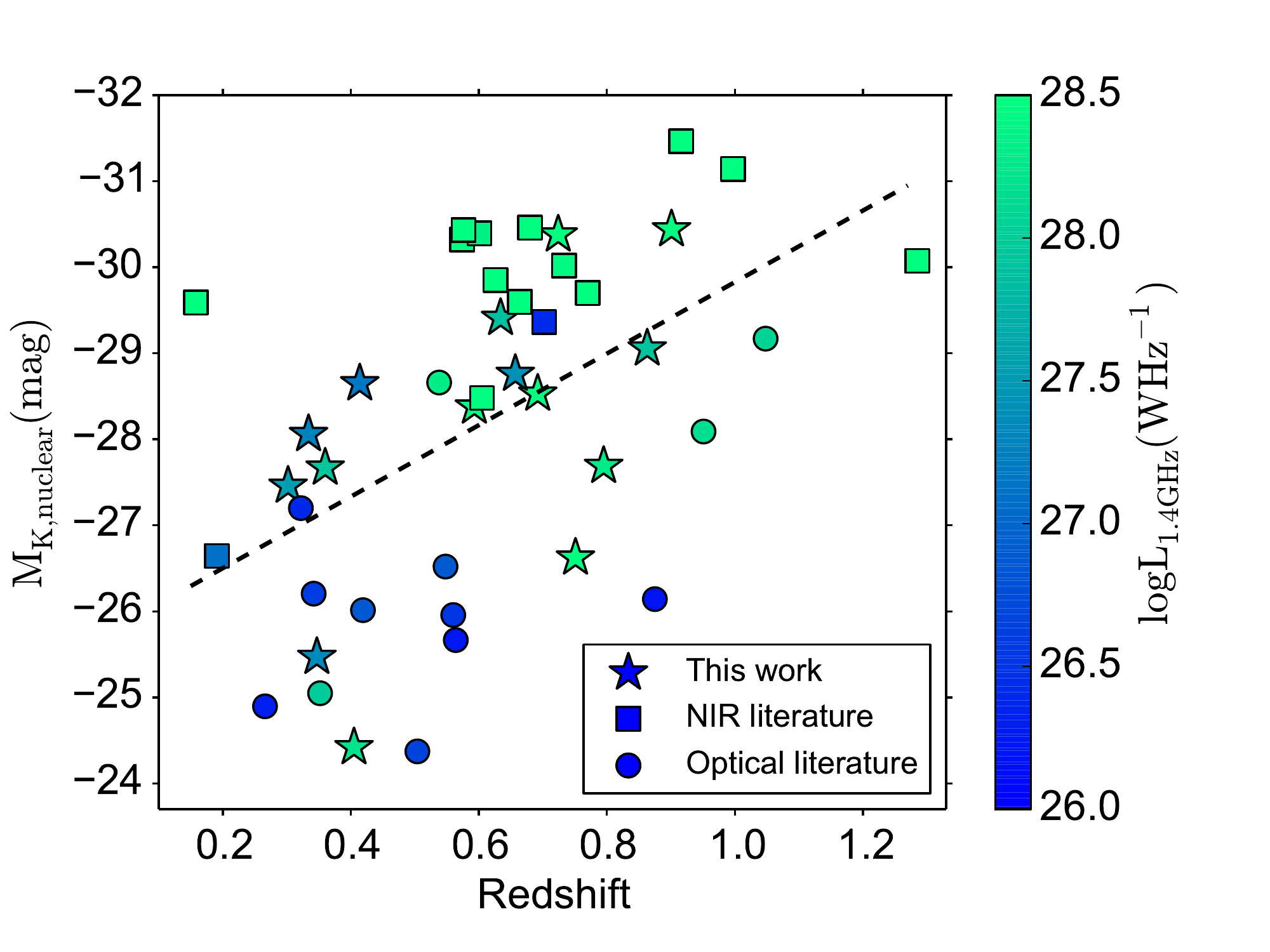} \\
    \end{tabular}
    \caption{Plot of the absolute $K-$ band magnitude of the bulge versus redshift (top panel) and absolute $K-$ band magnitude of the nucleus versus redshift (lower panel) for HLBs. The colour bars represent the 1.4GHz luminosities. Symbols are explained in the figure. The dashed lines are the best linear fits in each plot. It can be seen that larger radio luminosities are linked to brighter nucleus and bulges.}
\label{fig:z_host-nuc}
\end{figure}

In the other hand, for HLBs (blue symbols in Figure \ref{fig:host_agn}), a correlation between \mnuc and \mbulge arises from a visual inspection, and is confirmed with a partial correlation test (see table \ref{table:correlations}). It is pertinent to ask whether the correlation gleaned might be induced due to selection effects (if only the brightest bulges can be detected at high redshifts). To address this, in Figure \ref{fig:z_host-nuc} we plot the redshift against the bulge magnitude (top panel) and the redshift against the nuclear magnitude (lower panel). A statistically significant correlation, albeit with a relatively large scatter, is found for these parameters (see table \ref{table:correlations}). The three samples occupy different locations in the plot, depending on their radio luminosities (represented by a colour bar), which hints at a bulge--nucleus connection.\smallskip

If an observational bias effect was responsible to induce an artificial $M_{nuclear}-M_{bulge}$ correlation, then we should observe this correlation only for bright nuclei (where a faint bulge might have not been detected). However, the correlation holds for the complete sample of HLBs, from faint to bright, even for galaxies with very faint nuclei ($M_{nucleus}\sim-24$), where a bulge of any brightness would be easily detected. We further investigate this in table \ref{table:correlations}, where we perform correlation tests for a sub--sample of the nearest HLBs ($z<0.6$), where no selection effects are significant. The correlations between $z-M_{bulge}$ and $z-M_{nuclear}$ vanish for this sub--sample. However the $M_{nuclear}-M_{bulge}$ (and $m_{nuclear}-m_{bulge}$) correlation remain, which suggests that the $M_{nuclear}-M_{bulge}$ correlation is not driven by selection effects.\smallskip

%%%%%%%%%%%%%%%%%%%%%%%%%%%%%%%%%%%%%%%%%%%%%%%%%%%%%%%%%%%%%%%
%%%%%%%%%%%%%%%%%%%%%%%%%%%%%%%%%%%%%%%%%%%%%%%%%%%%%%%%%%%%%%%%

%%%%%%%%%%%%%%%%%%%%%%%%%%%%%%%%%%%%%%%%%%%%%%%%%%%%%%%%%%%%%%
%%%%%%%%%%%%%%%%%%%%%%%%%%%%%%%%%%%%%%%%%%%%%%%%%%%%%%%%%%%%%%%
%%%%%%%%%%%%%%%%%%%%%%%%%%%%%%%%%%%%%%%%%%%%%%%%%%%%%%%%%%%%%%%%

The correlation between nuclear and bulge magnitudes is consistent with a scenario where AGN activity results in positive feedback on the star formation rate \citep{silk_2005, silk_2009,silk_2013}, wherein the more powerful the jet, the more significant the effect caused on its host galaxy. The influence of the jet could induce an increment (or quenching) of star formation leading to an enhanced (or diminished) density of old stellar population contributing to luminous (or faint) bulges \citep{granato_2004,scannapieco_2004,antonuccio_2008}.

It is worth noting that the overall observed behaviour in Figure \ref{fig:host_agn} shows similarities with the semi-analytical model presented in \citet[][see also \citealt{hickox_2014}]{gutcke_2015}. In this model, a slightly negative trend between the bolometric AGN luminosity and 60$\mu m$ FIR emission (purely reprocessed stellar emission by dust) is observed for galaxies hosting low--luminosity AGN (bolometric luminosities below $10^{43}$ ergs/s) with black holes accreting hot gas, where gas cooling is being suppressed by AGN jets (jet-mode AGN feedback). In the other hand, a positive trend is observed for high--luminosity AGN (bolometric luminosities above $10^{43}$ ergs/s), with black holes accreting cold gas, being mergers or disc instabilities the triggering mechanisms of cold-gas flows, which in turn increase starburst activity.\smallskip

\section{Summary}

The $J-$ and $H-$band images of radio-loud AGN, presented here, has yielded new host galaxy detections in AGN with prominent relativistic jets (i.e. $L_{1.4GHz} >10^{27}$ WHz$^{-1}$). We compile host galaxy and nuclear parameters for blazars from literature and, combined together with our new host galaxy detections, yields to a sample of 100 radio-loud AGN with host galaxy detections and a broad range of radio luminosities $L_{1.4GHz} \sim 10^{24} - 10^{29}$ WHz$^{-1}$. Our sample is divided into low--luminosity (57\%) and high--luminosity blazars (43\%), allowing an investigation of the correlation between their central engine modes and the properties of their host galaxies. Our main findings are summarized below:

\begin{enumerate}[1.]

\item The host galaxy imaging survey presented here yielded a total of 15 host galaxy detections out of 19 radio-loud AGN. This study has increased the number of detected and resolved FSRQ host galaxies by a factor of 2 \citep{kotilainen_1998, nilsson_2009}.\bigskip

\item The results from our 2D modeling of the surface brightness of the detected host galaxies in our sample is consistent with previous findings \citep[][and references therein]{falomo_2014}, namely that radio-loud AGN with prominent relativistic jets (BL Lacs and FSRQ) are hosted by luminous $ <M_{K,host}> \sim -26$ and bulge dominated $n \sim 4 $ galaxies that follow the $\mu_{e}$-$R_{eff}$ relation for ellipticals and classical bulges.\bigskip

\item When plotting $M_{nuclear}$ versus $M_{bulge}$, LLBs and HLBs follow different behaviours. While LLBs cover a narrow range of magnitudes, HLBs follow a statistically significant positive correlation. Such correlation could be interpreted on the context of AGN feedback, wherein the more powerful the jet, the more significant the effect caused on its host galaxy. The influence of the jet could induce an increment (or quenching) of star formation leading to an enhanced (or diminished) density of old stellar population contributing to a luminous (or faint) bulge.\bigskip
\end{enumerate}\bigskip
\noindent
\textbf{ACKNOWLEDGEMENTS}\smallskip

We thank the anonymous referee for her/his thorough review and constructive comments, Kari Nilsson for his contribution on the preparation of the observing proposal and Anlaug Amanda Djupvik for her assistance with the NOT observations. This research is based on observations made with the Nordic Optical Telescope, operated by the Nordic Optical Telescope Scientific Association at the Observatorio del Roque de los Muchachos, La Palma, Spain, of the Instituto de Astrof\'isica de Canarias. We acknowledge support by CONACyT research grant 151494 (Mexico) and CONACyT program for PhD studies.

\bibliographystyle{mn2e} 
\bibliography{ms_27aug15}

\begin{thebibliography}{73}
\expandafter\ifx\csname natexlab\endcsname\relax\def\natexlab#1{#1}\fi

\bibitem[{{Antonucci}(1993)}]{antonucci_1993}
{Antonucci} R., 1993, araa, 31, 473

\bibitem[{{Antonuccio-Delogu} \& {Silk}(2008)}]{antonuccio_2008}
{Antonuccio-Delogu} V., {Silk} J., 2008, mnras, 389, 1750

\bibitem[{{Bertin} \& {Arnouts}(1996)}]{bertin_1996}
{Bertin} E., {Arnouts} S., 1996, A\&ASS, 117, 393

\bibitem[{{Best} \& {Heckman}(2012{\natexlab{a}})}]{best_2012}
{Best} P.~N., {Heckman} T.~M., 2012{\natexlab{a}}, MNRAS, 421, 1569

\bibitem[{{Best} \& {Heckman}(2012{\natexlab{b}})}]{best_heckman_2012}
{Best} P.~N., {Heckman} T.~M., 2012{\natexlab{b}}, mnras, 421, 1569

\bibitem[{{Bower} {et~al}\mbox{.}(2006){Bower}, {Benson}, {Malbon}, {Helly},
  {Frenk}, {Baugh}, {Cole}, \& {Lacey}}]{bower_2006}
{Bower} R.~G., {Benson} A.~J., {Malbon} R., {Helly} J.~C., {Frenk} C.~S.,
  {Baugh} C.~M., {Cole} S., {Lacey} C.~G., 2006, MNRAS, 370, 645

\bibitem[{{Buttiglione} {et~al}\mbox{.}(2010){Buttiglione}, {Capetti},
  {Celotti}, {Axon}, {Chiaberge}, {Macchetto}, \& {Sparks}}]{buttiglione_2010}
{Buttiglione} S., {Capetti} A., {Celotti} A., {Axon} D.~J., {Chiaberge} M.,
  {Macchetto} F.~D., {Sparks} W.~B., 2010, A\&A, 509, A6

\bibitem[{{Cattaneo} \& {Best}(2009)}]{Cattaneo_2009}
{Cattaneo} A., {Best} P.~N., 2009, mnras, 395, 518

\bibitem[{{Cheung} {et~al}\mbox{.}(2003){Cheung}, {Urry}, {Scarpa}, \&
  {Giavalisco}}]{cheung_2003}
{Cheung} C.~C., {Urry} C.~M., {Scarpa} R., {Giavalisco} M., 2003, ApJ, 599, 155

\bibitem[{{Chilingarian} {et~al}\mbox{.}(2010){Chilingarian}, {Melchior}, \&
  {Zolotukhin}}]{chilingarian_2010}
{Chilingarian} I.~V., {Melchior} A.-L., {Zolotukhin} I.~Y., 2010, MNRAS, 405,
  1409

\bibitem[{{Chilingarian} \& {Zolotukhin}(2012)}]{chilingarian_2012}
{Chilingarian} I.~V., {Zolotukhin} I.~Y., 2012, MNRAS, 419, 1727

\bibitem[{{Croton} {et~al}\mbox{.}(2006){Croton}, {Springel}, {White}, {De
  Lucia}, {Frenk}, {Gao}, {Jenkins}, {Kauffmann}, {Navarro}, \&
  {Yoshida}}]{croton_2006}
{Croton} D.~J. {et~al.}, 2006, MNRAS, 365, 11

\bibitem[{{Emonts} {et~al}\mbox{.}(2005){Emonts}, {Morganti}, {Tadhunter},
  {Oosterloo}, {Holt}, \& {van der Hulst}}]{emonts_2005}
{Emonts} B.~H.~C., {Morganti} R., {Tadhunter} C.~N., {Oosterloo} T.~A., {Holt}
  J., {van der Hulst} J.~M., 2005, MNRAS, 362, 931

\bibitem[{{Fabian}(2012)}]{fabian_2012}
{Fabian} A.~C., 2012, ARA\&A, 50, 455

\bibitem[{{Falomo}(1996)}]{falomo_1996}
{Falomo} R., 1996, MNRAS, 283, 241

\bibitem[{{Falomo} \& {Kotilainen}(1999)}]{falomo_1999}
{Falomo} R., {Kotilainen} J.~K., 1999, A\&A, 352, 85

\bibitem[{{Falomo} {et~al}\mbox{.}(2014){Falomo}, {Pian}, \&
  {Treves}}]{falomo_2014}
{Falomo} R., {Pian} E., {Treves} A., 2014, A\&ARv, 22, 73

\bibitem[{{Falomo} {et~al}\mbox{.}(2000){Falomo}, {Scarpa}, {Treves}, \&
  {Urry}}]{falomo_2000}
{Falomo} R., {Scarpa} R., {Treves} A., {Urry} C.~M., 2000, ApJ, 542, 731

\bibitem[{{Ferrarese} \& {Merritt}(2000)}]{ferrarese_merrit_2000}
{Ferrarese} L., {Merritt} D., 2000, ApJL, 539, L9

\bibitem[{{Gebhardt} {et~al}\mbox{.}(2000){Gebhardt}, {Bender}, {Bower},
  {Dressler}, {Faber}, {Filippenko}, {Green}, {Grillmair}, {Ho}, {Kormendy},
  {Lauer}, {Magorrian}, {Pinkney}, {Richstone}, \& {Tremaine}}]{gebhardt_2000}
{Gebhardt} K. {et~al.}, 2000, ApJL, 539, L13

\bibitem[{{Giommi} {et~al}\mbox{.}(2013){Giommi}, {Padovani}, \&
  {Polenta}}]{giommi_2013}
{Giommi} P., {Padovani} P., {Polenta} G., 2013, MNRAS, 431, 1914

\bibitem[{{Giommi} {et~al}\mbox{.}(2012){Giommi}, {Padovani}, {Polenta},
  {Turriziani}, {D'Elia}, \& {Piranomonte}}]{giommi_2012}
{Giommi} P., {Padovani} P., {Polenta} G., {Turriziani} S., {D'Elia} V.,
  {Piranomonte} S., 2012, MNRAS, 420, 2899

\bibitem[{{Graham} \& {Driver}(2005)}]{graham_2005}
{Graham} A.~W., {Driver} S.~P., 2005, PASP, 22, 118

\bibitem[{{Granato} {et~al}\mbox{.}(2004){Granato}, {De Zotti}, {Silva},
  {Bressan}, \& {Danese}}]{granato_2004}
{Granato} G.~L., {De Zotti} G., {Silva} L., {Bressan} A., {Danese} L., 2004,
  apj, 600, 580

\bibitem[{{Greene} {et~al}\mbox{.}(2008){Greene}, {Ho}, \&
  {Barth}}]{greene_2008}
{Greene} J.~E., {Ho} L.~C., {Barth} A.~J., 2008, ApJ, 688, 159

\bibitem[{{G{\"u}ltekin} {et~al}\mbox{.}(2009){G{\"u}ltekin}, {Richstone},
  {Gebhardt}, {Lauer}, {Tremaine}, {Aller}, {Bender}, {Dressler}, {Faber},
  {Filippenko}, {Green}, {Ho}, {Kormendy}, {Magorrian}, {Pinkney}, \&
  {Siopis}}]{gultekin_2009}
{G{\"u}ltekin} K. {et~al.}, 2009, ApJ, 698, 198

\bibitem[{{Gutcke} {et~al}\mbox{.}(2015){Gutcke}, {Fanidakis}, {Macci{\`o}}, \&
  {Lacey}}]{gutcke_2015}
{Gutcke} T.~A., {Fanidakis} N., {Macci{\`o}} A.~V., {Lacey} C., 2015, mnras,
  451, 3759

\bibitem[{{Heckman} \& {Best}(2014)}]{heckman_2014}
{Heckman} T.~M., {Best} P.~N., 2014, ARA\&A, 52, 589

\bibitem[{{Heidt} {et~al}\mbox{.}(2004){Heidt}, {Tr{\"o}ller}, {Nilsson},
  {J{\"a}ger}, {Takalo}, {Rekola}, \& {Sillanp{\"a}{\"a}}}]{heidt_2004}
{Heidt} J., {Tr{\"o}ller} M., {Nilsson} K., {J{\"a}ger} K., {Takalo} L.,
  {Rekola} R., {Sillanp{\"a}{\"a}} A., 2004, A\&A, 418, 813

\bibitem[{{Hickox} {et~al}\mbox{.}(2014){Hickox}, {Mullaney}, {Alexander},
  {Chen}, {Civano}, {Goulding}, \& {Hainline}}]{hickox_2014}
{Hickox} R.~C., {Mullaney} J.~R., {Alexander} D.~M., {Chen} C.-T.~J., {Civano}
  F.~M., {Goulding} A.~D., {Hainline} K.~N., 2014, apj, 782, 9

\bibitem[{{Hovatta} {et~al}\mbox{.}(2009){Hovatta}, {Valtaoja}, {Tornikoski},
  \& {L{\"a}hteenm{\"a}ki}}]{hovatta_2009}
{Hovatta} T., {Valtaoja} E., {Tornikoski} M., {L{\"a}hteenm{\"a}ki} A., 2009,
  A\&A, 494, 527

\bibitem[{{Hyv{\"o}nen} {et~al}\mbox{.}(2007){Hyv{\"o}nen}, {Kotilainen},
  {Falomo}, {{\"O}rndahl}, \& {Pursimo}}]{Hyvonen_2007}
{Hyv{\"o}nen} T., {Kotilainen} J.~K., {Falomo} R., {{\"O}rndahl} E., {Pursimo}
  T., 2007, A\&A, 476, 723

\bibitem[{{Kormendy}(1977)}]{kormendy_1977}
{Kormendy} J., 1977, ApJ, 218, 333

\bibitem[{{Kotilainen} {et~al}\mbox{.}(2007){Kotilainen}, {Falomo}, {Labita},
  {Treves}, \& {Uslenghi}}]{kotilainen_2007}
{Kotilainen} J.~K., {Falomo} R., {Labita} M., {Treves} A., {Uslenghi} M., 2007,
  ApJ, 660, 1039

\bibitem[{{Kotilainen} {et~al}\mbox{.}(1998{\natexlab{a}}){Kotilainen},
  {Falomo}, \& {Scarpa}}]{kotilainen_1998}
{Kotilainen} J.~K., {Falomo} R., {Scarpa} R., 1998{\natexlab{a}}, A\&A, 332,
  503

\bibitem[{{Kotilainen} {et~al}\mbox{.}(1998{\natexlab{b}}){Kotilainen},
  {Falomo}, \& {Scarpa}}]{kotilainen_1998bllacs}
{Kotilainen} J.~K., {Falomo} R., {Scarpa} R., 1998{\natexlab{b}}, A\&A, 336,
  479

\bibitem[{{Kotilainen} {et~al}\mbox{.}(2005){Kotilainen}, {Hyv{\"o}nen}, \&
  {Falomo}}]{kotilainen_2005}
{Kotilainen} J.~K., {Hyv{\"o}nen} T., {Falomo} R., 2005, A\&A, 440, 831

\bibitem[{{Kotilainen} {et~al}\mbox{.}(2011){Kotilainen}, {Hyv{\"o}nen},
  {Falomo}, {Treves}, \& {Uslenghi}}]{kotilainen_2011}
{Kotilainen} J.~K., {Hyv{\"o}nen} T., {Falomo} R., {Treves} A., {Uslenghi} M.,
  2011, A\&A, 534, L2

\bibitem[{{Le{\'o}n-Tavares} {et~al}\mbox{.}(2013){Le{\'o}n-Tavares},
  {Chavushyan}, {Pati{\~n}o-{\'A}lvarez}, {Valtaoja}, {Arshakian},
  {Popovi{\'c}}, {Tornikoski}, {Lobanov}, {Carrami{\~n}ana}, {Carrasco}, \&
  {L{\"a}hteenm{\"a}ki}}]{leontavares_2013}
{Le{\'o}n-Tavares} J. {et~al.}, 2013, ApJL, 763, L36

\bibitem[{{Le{\'o}n Tavares} {et~al}\mbox{.}(2014){Le{\'o}n Tavares},
  {Kotilainen}, {Chavushyan}, {A{\~n}orve}, {Puerari}, {Cruz-Gonz{\'a}lez},
  {Pati{\~n}o-Alvarez}, {Ant{\'o}n}, {Carrami{\~n}ana}, {Carrasco}, {Guichard},
  {Karhunen}, {Olgu{\'{\i}}n-Iglesias}, {Sanghvi}, \&
  {Valdes}}]{leontavares_2014}
{Le{\'o}n Tavares} J. {et~al.}, 2014, ApJ, 795, 58

\bibitem[{{Le{\'o}n-Tavares}
  {et~al}\mbox{.}(2011{\natexlab{a}}){Le{\'o}n-Tavares}, {Valtaoja},
  {Chavushyan}, {Tornikoski}, {A{\~n}orve}, {Nieppola}, \&
  {L{\"a}hteenm{\"a}ki}}]{leontavares_2011_mbh}
{Le{\'o}n-Tavares} J., {Valtaoja} E., {Chavushyan} V.~H., {Tornikoski} M.,
  {A{\~n}orve} C., {Nieppola} E., {L{\"a}hteenm{\"a}ki} A., 2011{\natexlab{a}},
  MNRAS, 411, 1127

\bibitem[{{Le{\'o}n-Tavares}
  {et~al}\mbox{.}(2011{\natexlab{b}}){Le{\'o}n-Tavares}, {Valtaoja},
  {Tornikoski}, {L{\"a}hteenm{\"a}ki}, \& {Nieppola}}]{leontavares_2011}
{Le{\'o}n-Tavares} J., {Valtaoja} E., {Tornikoski} M., {L{\"a}hteenm{\"a}ki}
  A., {Nieppola} E., 2011{\natexlab{b}}, A\&A, 532, A146

\bibitem[{{Lister} {et~al}\mbox{.}(2009){Lister}, {Cohen}, {Homan}, {Kadler},
  {Kellermann}, {Kovalev}, {Ros}, {Savolainen}, \& {Zensus}}]{mojave}
{Lister} M.~L. {et~al.}, 2009, AJ, 138, 1874

\bibitem[{{Magorrian} {et~al}\mbox{.}(1998){Magorrian}, {Tremaine},
  {Richstone}, {Bender}, {Bower}, {Dressler}, {Faber}, {Gebhardt}, {Green},
  {Grillmair}, {Kormendy}, \& {Lauer}}]{magorrian_1998}
{Magorrian} J. {et~al.}, 1998, AJ, 115, 2285

\bibitem[{{Meisner} \& {Romani}(2010)}]{meisner_2010}
{Meisner} A.~M., {Romani} R.~W., 2010, ApJ, 712, 14

\bibitem[{{Morganti} {et~al}\mbox{.}(2013){Morganti}, {Fogasy}, {Paragi},
  {Oosterloo}, \& {Orienti}}]{morganti_2013}
{Morganti} R., {Fogasy} J., {Paragi} Z., {Oosterloo} T., {Orienti} M., 2013,
  Science, 341, 1082

\bibitem[{{Nesvadba} {et~al}\mbox{.}(2008){Nesvadba}, {Lehnert}, {De Breuck},
  {Gilbert}, \& {van Breugel}}]{nesvadba_2008}
{Nesvadba} N.~P.~H., {Lehnert} M.~D., {De Breuck} C., {Gilbert} A.~M., {van
  Breugel} W., 2008, A\&A, 491, 407

\bibitem[{{Nieppola} {et~al}\mbox{.}(2011){Nieppola}, {Tornikoski}, {Valtaoja},
  {Le{\'o}n-Tavares}, {Hovatta}, {L{\"a}hteenm{\"a}ki}, \&
  {Tammi}}]{nieppola_2011}
{Nieppola} E., {Tornikoski} M., {Valtaoja} E., {Le{\'o}n-Tavares} J., {Hovatta}
  T., {L{\"a}hteenm{\"a}ki} A., {Tammi} J., 2011, A\&A, 535, A69

\bibitem[{{Nilsson} {et~al}\mbox{.}(2003){Nilsson}, {Pursimo}, {Heidt},
  {Takalo}, {Sillanp{\"a}{\"a}}, \& {Brinkmann}}]{nilsson_2003}
{Nilsson} K., {Pursimo} T., {Heidt} J., {Takalo} L.~O., {Sillanp{\"a}{\"a}} A.,
  {Brinkmann} W., 2003, A\&A, 400, 95

\bibitem[{{Nilsson} {et~al}\mbox{.}(2008){Nilsson}, {Pursimo},
  {Sillanp{\"a}{\"a}}, {Takalo}, \& {Lindfors}}]{nilsson_2008}
{Nilsson} K., {Pursimo} T., {Sillanp{\"a}{\"a}} A., {Takalo} L.~O., {Lindfors}
  E., 2008, A\&A, 487, L29

\bibitem[{{Nilsson} {et~al}\mbox{.}(2009){Nilsson}, {Pursimo}, {Villforth},
  {Lindfors}, \& {Takalo}}]{nilsson_2009}
{Nilsson} K., {Pursimo} T., {Villforth} C., {Lindfors} E., {Takalo} L.~O.,
  2009, A\&A, 505, 601

\bibitem[{{Nilsson} {et~al}\mbox{.}(2012){Nilsson}, {Pursimo}, {Villforth},
  {Lindfors}, {Takalo}, \& {Sillanp{\"a}{\"a}}}]{nilsson_2012}
{Nilsson} K., {Pursimo} T., {Villforth} C., {Lindfors} E., {Takalo} L.~O.,
  {Sillanp{\"a}{\"a}} A., 2012, A\&A, 547, A1

\bibitem[{{O'Dowd} \& {Urry}(2005)}]{odowd_2005}
{O'Dowd} M., {Urry} C.~M., 2005, ApJ, 627, 97

\bibitem[{{Pahre} {et~al}\mbox{.}(1995){Pahre}, {Djorgovski}, \& {de
  Carvalho}}]{pahre_1995}
{Pahre} M.~A., {Djorgovski} S.~G., {de Carvalho} R.~R., 1995, ApJL, 453, L17

\bibitem[{{Peng} {et~al}\mbox{.}(2011){Peng}, {Ho}, {Impey}, \&
  {Rix}}]{peng_2011}
{Peng} C.~Y., {Ho} L.~C., {Impey} C.~D., {Rix} H.-W., 2011, {GALFIT: Detailed
  Structural Decomposition of Galaxy Images}. Astrophysics Source Code Library

\bibitem[{{Recillas-Cruz} {et~al}\mbox{.}(1990){Recillas-Cruz}, {Serrano},
  {Cruz-Gonzalez}, \& {Carrasco}}]{recillas_1990}
{Recillas-Cruz} E., {Serrano} P.~G.~A., {Cruz-Gonzalez} I., {Carrasco} L.,
  1990, A\&A, 229, 64

\bibitem[{{Romanishin}(1987)}]{romanishin_1987}
{Romanishin} W., 1987, ApJ, 320, 586

\bibitem[{{Sbarufatti} {et~al}\mbox{.}(2006){Sbarufatti}, {Falomo}, {Treves},
  \& {Kotilainen}}]{sbarufatti_2006}
{Sbarufatti} B., {Falomo} R., {Treves} A., {Kotilainen} J., 2006, A\&A, 457, 35

\bibitem[{{Sbarufatti} {et~al}\mbox{.}(2005){Sbarufatti}, {Treves}, \&
  {Falomo}}]{sbarufatti_2005}
{Sbarufatti} B., {Treves} A., {Falomo} R., 2005, ApJ, 635, 173

\bibitem[{{Scannapieco} \& {Oh}(2004)}]{scannapieco_2004}
{Scannapieco} E., {Oh} S.~P., 2004, apj, 608, 62

\bibitem[{{Scarpa} {et~al}\mbox{.}(2000){Scarpa}, {Urry}, {Falomo}, {Pesce}, \&
  {Treves}}]{scarpa_2000}
{Scarpa} R., {Urry} C.~M., {Falomo} R., {Pesce} J.~E., {Treves} A., 2000, ApJ,
  532, 740

\bibitem[{{Silk}(2005)}]{silk_2005}
{Silk} J., 2005, mnras, 364, 1337

\bibitem[{{Silk}(2013)}]{silk_2013}
{Silk} J., 2013, apj, 772, 112

\bibitem[{{Silk} \& {Norman}(2009)}]{silk_2009}
{Silk} J., {Norman} C., 2009, apj, 700, 262

\bibitem[{{Stadnik} \& {Romani}(2014)}]{stadnik_2014}
{Stadnik} M., {Romani} R.~W., 2014, ApJ, 784, 151

\bibitem[{{Stickel} {et~al}\mbox{.}(1991){Stickel}, {Padovani}, {Urry},
  {Fried}, \& {Kuehr}}]{stickel_1991}
{Stickel} M., {Padovani} P., {Urry} C.~M., {Fried} J.~W., {Kuehr} H., 1991,
  apj, 374, 431

\bibitem[{{Tadhunter} {et~al}\mbox{.}(2014){Tadhunter}, {Morganti}, {Rose},
  {Oonk}, \& {Oosterloo}}]{tadhunter_2014}
{Tadhunter} C., {Morganti} R., {Rose} M., {Oonk} J.~B.~R., {Oosterloo} T.,
  2014, Nature, 511, 440

\bibitem[{{Teraesranta} {et~al}\mbox{.}(1998){Teraesranta}, {Tornikoski},
  {Mujunen}, {Karlamaa}, {Valtonen}, {Henelius}, {Urpo}, {Lainela}, {Pursimo},
  {Nilsson}, {Wiren}, {Laehteenmaeki}, {Korpi}, {Rekola}, {Heinaemaeki},
  {Hanski}, {Nurmi}, {Kokkonen}, {Keinaenen}, {Joutsamo}, {Oksanen},
  {Pietilae}, {Valtaoja}, {Valtonen}, \& {Koenoenen}}]{terasranta_1998}
{Teraesranta} H. {et~al.}, 1998, A\&AS, 132, 305

\bibitem[{{Terasranta} {et~al}\mbox{.}(1992){Terasranta}, {Tornikoski},
  {Valtaoja}, {Urpo}, {Nesterov}, {Lainela}, {Kotilainen}, {Wiren}, {Laine},
  {Nilsson}, \& {Valtonen}}]{terasranta_1992}
{Terasranta} H. {et~al.}, 1992, A\&AS, 94, 121

\bibitem[{{Tremaine} {et~al}\mbox{.}(2002){Tremaine}, {Gebhardt}, {Bender},
  {Bower}, {Dressler}, {Faber}, {Filippenko}, {Green}, {Grillmair}, {Ho},
  {Kormendy}, {Lauer}, {Magorrian}, {Pinkney}, \& {Richstone}}]{tremaine_2002}
{Tremaine} S. {et~al.}, 2002, ApJ, 574, 740

\bibitem[{{Treves} {et~al}\mbox{.}(2007){Treves}, {Falomo}, \&
  {Uslenghi}}]{treves_2007}
{Treves} A., {Falomo} R., {Uslenghi} M., 2007, A\&A, 473, L17

\bibitem[{{Urry} \& {Padovani}(1995)}]{urry_padovani_1995}
{Urry} C.~M., {Padovani} P., 1995, pasp, 107, 803

\bibitem[{{Urry} {et~al}\mbox{.}(2000){Urry}, {Scarpa}, {O'Dowd}, {Falomo},
  {Pesce}, \& {Treves}}]{urry_2000}
{Urry} C.~M., {Scarpa} R., {O'Dowd} M., {Falomo} R., {Pesce} J.~E., {Treves}
  A., 2000, ApJ, 532, 816

\end{thebibliography}
\appendix{}

\newpage

\section{Host galaxy and nuclear  properties for the compiled  sample of AGNs}
%%%%%%%%%%%%%%%%%%
%%%%%%%%%%%%%%%%%%
\begin{landscape}
\begin{table}
 \centering
 \begin{minipage}{215mm}
  \caption{Best-fit parameters of the morphological fittings and general properties of the combined sample (This work and literature compilation).}
\label{table:compilation}
  \begin{tabular}{@{}llcccccccccccccc@{}}
  \hline
Name	&	$z$	&	RA	&	DEC	&	Type	&	$S_{1.4GHz}$  & $logL_{1.4GHz}$& Filter   & $m_{host}$ &$m_{nuclear}$ & $MK_{host}$  &$MK_{nuclear}$&Re     &$\mu_{e}$			&ref 	\\
    		&   		 &  		 &    		&  		 &  	Jy		    &	$W Hz^{-1}$  & 	     &		           &			     &			      &			 &kpc   & $mag/arcsec^{2}$     &     \\
(1)	&  	(2)	 &  	(3)	&    	(4)	&  	(5)	 &  	(6)		    &	(7)  & (8) & (9)           &		(10)	     &  (11)	 	      &	(12)		 &(13)   & (14)     &   (15)  \\
\hline
0003$-$066	&	0.347	&	00 06 13.8	&	$-$06 23 35.3	&	FSRQ		&	1.68	&	26.70	&	J	&	16.68	&	17.29	&	-25.44	&	-25.47	&	9.77		&	18.7	0	&	7	\\
0048$-$097	&	0.634	&	00 50 41.3	&	$-$09 29 05.2	&	FSRQ		&	0.89	&	26.97	&	J	&	17.34	&	14.91	&	-26.34	&	-29.41	&	17.60	&	19.60	&	7	\\
0057$-$338	&	0.875	&	01 00 09.3	&	$-$33 37 32.0	&	BLLAC	&	0.07	&	26.15	&	R	&	20.93	&	20.30	&	-25.51	&	-26.14	&	2.63	&	15.54	&	9	\\
0120$+$340	&	0.272	&	01 23 08.6	&	$+$34 20 48.6	&	BLLAC	&	0.05	&	24.91	&	R	&	17.24	&	16.39	&	-26.17	&	-27.27	&	14.14	&	19.02	&	1	\\
0123$+$343	&	0.272	&	01 23 08.0	&	$+$34 20 50.0	&	BLLAC	&	0.05	&	24.91	&	R	&	18.20	&	17.98	&	-25.21	&	-25.68	&	7.07	&	19.95	&	3	\\
0138$-$097	&	0.733	&	01 41 25.8	&	$-$09 28 43.6	&	BLLAC	&	0.66	&	26.96	&	K	&	16.17	&	13.25	&	-27.10	&	-30.02	&	2.98	&	17.58	&	2	\\
0158$+$003	&	0.299	&	02 01 06.1	&	$+$00 34 00.1	&	BLLAC	&	0.01	&	24.46	&	K	&	14.60	&	15.90	&	-26.35	&	-25.05	&	3.56	&	17.00	&	6	\\
0202$+$149	&	0.405	&	02 04 50.4	&	$+$15 14 11.0	&	FSRQ		&	3.46	&	27.15	&	J	&	18.71	&	18.73	&	-23.81	&	-24.42	&	8.12	&	19.85	&	7	\\
0205$+$350	&	0.318	&	02 08 40.0	&	$+$35 23 20.0	&	BLLAC	&	0.00	&	24.00	&	R	&	19.03	&	17.53	&	-24.77	&	-26.52	&	5.56	&	18.29	&	1	\\
0229$+$200	&	0.139	&	02 32 48.6	&	$+$20 17 17.4	&	BLLAC	&	0.06	&	24.45	&	R	&	15.76	&	18.25	&	-26.03	&	-23.79	&	9.81	&	18.36	&	1	\\
0306$+$102	&	0.863	&	03 09 03.6	&	$+$10 29 16.3	&	FSRQ 	&	0.51	&	27.00	&	H	&	16.93	&	15.75	&	-26.98	&	-29.06	&	13.47	&	19.70	&	7	\\
0326$+$024	&	0.147	&	03 26 13.0	&	$+$02 25 10.0	&	BLLAC	&	0.07	&	24.54	&	R	&	17.05	&	16.86	&	-24.87	&	-25.31	&	6.94	&	19.63	&	3	\\
0331$-$365	&	0.308	&	03 33 12.2	&	$-$36 19 46.6	&	BLLAC	&	0.01	&	24.51	&	K	&	14.50	&	16.30	&	-26.52	&	-24.72	&	6.35	&	18.10	&	6	\\
0347$-$121	&	0.185	&	03 49 23.1	&	$-$11 59 27.0	&	BLLAC	&	0.01	&	23.86	&	K	&	14.20	&	15.80	&	-25.57	&	-23.97	&	1.86	&	16.20	&	6	\\
0350$-$371	&	0.165	&	03 51 54.5	&	$-$37 03 44.1	&	BLLAC	&	0.03	&	24.28	&	K	&	13.70	&	15.40	&	-25.79	&	-24.09	&	3.39	&	17.30	&	6	\\
0405$-$123	&	0.574	&	04 07 48.4	&	$-$12 11 36.6	&	FSRQ	&	3.27	&	27.44	&	H	&	15.60	&	13.40	&	-27.22	&	-30.32	&	7.86	&		&	4	\\
0406$+$121	&	0.504	&	04 09 22.0	&	$+$12 17 39.8	&	BLLAC	&	0.53	&	26.53	&	R	&	20.95	&	20.61	&	-24.03	&	-24.37	&	3.31	&	17.06	&	9	\\
0414$+$000	&	0.287	&	04 16 52.4	&	$+$01 05 23.9	&	BLLAC	&	0.13	&	25.42	&	R	&	17.47	&	16.97	&	-26.07	&	-26.82	&	4.32	&	16.35	&	1	\\
0419$+$194	&	0.512	&	04 22 18.3	&	$+$19 50 55.8	&	BLLAC	&	0.01	&	24.78	&	R	&	21.05	&	19.53	&	-23.97	&	-25.74	&	2.47	&	17.01	&	5	\\
0420$-$014	&	0.916	&	04 23 15.8	&	$-$01 20 33.0	&	FSRQ	&	1.92	&	27.62	&	H	&	17.00	&	13.50	&	-27.06	&	-31.46	&	19.18	&		&	4	\\
0502$+$675	&	0.314	&	05 07 56.2	&	$+$67 37 24.4	&	BLLAC	&	0.03	&	24.80	&	R	&	18.19	&	16.83	&	-25.58	&	-27.19	&	1.61	&	14.42	&	1	\\
0506$-$039	&	0.304	&	05 09 38.1	&	$-$04 00 45.5	&	BLLAC	&	0.07	&	25.21	&	R	&	18.35	&	18.73	&	-25.34	&	-25.21	&	7.19	&	18.51	&	5	\\
0607$+$710	&	0.267	&	06 13 42.8	&	$+$71 07 29.3	&	BLLAC	&	0.03	&	24.67	&	R	&	17.83	&	18.23	&	-25.54	&	-25.39	&	9.85	&	19.06	&	5	\\
0654$+$427	&	0.126	&	06 54 43.0	&	$+$42 47 58.0	&	BLLAC	&	0.19	&	24.84	&	R	&	15.98	&	17.55	&	-25.57	&	-24.25	&	6.32	&	19.95	&	3	\\
0706$+$591	&	0.125	&	07 10 30.0	&	$+$59 08 20.4	&	BLLAC	&	0.13	&	24.66	&	R	&	15.94	&	17.53	&	-25.60	&	-24.26	&	6.83	&	18.40	&	5	\\
0736$+$017	&	0.191	&	07 39 18.0	&	$+$01 37 04.6	&	FSRQ	&	2.48	&	26.33	&	H	&	14.30	&	14.30	&	-25.74	&	-26.64	&	2.39	&		&	4	\\
0737$+$744	&	0.315	&	07 43 59.0	&	$+$74 33 50.0	&	BLLAC	&	0.02	&	24.75	&	R	&	18.01	&	17.88	&	-25.77	&	-26.15	&	9.67	&	18.71	&	5	\\
0754$+$100	&	0.266	&	07 57 06.6	&	$+$09 56 34.8	&	BLLAC	&	1.00	&	26.24	&	R	&	18.48	&	18.46	&	-24.88	&	-24.90	&	10.06	&	19.91	&	9	\\
0806$+$524	&	0.138	&	08 09 49.1	&	$+$52 18 58.2	&	BLLAC	&	0.17	&	24.87	&	R	&	16.62	&	15.98	&	-25.15	&	-26.04	&	3.53	&	17.39	&	5	\\
0820$+$255	&	0.951	&	08 23 24.0	&	$+$22 23 03.0	&	BLLAC	&	2.27	&	27.73	&	R	&	19.12	&	18.83	&	-27.55	&	-28.09	&	11.07	&		&	8	\\
0828$+$493	&	0.548	&	08 32 23.2	&	$+$49 13 21.0	&	BLLAC	&	0.67	&	26.71	&	R	&	20.26	&	18.93	&	-24.94	&	-26.52	&	4.16	&	17.05	&	5	\\
0829$+$046	&	0.180	&	08 31 48.8	&	$+$04 29 39.0	&	BLLAC	&	0.91	&	25.84	&	R	&	16.94	&	15.88	&	-25.46	&	-26.77	&	13.05	&	19.87	&	5	\\
0916$+$526	&	0.190	&	09 16 51.0	&	$+$52 38 28.0	&	BLLAC	&	0.14	&	25.08	&	R	&	16.21	&	18.42	&	-26.32	&	-24.36	&	15.85	&	20.20	&	3	\\
0922$+$749	&	0.638	&	09 28 02.9	&	$+$74 47 19.1	&	BLLAC	&	0.09	&	26.00	&	R	&	20.25	&	20.13	&	-25.35	&	-25.72	&	5.84	&	17.09	&	5	\\
0927$+$500	&	0.188	&	09 30 37.5	&	$+$49 50 25.5	&	BLLAC	&	0.02	&	24.25	&	R	&	17.62	&	17.48	&	-24.89	&	-25.28	&	6.29	&	18.85	&	5	\\
0928$+$747	&	0.638	&	09 28 03.0	&	$+$74 47 19.0	&	BLLAC	&	0.04	&	25.66	&	R	&	20.12	&	20.25	&	-25.48	&	-25.60	&	8.94	&	21.04	&	3	\\
0930$+$393	&	0.638	&	09 30 56.8	&	$+$39 33 35.9	&	BLLAC	&	0.01	&	25.06	&	R	&	19.43	&	19.62	&	-26.17	&	-25.98	&	10.31	&	19.08	&	9	\\
0957$+$227	&	0.419	&	10 00 21.8	&	$+$22 33 18.7	&	BLLAC	&	1.12	&	26.69	&	R	&	18.63	&	18.74	&	-25.87	&	-26.01	&	18.23	&	21.72	&	3	\\
1009$+$427	&	0.364	&	10 12 44.2	&	$+$42 29 57.0	&	BLLAC	&	0.08	&	25.42	&	R	&	18.13	&	17.90	&	-26.01	&	-26.49	&	19.25	&	21.66	&	3	\\
\hline
\end{tabular}
\end{minipage}
\end{table}
\end{landscape}

\begin{landscape}
\begin{table}
 \centering
 \begin{minipage}{215mm}
  \begin{tabular}{@{}llcccccccccccccc@{}}
  \hline
Name	&	$z$	&	RA	&	DEC	&	Type	&	$S_{1.4GHz}$  & $logL_{1.4GHz}$& Filter   & $m_{host}$ &$m_{nuclear}$ & $MK_{host}$  &$MK_{nuclear}$&Re     &$\mu_{e}$			&ref	\\
    		&   		 &  		 &    		&  		 &  	Jy		    &	$W Hz^{-1}$  & 	     &		           &			     &			      &			 &kpc   & $mag/arcsec^{2}$     &     \\
(1)	&  	(2)	 &  	(3)	&    	(4)	&  	(5)	 &  	(6)		    &	(7)  & (8) & (9)           &		(10)	     &  (11)	 	      &	(12)		 &(13)   & (14)     &   (15)  \\
\hline
1011$+$496	&	0.212	&	10 15 04.1	&	$+$49 26 00.7	&	BLLAC	&	0.38	&	25.61	&	R	&	17.12	&	16.07	&	-25.68	&	-26.98	&	9.43	&	18.94	&	1	\\
1028$+$511	&	0.361	&	10 31 18.5	&	$+$50 53 35.8	&	BLLAC	&	0.04	&	25.08	&	R	&	18.55	&	16.48	&	-25.57	&	-27.89	&	9.07	&	18.66	&	5	\\
1040$+$224	&	0.560	&	10 43 09.0	&	$+$24 08 35.0	&	BLLAC	&	0.33	&	26.42	&	R	&	19.74	&	19.30	&	-25.52	&	-25.96	&	5.18	&	17.51	&	9	\\
1053$+$494	&	0.140	&	10 53 44.0	&	$+$49 29 56.0	&	BLLAC	&	0.06	&	24.47	&	R	&	15.33	&	17.79	&	-26.47	&	-24.26	&	17.52	&	19.96	&	3	\\
1133$+$704	&	0.136	&	11 36 26.4	&	$+$70 09 27.3	&	BLLAC	&	0.33	&	25.15	&	R	&	16.10	&	17.91	&	-25.63	&	-24.07	&	7.95	&	19.03	&	3	\\
1144$-$379	&	1.048	&	11 47 01.0	&	$-$02 04 00	&	BLLAC	&	1.60	&	27.66	&	R	&	21.18	&	18.00	&	-25.75	&	-29.17	&	20.20	&		&	8	\\
1149$+$246	&	0.402	&	11 49 30.0	&	$+$24 39 26.0	&	BLLAC	&	0.03	&	25.06	&	R	&	18.74	&	19.28	&	-25.66	&	-25.37	&	5.93	&	19.48	&	3	\\
1150$+$497	&	0.334	&	11 53 24.4	&	$+$49 31 08.8	&	FSRQ	&	1.68	&	26.67	&	J	&	16.07	&	14.60	&	-25.95	&	-28.06	&	4.33 	&	15.10	&	7	\\
1156$+$295	&	0.724	&	11 59 31.8	&	$+$29 14 43.8	&	FSRQ	&	1.85	&	27.40	&	J	&	17.04	&	14.29	&	-26.90	&	-30.38	&	8.47 	&	18.40	&	7	\\
1212$+$078	&0.130	&	12 15 10.9	&	$+$07 32 04.7	&	BLLAC	&	0.11	&	24.64	&	R	&	15.86	&	17.26	&	-25.77	&	-24.62	&	12.77	&	19.53	&	1	\\
1218$+$304	&	0.182	&	12 21 21.0	&	$+$30 10 37.0	&	BLLAC	&	0.06	&	24.69	&	R	&	16.80	&	16.32	&	-25.63	&	-26.36	&	6.65	&	18.18	&	1	\\
1219$+$285	&	0.102	&	12 21 31.6	&	$+$28 13 58.5	&	BLLAC	&	0.73	&	25.24	&	R	&	16.60	&	14.26	&	-24.46	&	-27.05	&	3.94	&	18.66	&	3	\\
1221$+$245	&	0.218	&	12 24 24.1	&	$+$24 36 23.5	&	BLLAC	&	0.03	&	24.47	&	R	&	18.63	&	16.89	&	-24.24	&	-26.23	&	4.41	&	18.69	&	5	\\
1226$+$023	&	0.158	&	12 29 06.6	&	$+$02 03 08.5	&	FSRQ	&	44.73	&	27.42	&	H	&	13.30	&	10.90	&	-26.29	&	-29.59	&	12.28	&		&	4	\\
1229$+$643	&	0.164	&	12 31 31.3	&	$+$64 14 18.3	&	BLLAC	&	0.06	&	24.56	&	R	&	16.38	&	18.03	&	-25.80	&	-24.40	&	5.63	&	17.72	&	5	\\
1235$+$632	&	0.297	&	12 37 39.0	&	$+$62 58 42.8	&	BLLAC	&	0.01	&	24.41	&	R	&	18.01	&	19.20	&	-25.62	&	-24.68	&	6.15	&	17.90	&	1	\\
1253$-$055	&	0.538	&	12 56 11.1 	&	$-$05 47 22		&	FSRQ		&	9.71		&	27.86		&	I	&	18.43	&	16.13	&	-26.02		&	-28.66	&	17.12		&	21.70			&	10	\\
1255$+$244	&	0.141	&	12 57 31.9	&	$+$24 12 40.2	&	BLLAC	&	0.01	&	23.69	&	R	&	16.64	&	18.27	&	-25.18	&	-23.80	&	4.72	&	17.93	&	1	\\
1308$+$326	&	0.997	&	13 10 28.6	&	$+$32 20 43.7	&	BLLAC	&	1.27	&	27.52	&	K	&	15.49	&	12.95	&	-28.60	&	-31.14	&	5.76	&	18.18	&	2	\\
1407$+$595	&	0.495	&	14 09 23.4	&	$+$59 39 40.7	&	BLLAC	&	0.04	&	25.36	&	R	&	19.04	&	18.84	&	-25.90	&	-26.35	&	10.63	&	18.31	&	5	\\
1418$+$546	&	0.151	&	14 19 46.5	&	$+$54 23 14.7	&	BLLAC	&	0.79	&	25.62	&	R	&	16.18	&	15.51	&	-25.80	&	-26.72	&	11.83	&	19.62	&	3	\\
1421$+$582	&	0.638	&	14 22 38.8	&	$+$58 01 55.5	&	BLLAC	&	0.01	&	25.14	&	K	&	16.20	&	15.61	&	-26.70	&	-27.29	&	4.81	&	18.83	&	2	\\
1426$+$428	&	0.129	&	14 28 32.6	&	$+$42 40 20.6	&	BLLAC	&	0.03	&	24.09	&	R	&	16.14	&	17.38	&	-25.47	&	-24.48	&	5.18	&	17.92	&	5	\\
1427$+$541	&	0.105	&	14 27 30.0	&	$+$54 09 23.0	&	BLLAC	&	0.04	&	24.05	&	R	&	14.95	&	19.21	&	-26.18	&	-22.17	&	16.94	&	20.21	&	3	\\
1428$+$370	&	0.564	&	14 30 40.0	&	$+$36 49 03.0	&	BLLAC	&	0.19	&	26.18	&	R	&	20.05	&	19.61	&	-25.23	&	-25.67	&	2.47	&	16.24	&	9	\\
1440$+$122	&	0.162	&	14 42 48.2	&	$+$12 00 40.3	&	BLLAC	&	0.06	&	24.57	&	R	&	16.71	&	16.93	&	-25.44	&	-25.47	&	10.87	&	19.51	&	5	\\
1443$+$634	&	0.299	&	14 44 36.0	&	$+$63 36 20.0	&	BLLAC	&	0.02	&	24.62	&	R	&	18.11	&	19.38	&	-25.54	&	-24.52	&	12.89	&	19.58	&	1	\\
1458$+$224	&	0.235	&	15 01 01.8	&	$+$22 38 06.3	&	BLLAC	&	0.03	&	24.63	&	R	&	17.80	&	15.78	&	-25.25	&	-27.52	&	11.95	&	19.82	&	5	\\
1510$-$089	&	0.360	&	15 12 50.5	&	$-$09 05 59.8	&	FSRQ		&	3.17	&	27.01	&	J	&	16.32	&	15.19	&	-25.90	&	-27.67	&	10.03	&	17.70	&	7	\\
1517$+$656	&	0.702	&	15 17 47.6	&	$+$65 25 23.9	&	BLLAC	&	0.08	&	26.00	&	K	&	14.96	&	13.79	&	-28.19	&	-29.36	&	6.44	&	18.12	&	2	\\
1533$+$535	&	0.890	&	15 35 00.8	&	$+$53 20 37.3		&	BLLAC	&	0.02	&	25.57	&	K	&	16.78	&	16.67	&	-27.01	&	-27.12	&	3.73	&	17.93	&	2	\\
1534$+$014	&	0.312	&	15 36 46.0	&	$+$01 38 00.1	&	BLLAC	&	0.11	&	25.41	&	R	&	18.16	&	19.08	&	-25.59	&	-24.92	&	9.15	&	18.77	&	5	\\
1534$+$372	&	0.143	&	15 34 47.0	&	$+$37 15 54.0	&	BLLAC	&	0.02	&	24.02	&	R	&	17.18	&	18.21	&	-24.67	&	-23.89	&	5.02	&	19.28	&	3	\\
1538$+$149	&	0.605	&	15 40 49.4	&	$+$14 47 45.8	&	BLLAC	&	1.28	&	27.08	&	K	&	15.31	&	14.28	&	-27.45	&	-28.48	&	11.75	&	19.93	&	2	\\
1546$+$027	&	0.414	&	15 49 29.4	&	$+$02 37 01.1	&	FSRQ		&	0.95	&	26.61	&	J	&	16.45	&	14.56	&	-26.12	&	-28.65	&	6.58	&	17.30	&	7	\\
1552$+$202	&	0.222	&	15 54 24.1	&	$+$20 11 25.4	&	BLLAC	&	0.08	&	24.97	&	R	&	16.87	&	17.97	&	-26.04	&	-25.19	&	8.41	&	18.14	&	1	\\
1641$+$399	&	0.593	&	16 42 58.8	&	$+$39 48 36.9	&	FSRQ		&	6.78	&	27.79	&	J	&	16.04	&	15.76	&	-27.46	&	-28.38	&	9.10	&	17.50	&	7	\\
\hline
\end{tabular}
\end{minipage}
\end{table}
\end{landscape}

\begin{landscape}
\begin{table}
 \centering
 \begin{minipage}{215mm}
\label{log}
  \begin{tabular}{@{}lccccccccccccccc@{}}
  \hline
Name	&	z	&	RA	&	DEC	&	Type	&	$S_{1.4GHz}$  & $logL_{1.4GHz}$ & Filter  & $m_{host}$ &$m_{nuclear}$ & $MK_{host}$  &$MK_{nuclear}$ &Re     &$\mu_{e}$ 		&ref\\
 	&  		 &  	&    	&  	 &  	Jy 		    &	$W Hz^{-1}$  &      &		           &			     &	 	      &			 &kpc   & $mag/arcsec^{2}$     &     \\
(1)	&  	(2)	 &  	(3)	&    	(4)	&  	(5)	 &  	(6)		    &	(7)  & (8) & (9)           &		(10)	     &  (11)	 	      &	(12)		 &(13)   & (14)     &   (15)  \\
\hline
1642$+$690	&	0.751	&	16 42 07.8	&	$+$68 56 39.7	&	FSRQ	&	1.62	&	27.37	&	J	&	17.61	&	18.15	&	-26.52	&	-26.62	&	14.68	&	20.80	&	7	\\
1749$+$096	&	0.322	&	17 51 32.8	&	$+$09 39 00.7	&	BLLAC	&	0.81	&	26.32	&	R	&	18.82	&	16.88	&	-25.01	&	-27.20	&	14.03	&	20.26	&	5	\\
1749$+$701	&0.770	&	17 48 32.8	&	$+$70 05 50.7	&	BLLAC	&	0.93	&	27.16	&	K	&	15.42	&	13.70	&	-27.98	&	-29.70	&	2.96	&	16.83	&	2	\\
1757$+$703	&	0.407	&	17 57 13.0	&	$+$70 33 37.6	&	BLLAC	&	0.01	&	24.65	&	R	&	19.58	&	18.43	&	-24.85	&	-26.25	&	4.62	&	17.81	&	5	\\
1803$+$784	& 0.680 &	18 00 45.6	&	$+$78 28 04.0	&	BLLAC	&	1.52	&	27.26	&	K	&	14.14	&	12.61	&	-28.93	&	-30.46	&	5.58	&	17.03	&	2	\\
1823$+$568	&	0.664	&	18 24 07.0	&	$+$56 51 01.4	&	BLLAC	&	1.45	&	27.22	&	K	&	14.68	&	13.41	&	-28.33	&	-29.60	&	3.50	&	16.58	&	2	\\
1828$+$487	&	0.692	&	18 29 31.7	&	$+$48 44 46.1	&	FSRQ	&	14.61	&	28.26	&	J	&	16.91	&	16.03	&	-27.01	&	-28.53	&	4.51	&	16.4	0&	7	\\
1841$+$591	&0.530	&	18 41 20.0	&	$+$59 06 08.0	&	BLLAC	&	0.02	&	25.15	&	R	&	19.04	&	19.63	&	-26.07	&	-25.73	&	13.22	&	21.07	&	3	\\
1849$+$670	&	0.657	&	18 49 16.0	&	$+$67 05 41.6	&	FSRQ	&	0.50	&	26.75	&	J	&	16.57	&	15.66	&	-27.21	&	-28.76	&	17.89	&	19.95	&	7	\\
1853$+$671	&	0.212	&	18 53 52.0	&	$+$67 13 55.7	&	BLLAC	&	0.01	&	24.07	&	R	&	18.19	&	19.48	&	-24.61	&	-23.57	&	5.18	&	18.67	&	5	\\
1921$-$293	&	0.352	&	19 24 51.0	&	$-$29 14 30.1	&	BLLAC	&	12.95	&	27.60	&	R	&	19.15	&	19.01	&	-24.91	&	-25.05	&	8.18	&	19.11	&	9	\\
1928$+$738	&	0.302	&	19 27 48.4	&	$+$73 58 01.5	&	FSRQ	&	2.86	&	26.81	&	J	&	16.38	&	14.75	&	-25.39	&	-27.46	&	15.71	&	19.40	&	7	\\
1954$-$388	&	0.626	&	19 57 59.8	&	$-$38 45 06.3	&	FSRQ	&	1.51	&	27.18	&	H	&	16.30	&	14.10	&	-26.75	&	-29.85	&	4.43	&		&	4	\\
2007$+$777	&	0.342	&	20 05 30.9	&	$+$77 52 43.1	&	BLLAC	&	1.00	&	26.46	&	R	&	19.03	&	18.03	&	-24.96	&	-26.21	&	16.06	&	20.56	&	5	\\
2032$+$107	&	0.601	&	20 35 22.3	&	$+$10 56 06.7	&	BLLAC	&	0.93	&	26.94	&	K	&	15.50	&	12.35	&	-27.24	&	-30.39	&	5.02	&	18.28	&	2	\\
2131$-$021	&	1.285	&	21 34 10.0 	&	$-$01 53 17.2	&	BLLAC	&	1.92	&	27.91	&	K	&	16.52	&	14.70	&	-28.25	&	-30.07	&	3.35	&	17.93	&	2	\\
2143$+$070	&	0.237	&	21 45 52.3	&	$+$07 19 27.2	&	BLLAC	&	0.10	&	25.14	&	R	&	17.89	&	18.21	&	-25.18	&	-25.11	&	7.89	&	18.98	&	5	\\
2216$-$038	&	0.901	&	22 18 52.0	&	$-$03 35 36.8	&	FSRQ	&	1.53	&	27.51	&	H	&	17.52	&	14.48	&	-26.50	&	-30.44	&	20.42	&	19.80	&	7	\\
2234$+$282	&	0.795	&	22 36 22.4	&	$+$28 28 57.4	&	FSRQ	&	0.96	&	27.20	&	H	&	18.18	&	16.90	&	-25.51	&	-27.69	&	8.91	&	18.70	&	7	\\
2254$+$074	&0.190	&	22 57 17.3	&	$+$07 43 12.3	&	BLLAC	&	0.38	&	25.51	&	R	&	16.61	&	16.94	&	-25.92	&	-25.84	&	15.53	&	19.78	&	5	\\
2326$+$174	&	0.213	&	23 29 03.3	&	$+$17 43 30.5	&	BLLAC	&	0.03	&	24.43	&	R	&	17.56	&	17.63	&	-25.25	&	-25.43	&	6.24	&	18.43	&	5	\\
2343$-$151	&	0.226	&	23 45 38.4	&	$-$14 49 28.7	&	BLLAC	&	0.01	&	24.09	&	R	&	17.18	&	20.36	&	-25.77	&	-22.84	&	5.80	&	17.67	&	1	\\
2345$-$167	&	0.576	&	23 48 02.6	&	$-$16 31 12.0	&	FSRQ	&	1.71	&	27.16	&	H	&	15.70	&	13.30	&	-27.13	&	-30.43	&	5.90	&		&	4	\\
2356$-$309	&	0.165	&	23 59 07.9	&	$-$30 37 40.6	&	BLLAC	&	0.05	&	24.52	&	K	&	14.10	&	14.50	&	-25.39	&	-24.99	&	3.11	&	17.50	&	6	\\
\hline
\end{tabular} 
Column (1) gives the galaxy name;
(2) the reported redshifts for the sources. BL Lac objects redshifts were derived from weak stellar absorption features or weak emission lines;
(3) and (4) the J2000 right ascension and declination;
(5) object type: BLLAC=BL Lac type object; FSRQ= Flat Spectrum Radio Quasar;
(6)  1.4GHz flux density as retrieved from NASA/IPAC Extragalactic Database (NED)\footnote{\href{http://ned.ipac.caltech.edu/}{http://ned.ipac.caltech.edu/}};
(7) luminosity at $1.4_{GHz}$;
(8)  observed filter;
(9) and (10) the host and nuclear apparent magnitudes in the observed band;
(11) and (12) the host and nuclear absolute magnitudes transformed to K--band assuming the colours in section \ref{sec:sample};
(13)  effective radius converted to our adopted cosmology;
(14)  surface brightness at effective radius in K--band;
(15)  references for fitting parameters 1= \citet{falomo_1999}; 2=\citet{kotilainen_2005}; 3=\citet{nilsson_2003}; 4=\citet{kotilainen_1998}; 5=\citet{urry_2000}; 6=\citet{cheung_2003}; 7=This work; 8=\citet{odowd_2005}; 9=\citet{heidt_2004}; 10=\citet{nilsson_2009}
\end{minipage}
\end{table}
\end{landscape}

\section{Surface Brightness decomposition and radial profiles for the sample of sources observed with NOTcam}
\begin{figure*}
 \begin{minipage}{180mm}
\begin{tabular}{@{}ll@{}}
	\includegraphics*[width=0.46\textwidth]{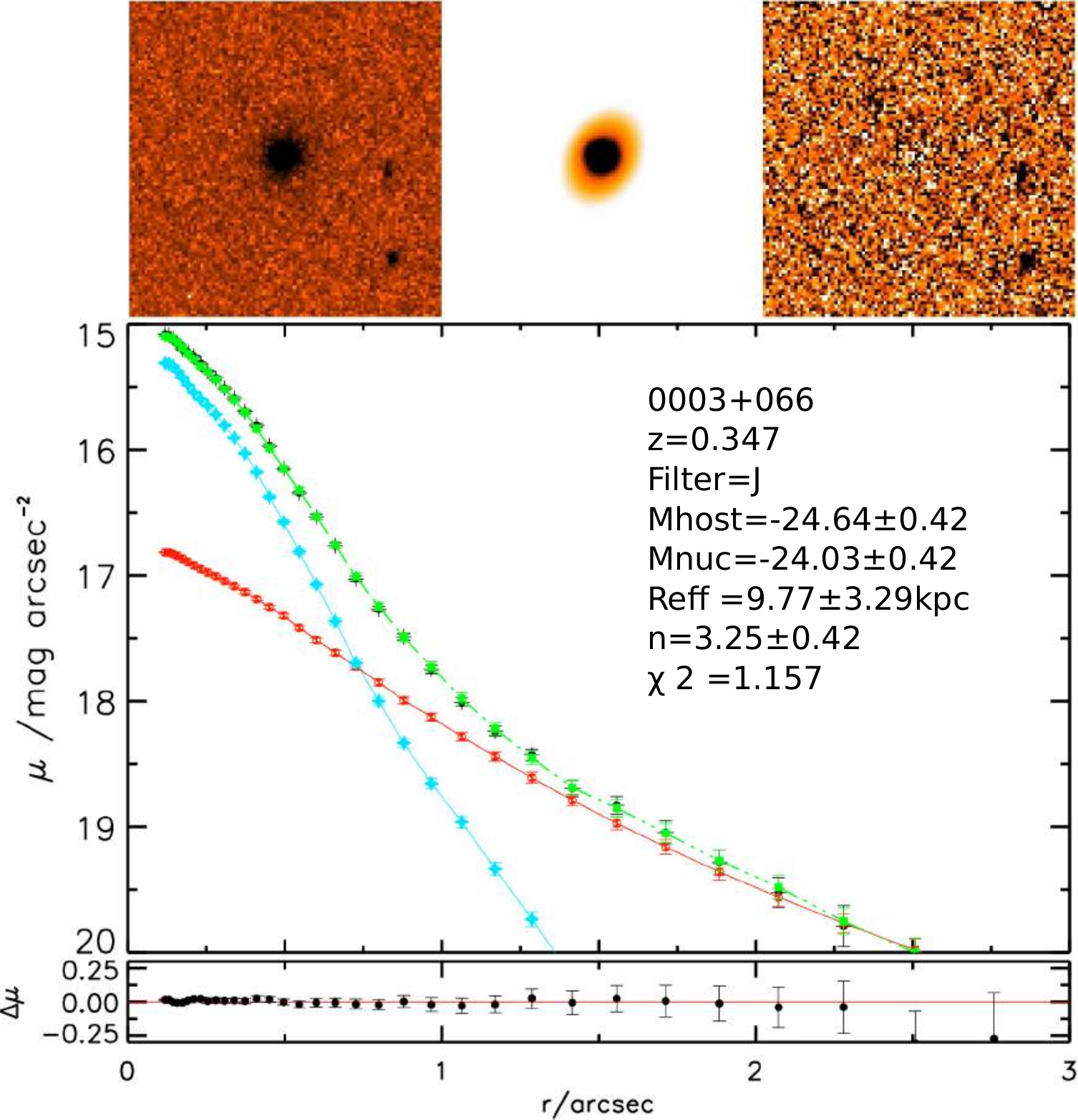}    &     \includegraphics*[width=0.46\textwidth]{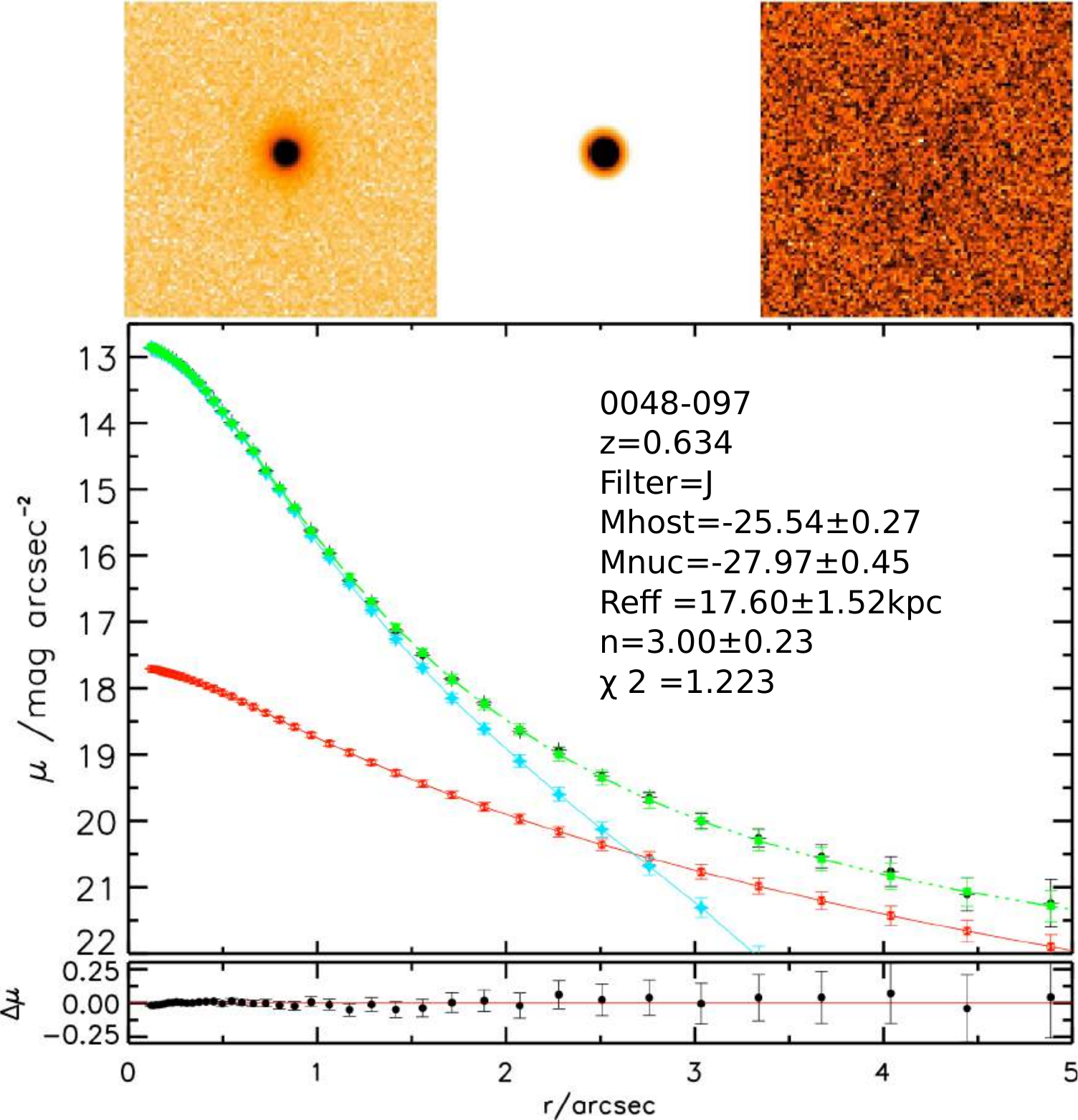} \\
	\includegraphics*[width=0.46\textwidth]{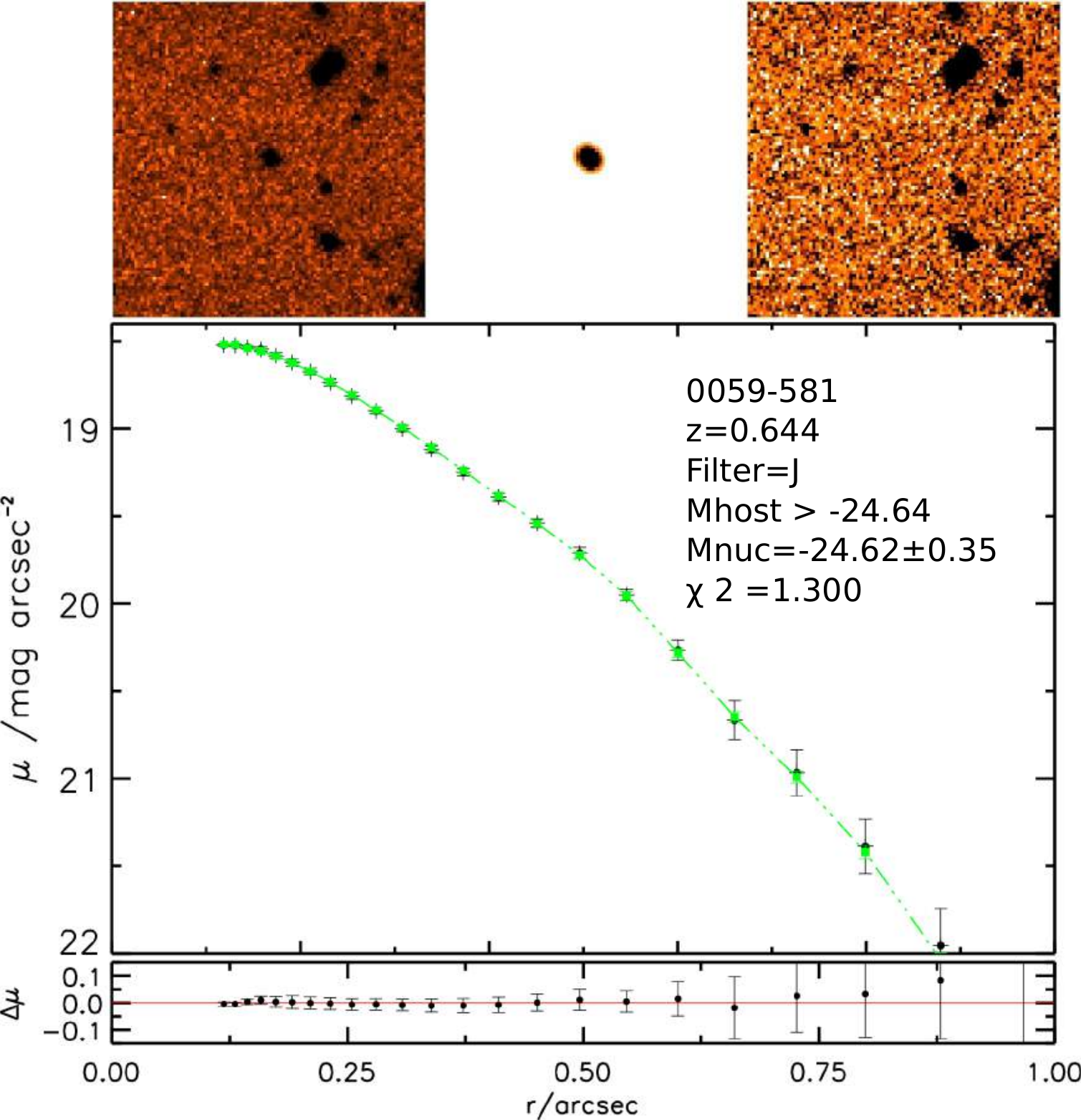}    &     \includegraphics*[width=0.46\textwidth]{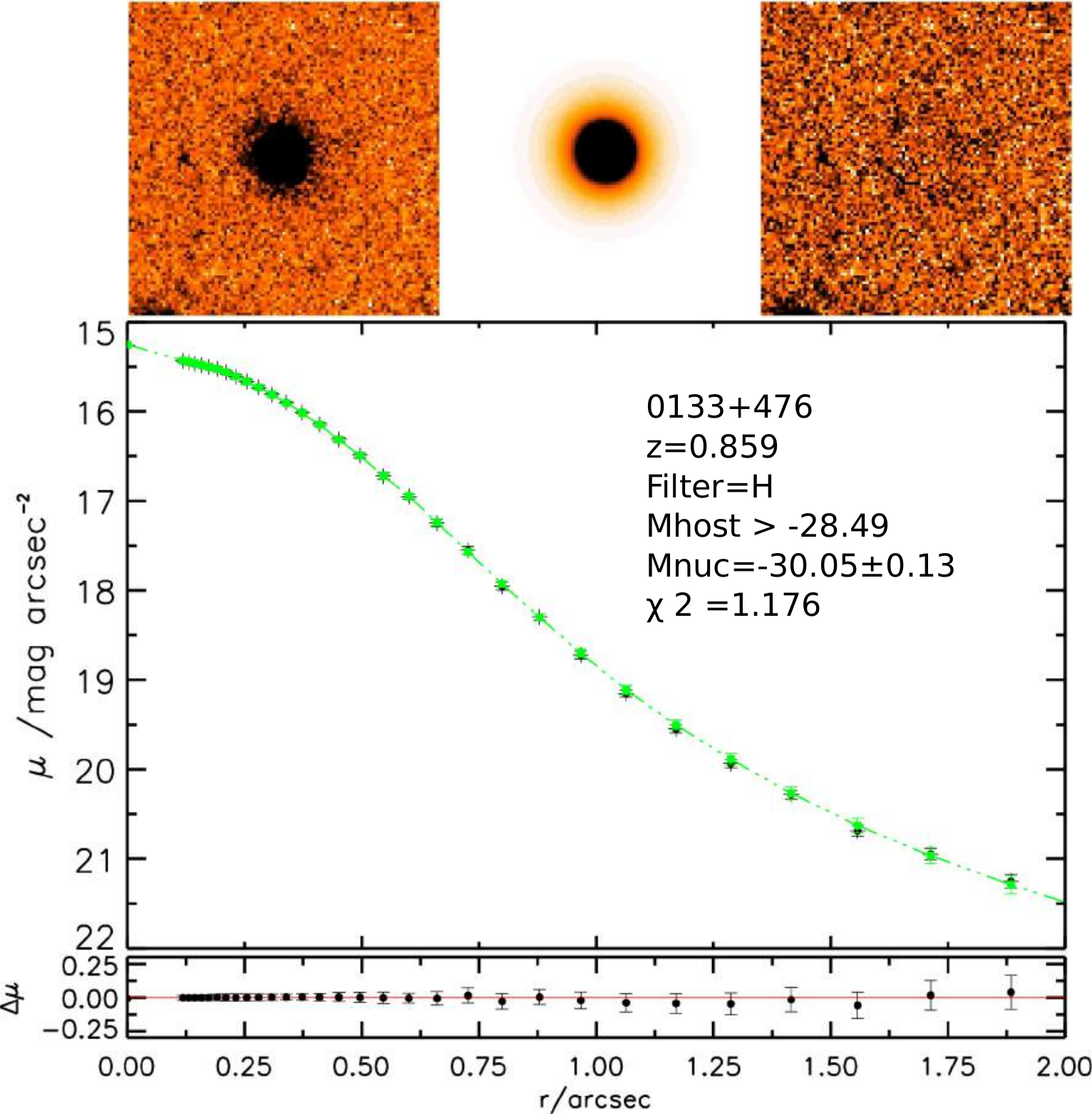} \\   
   \end{tabular}
  \caption{Top left sub--panel shows the reduced observed image. Middle top sub--panel shows the best--fit model. Right top sub--panel shows the model subtracted residuals. In the middle panel we plot the observed azimuthally averaged radial surface brightness profiles for each blazar (solid black data point), overlaid with the scaled PSF model (cian rhombus), the S\'ersic model convolved with the PSF (red circles) and the fitted PSF+S\'ersic model profile (green squares). The main fitting parameters are shown in the plot of each galaxy. In the bottom panel we plot the residuals of the model.}
   \label{figureA1}
 \end{minipage}
\end{figure*}

\begin{figure*}
 \begin{minipage}{180mm}
\begin{tabular}{@{}ll@{}}
	\includegraphics*[width=0.46\textwidth]{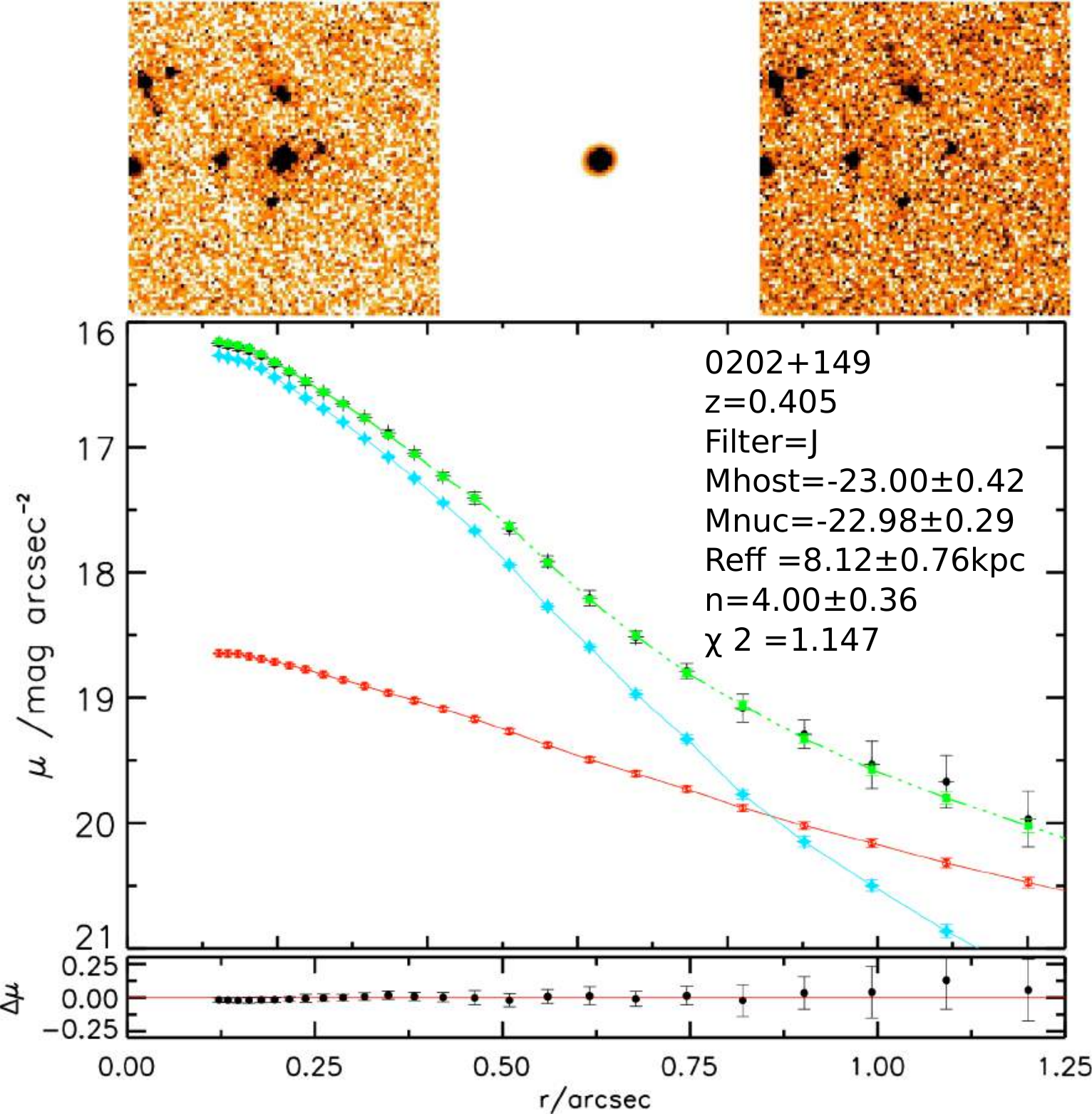}    &     \includegraphics*[width=0.46\textwidth]{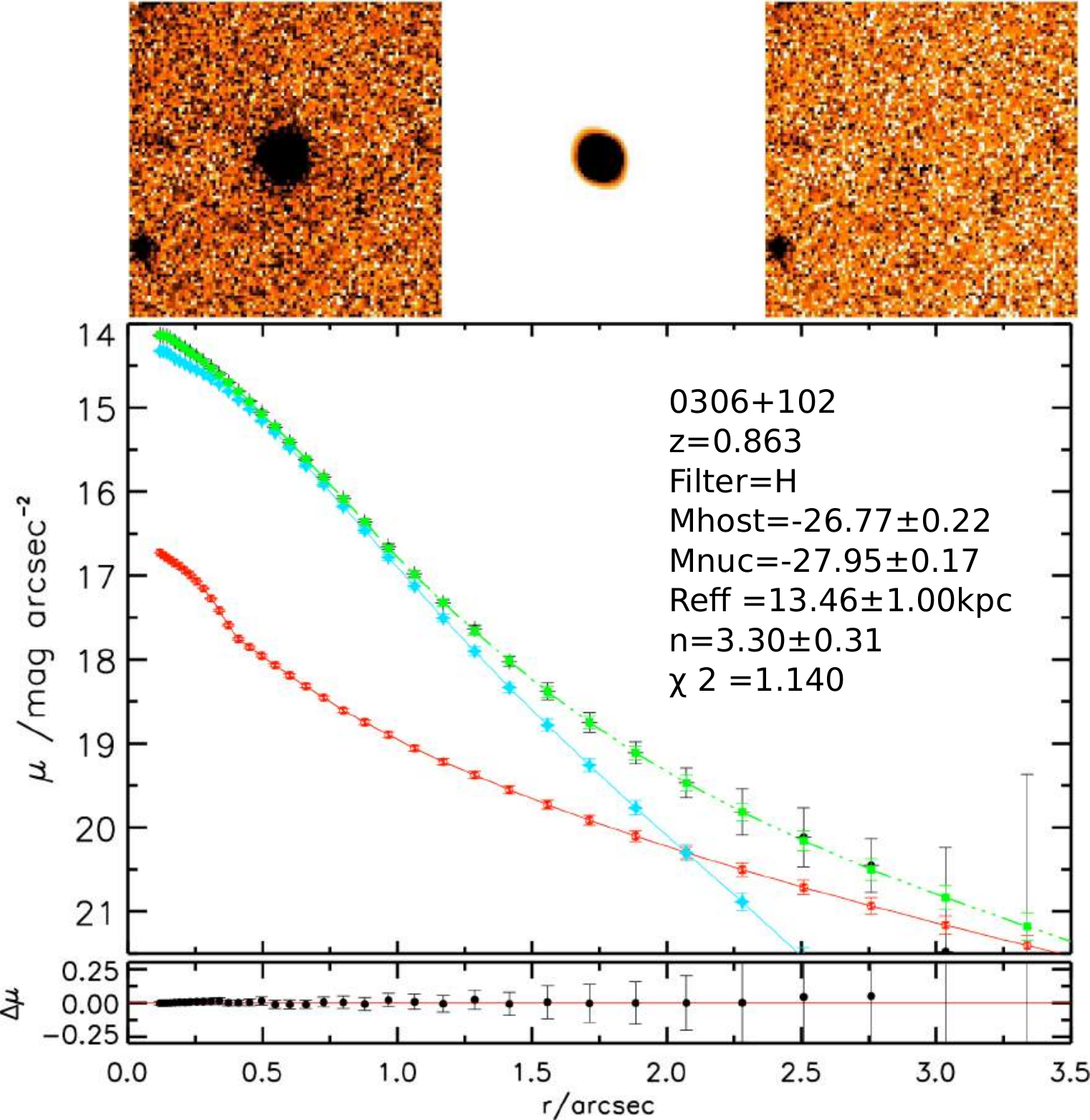} \\
	\includegraphics*[width=0.46\textwidth]{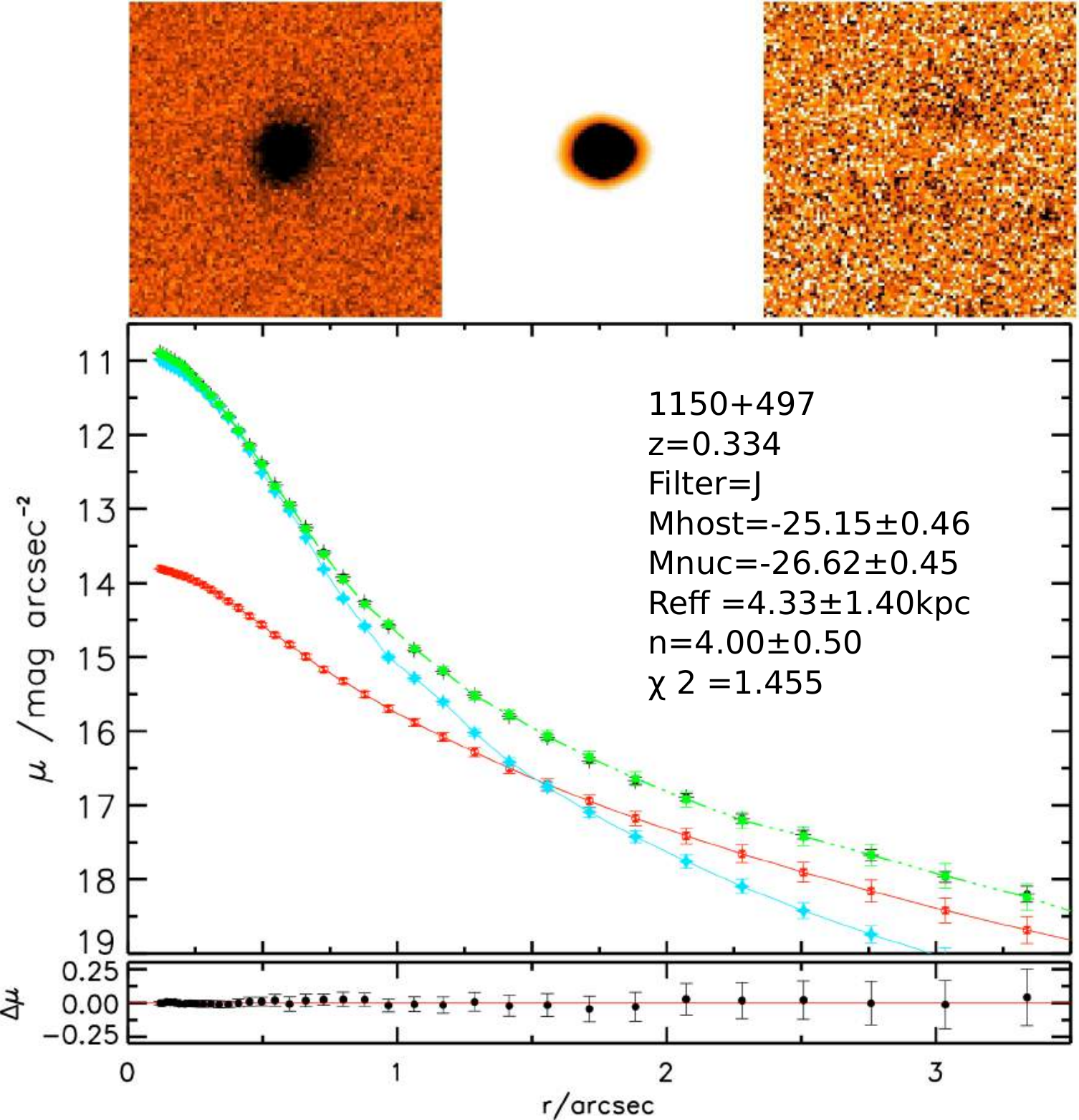}    &     \includegraphics*[width=0.46\textwidth]{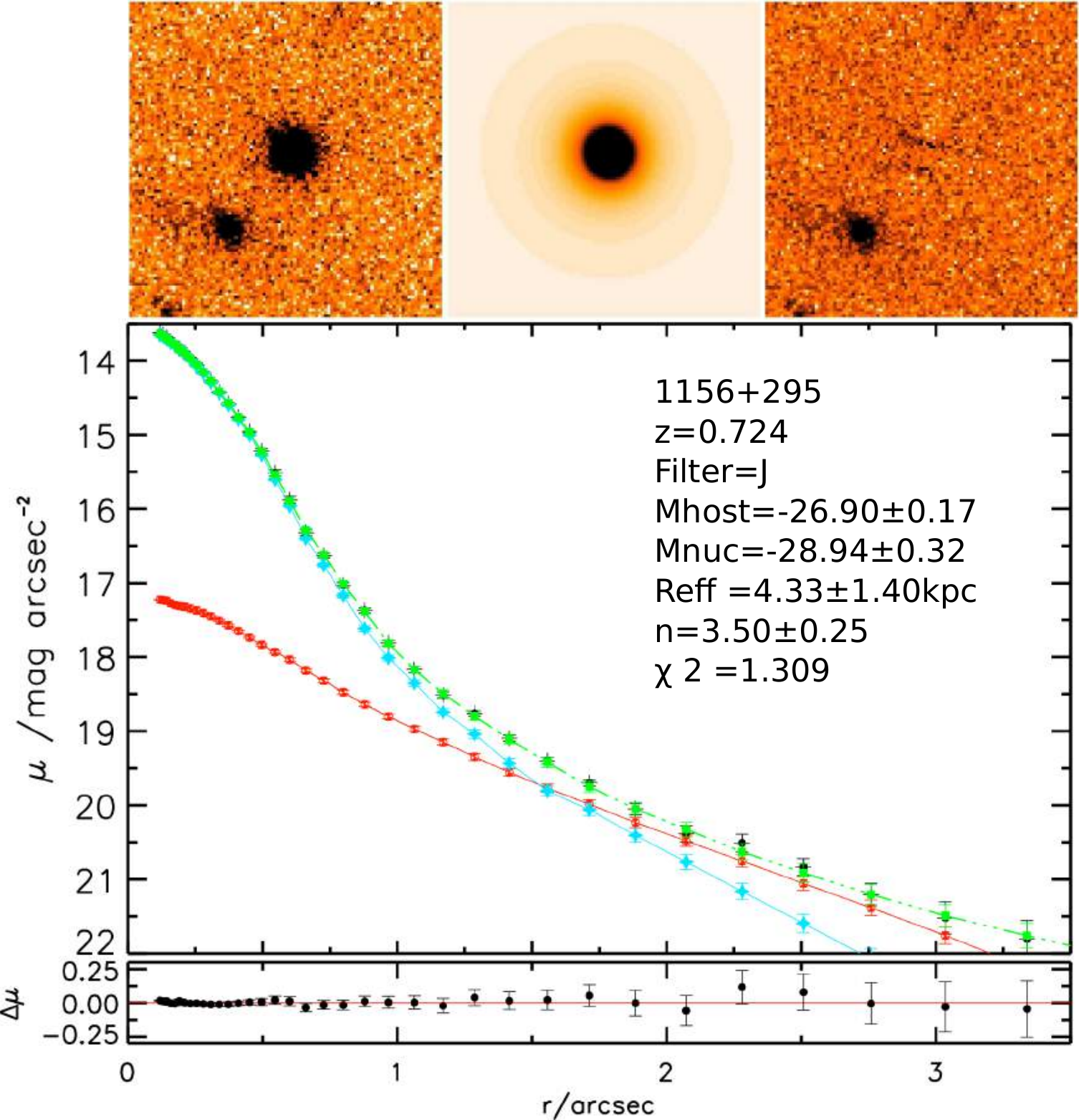} \\
   \end{tabular}
 \end{minipage}
 \textbf{Figure B1.} Continued...
\end{figure*}

\begin{figure*}
 \begin{minipage}{180mm}
\begin{tabular}{@{}ll@{}}
	\includegraphics*[width=0.46\textwidth]{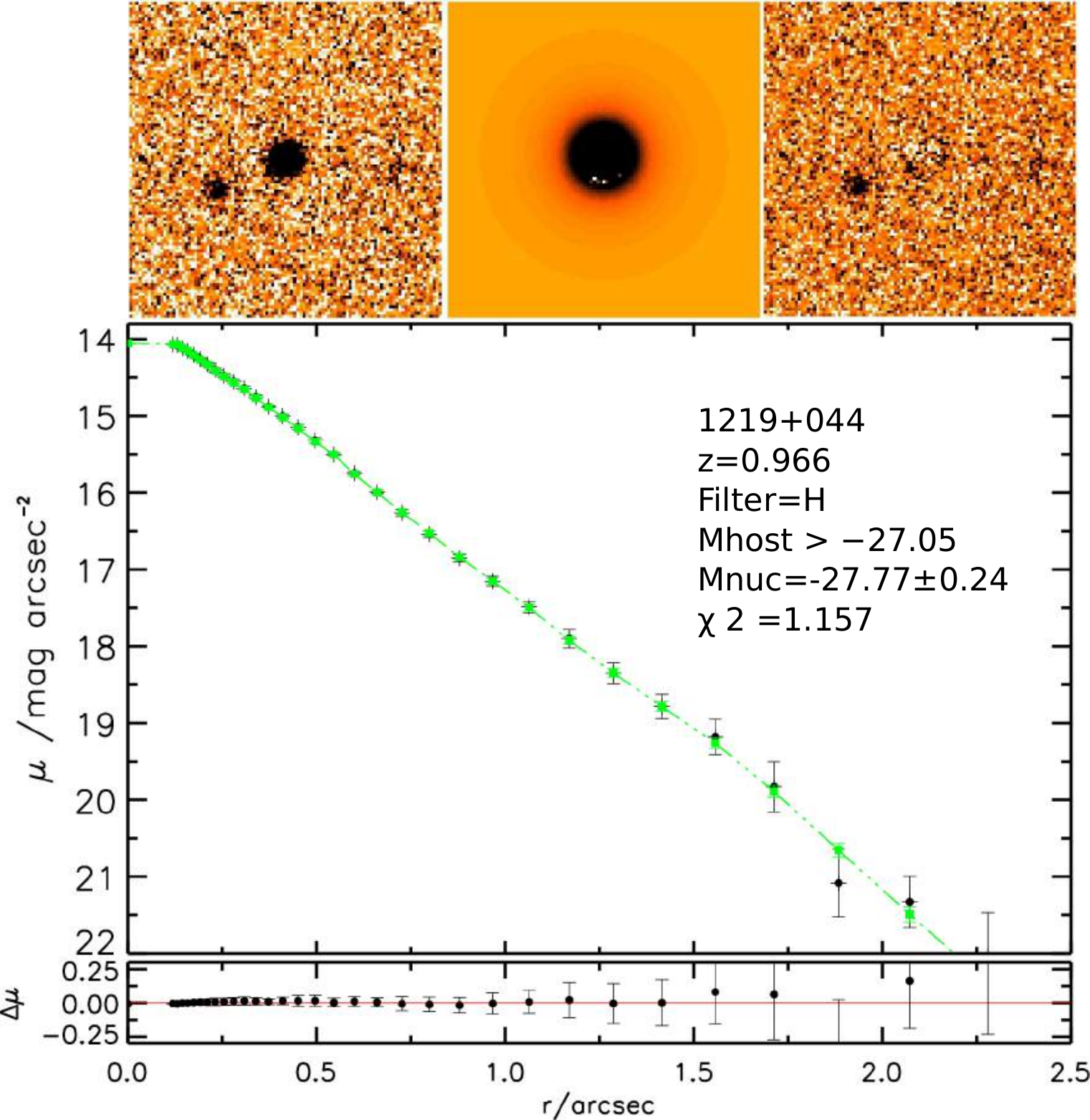}    &     \includegraphics*[width=0.46\textwidth]{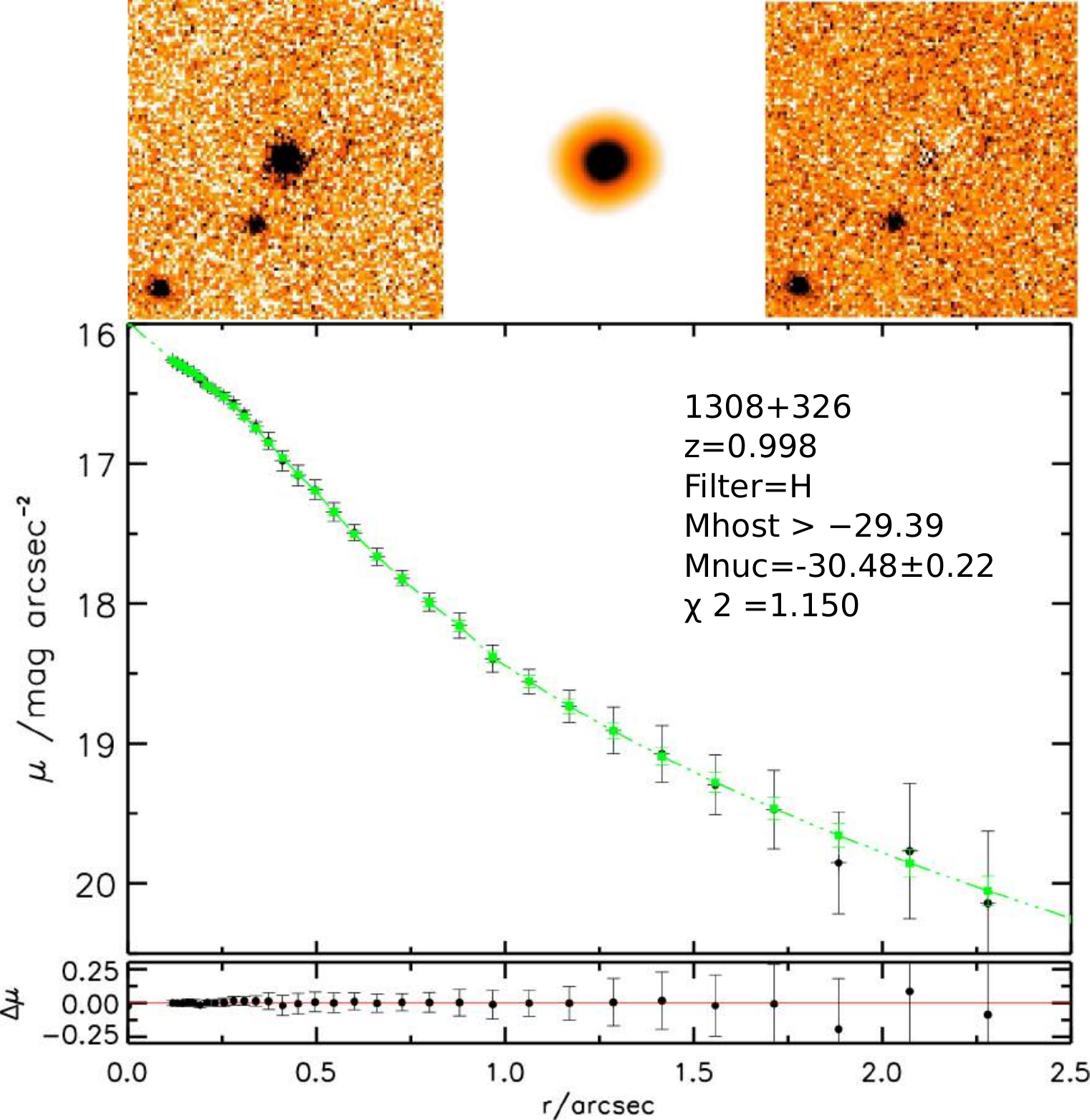} \\
	\includegraphics*[width=0.46\textwidth]{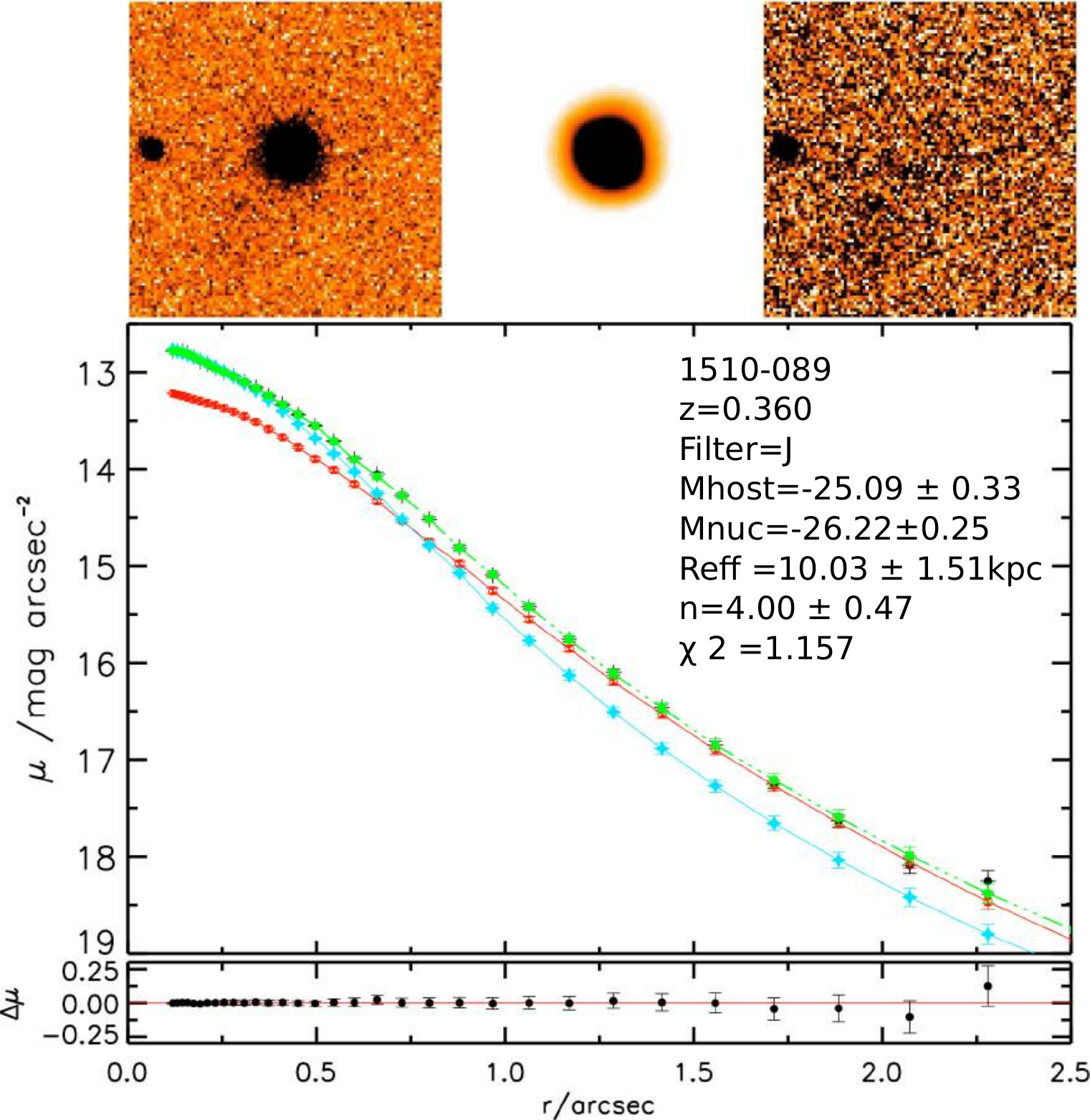}    &     \includegraphics*[width=0.46\textwidth]{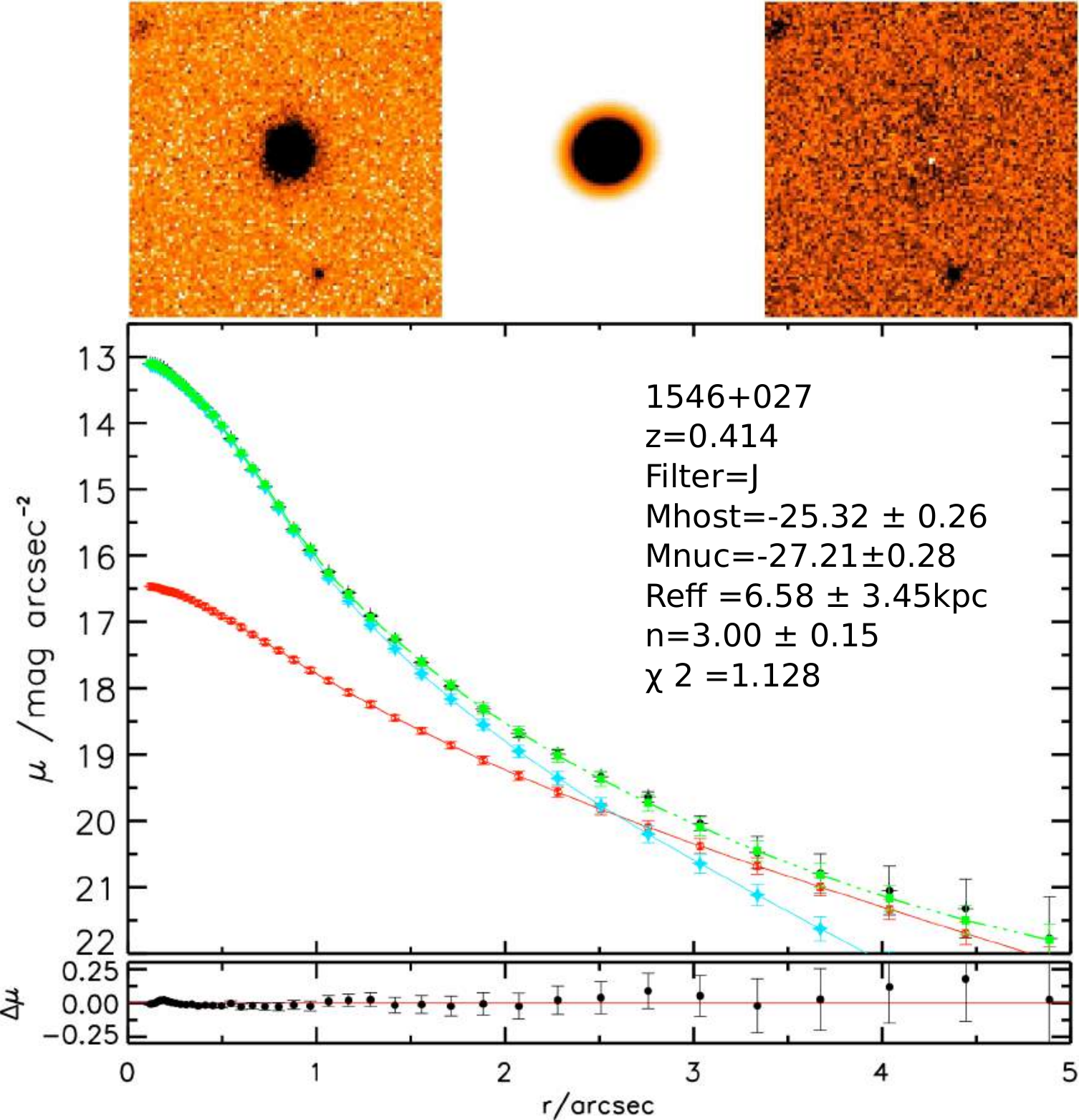} \\
   \end{tabular}
 \end{minipage}
 \textbf{Figure B1.} Continued...
\end{figure*}

\begin{figure*}
 \begin{minipage}{180mm}
\begin{tabular}{@{}ll@{}}
	\includegraphics*[width=0.46\textwidth]{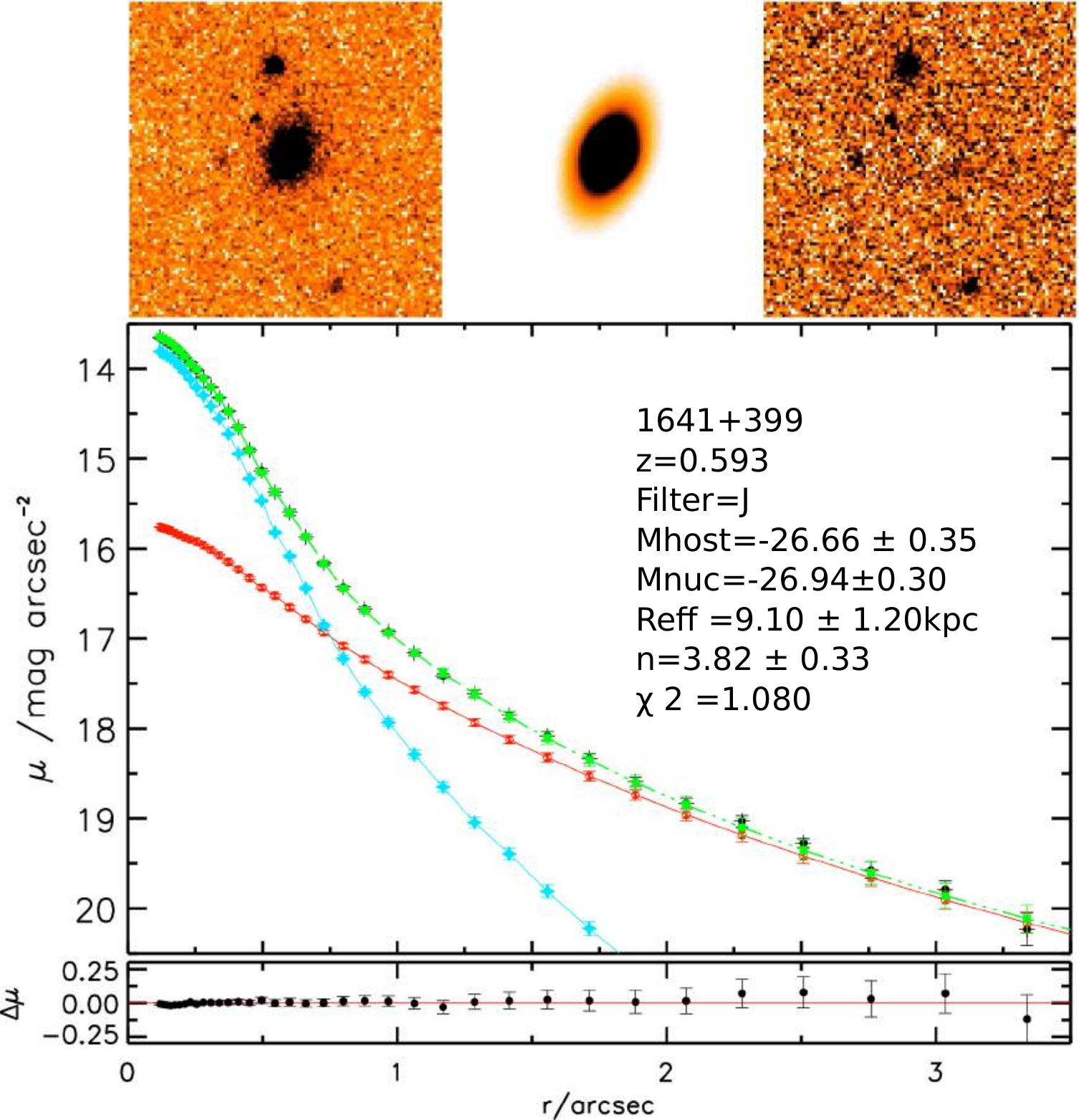}    &     \includegraphics*[width=0.46\textwidth]{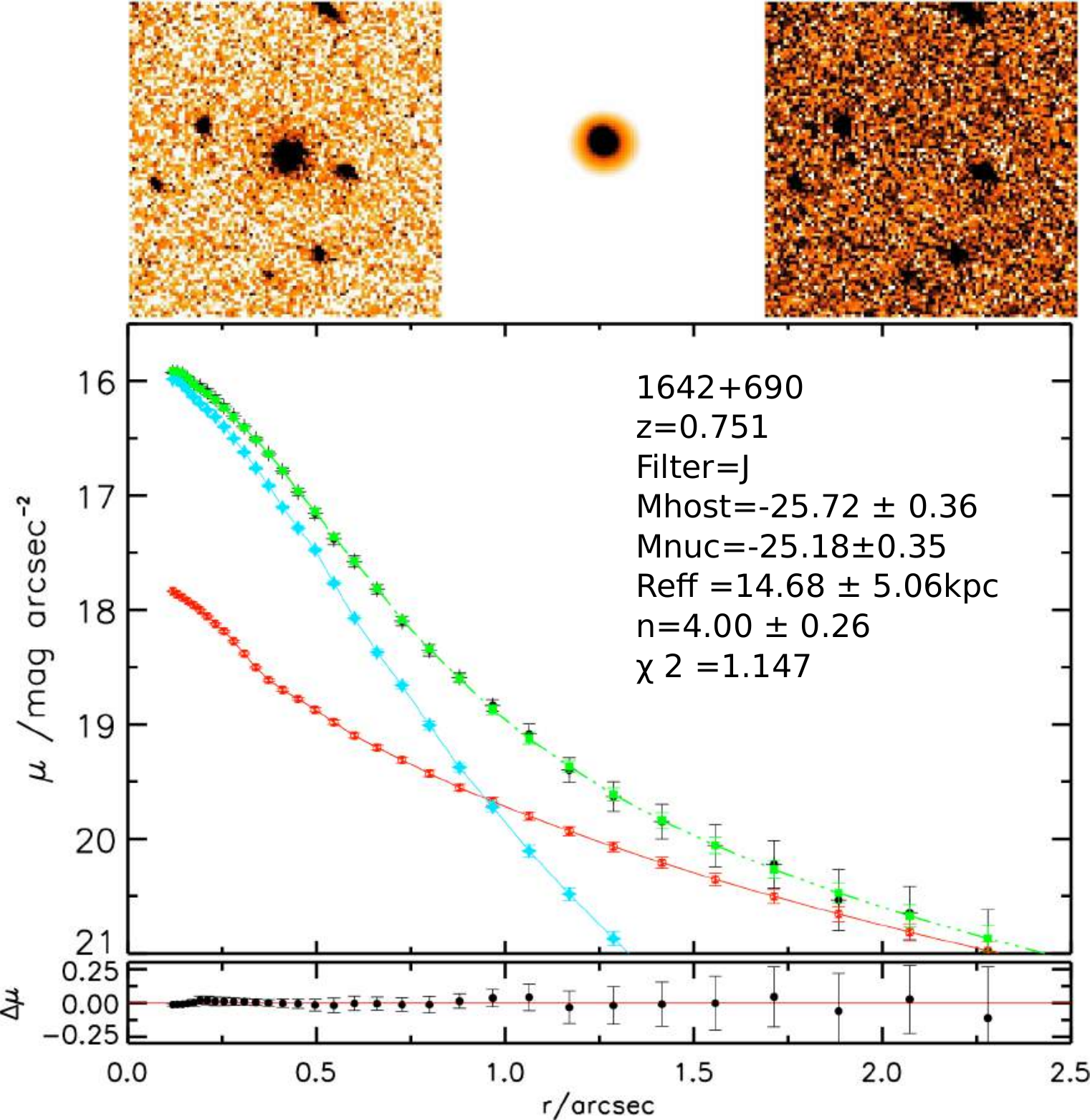} \\
	\includegraphics*[width=0.46\textwidth]{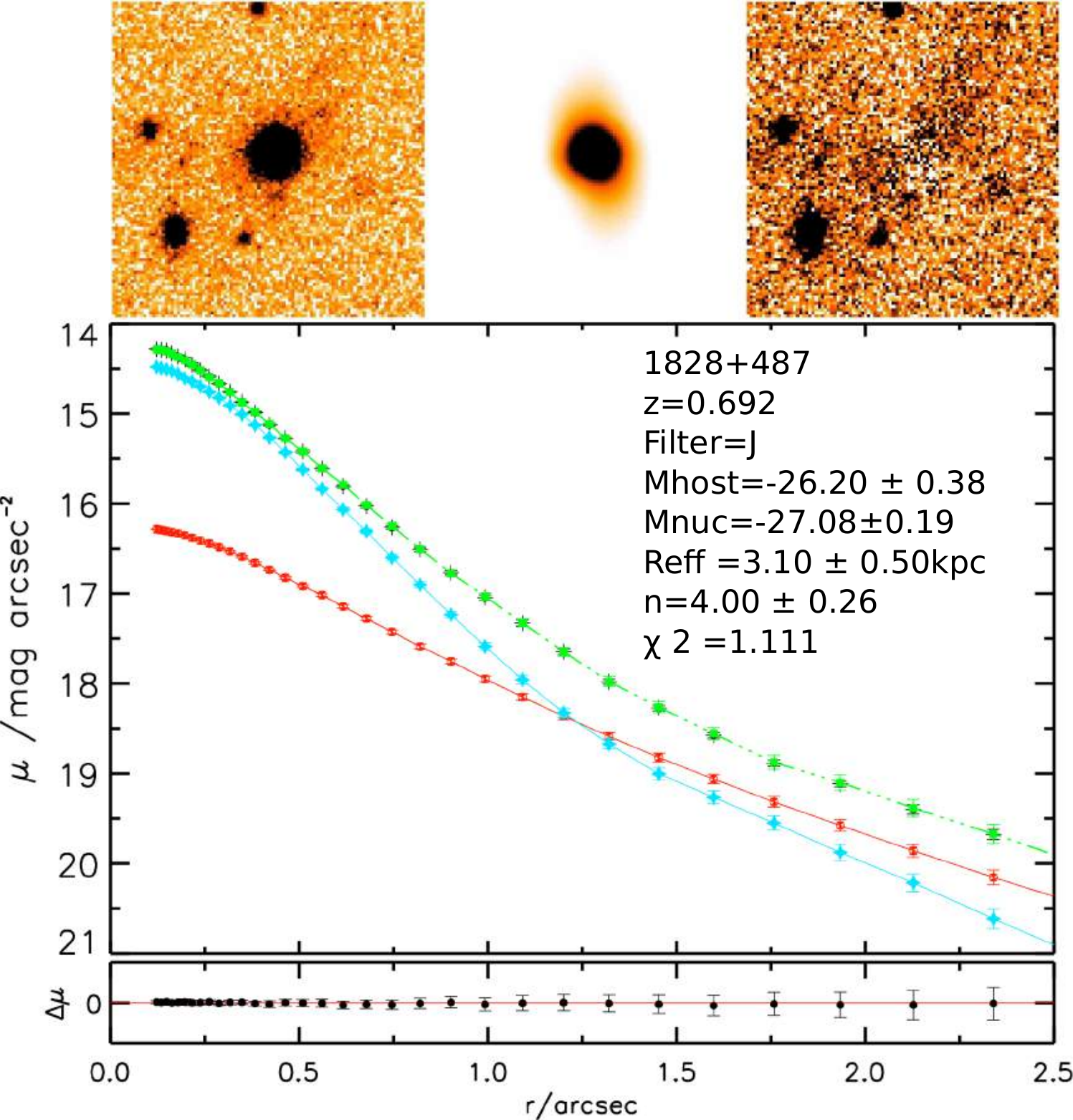}    &     \includegraphics*[width=0.46\textwidth]{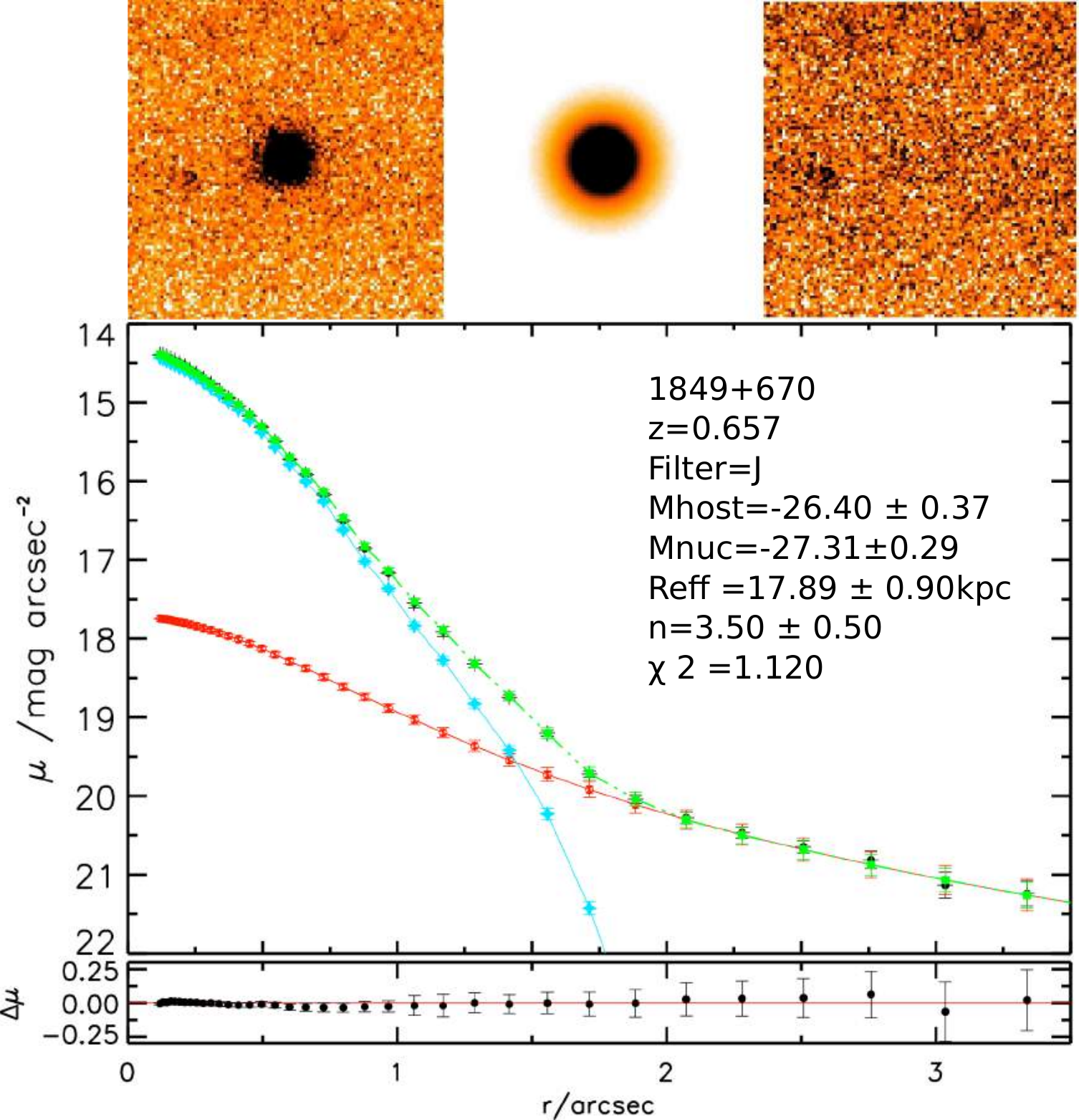} \\
   \end{tabular}
 \end{minipage}
 \textbf{Figure B1.} Continued...
\end{figure*}

\begin{figure*}
 \begin{minipage}{180mm}
\begin{tabular}{@{}ll@{}}
	\includegraphics*[width=0.46\textwidth]{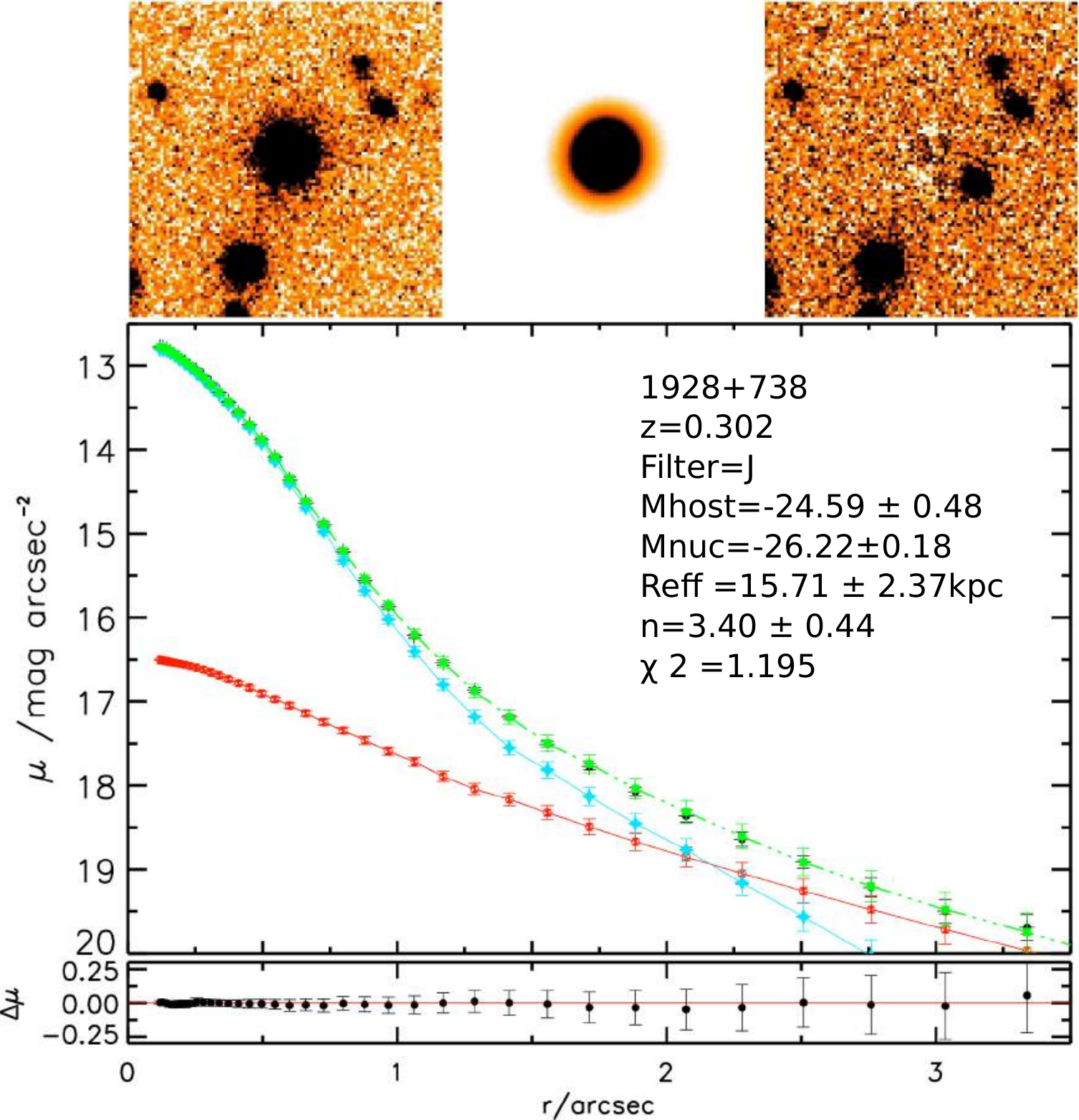}    &     \includegraphics*[width=0.46\textwidth]{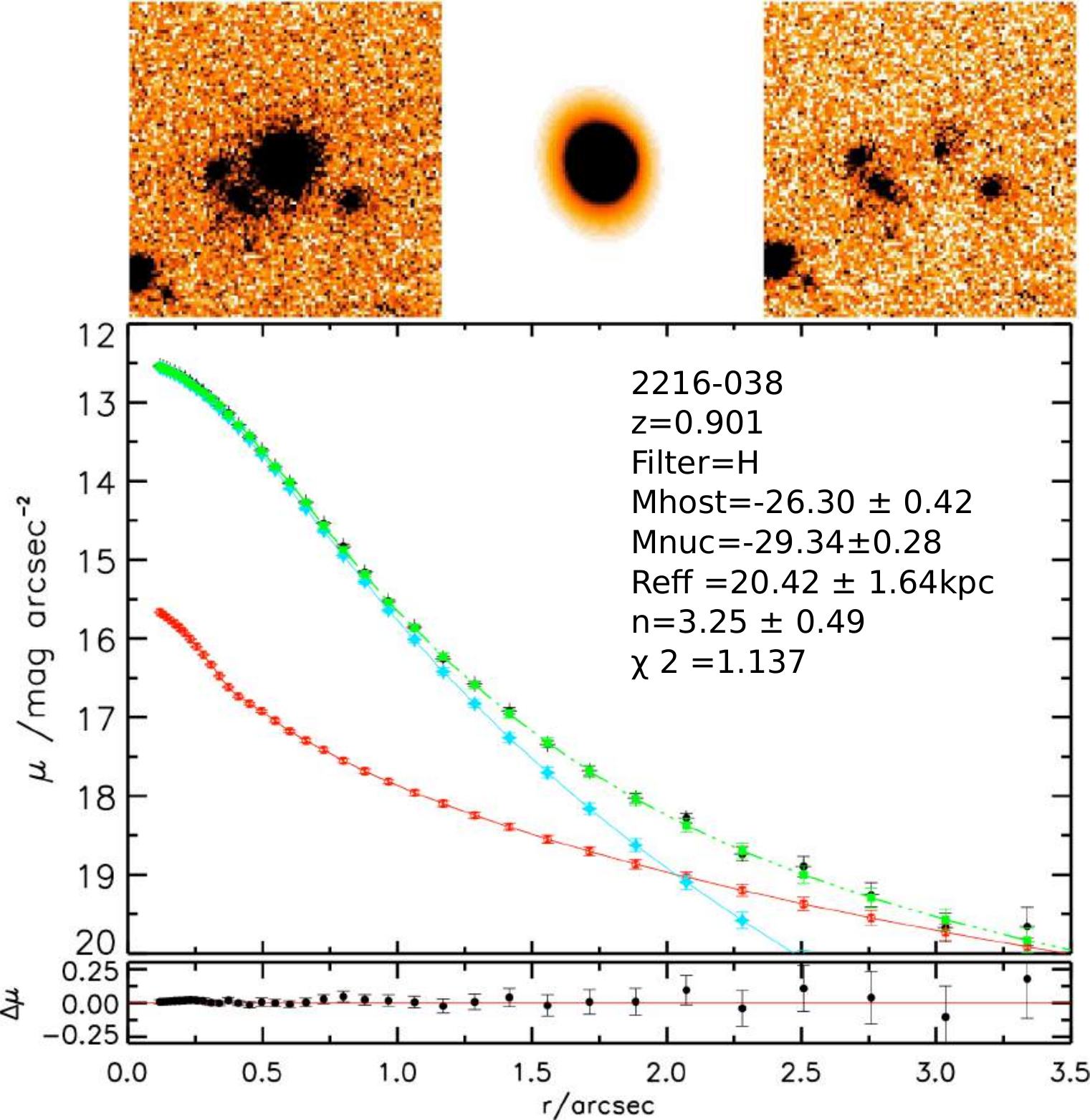} \\
	\includegraphics*[width=0.46\textwidth]{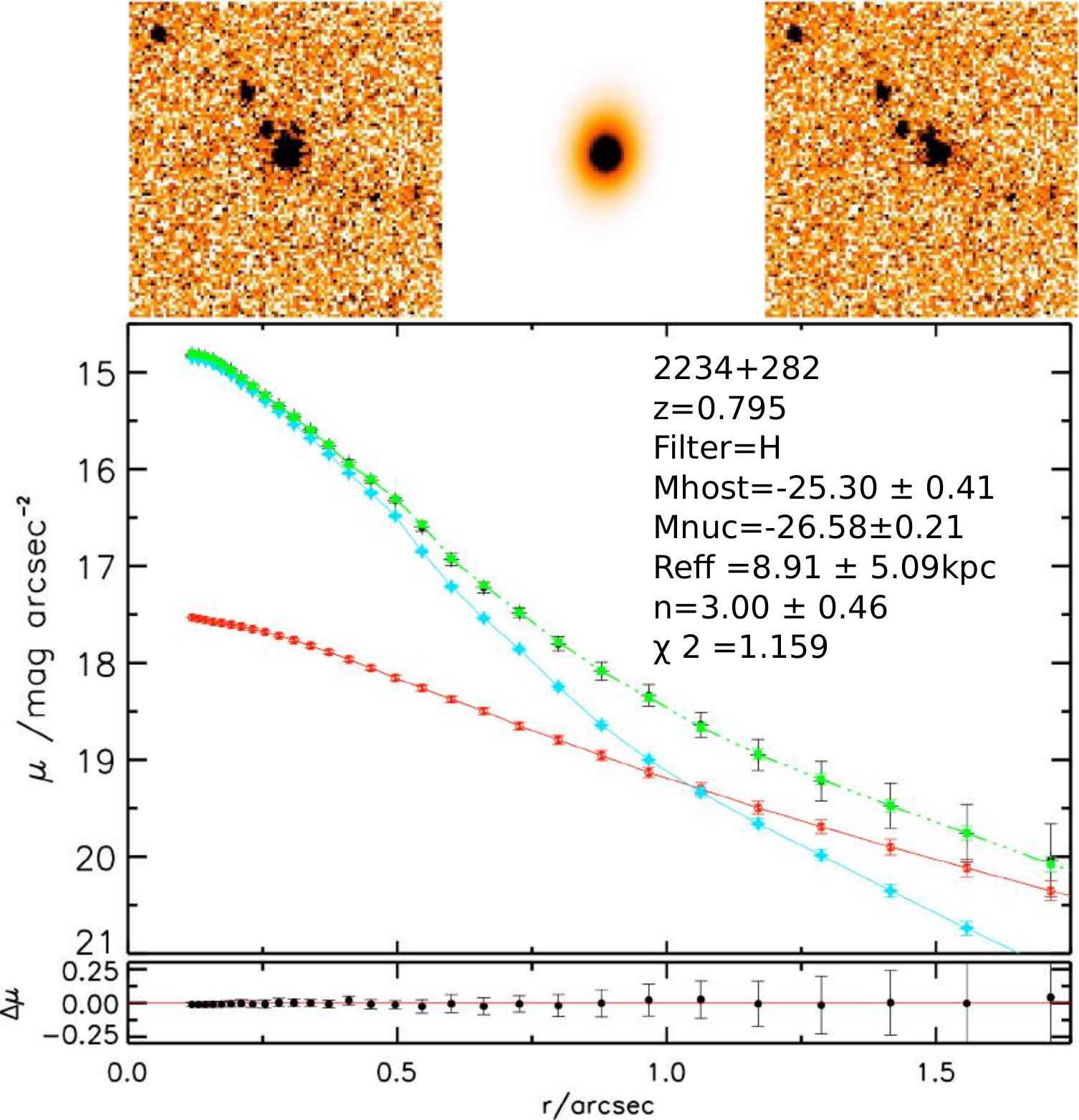}    &    \\
   \end{tabular}
 \end{minipage}
 \textbf{Figure B1.} Continued...
\end{figure*}

%%%%%%%%%%%%%%%%%%%%%%%%%%%%%%%%%%%%%%%%%%%%%%%%%%%%%%%%%%%%%%%%%%%%%%%
%%%%%%%%%%%%%%%%%%%%%%%%%%%%%%%%%%%%%%%%%%%%%%%%%%%%%%%%%%%%%%%%%%%%%%%%%

%%%%%%%%%%%%%%%%%%
%%%%%%%%%%%%%%%%%%
\begin{figure*}
 \begin{minipage}{180mm}
\begin{tabular}{@{}llll@{}}
		\bf{0003$-$066}  									 &  \bf{0048$-$097}  										&  \bf{0059$+$581}											&	\bf{0133$+$476}								\\
	\includegraphics[width=0.23\textwidth]{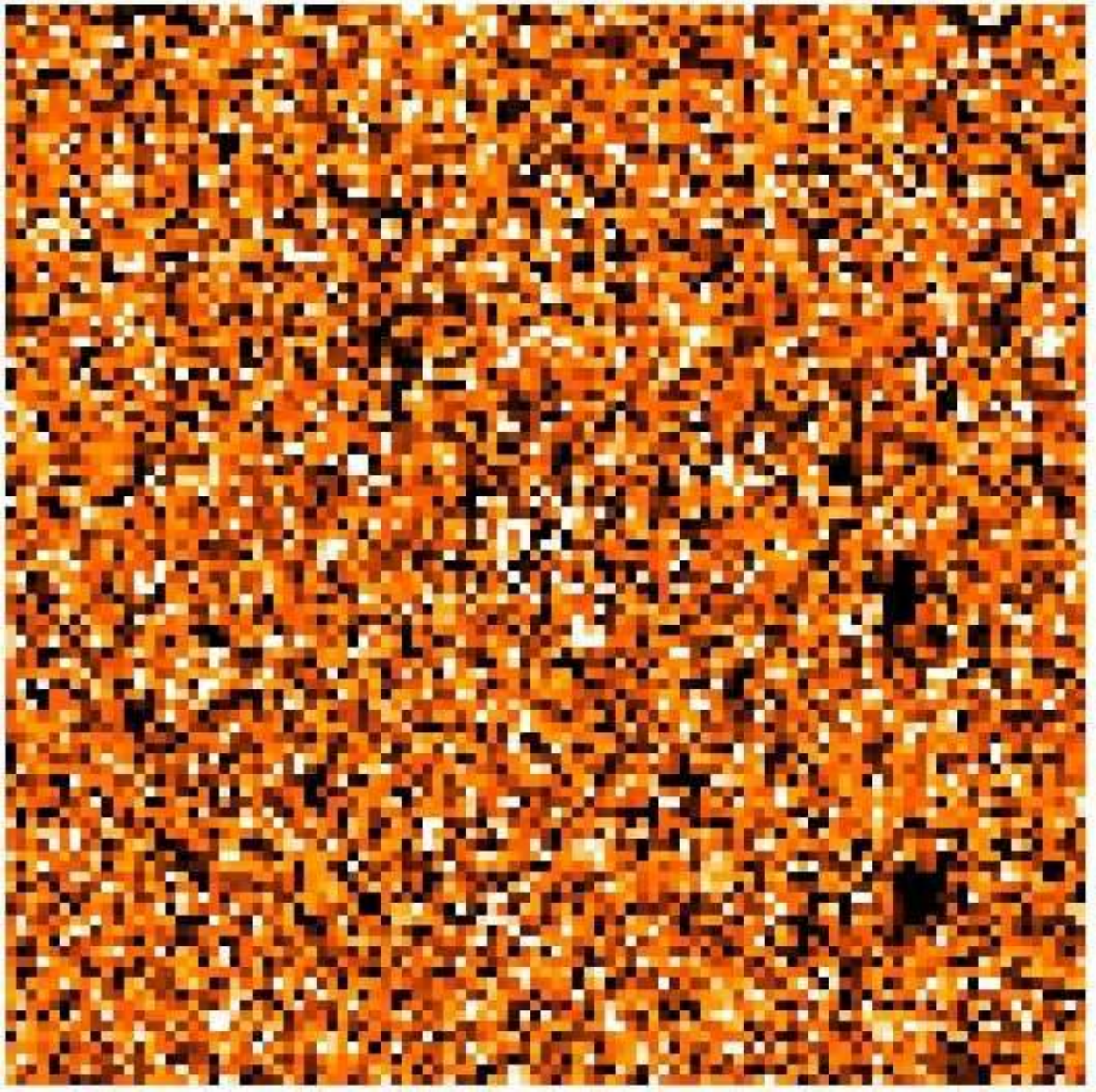}    &     \includegraphics[width=0.23\textwidth]{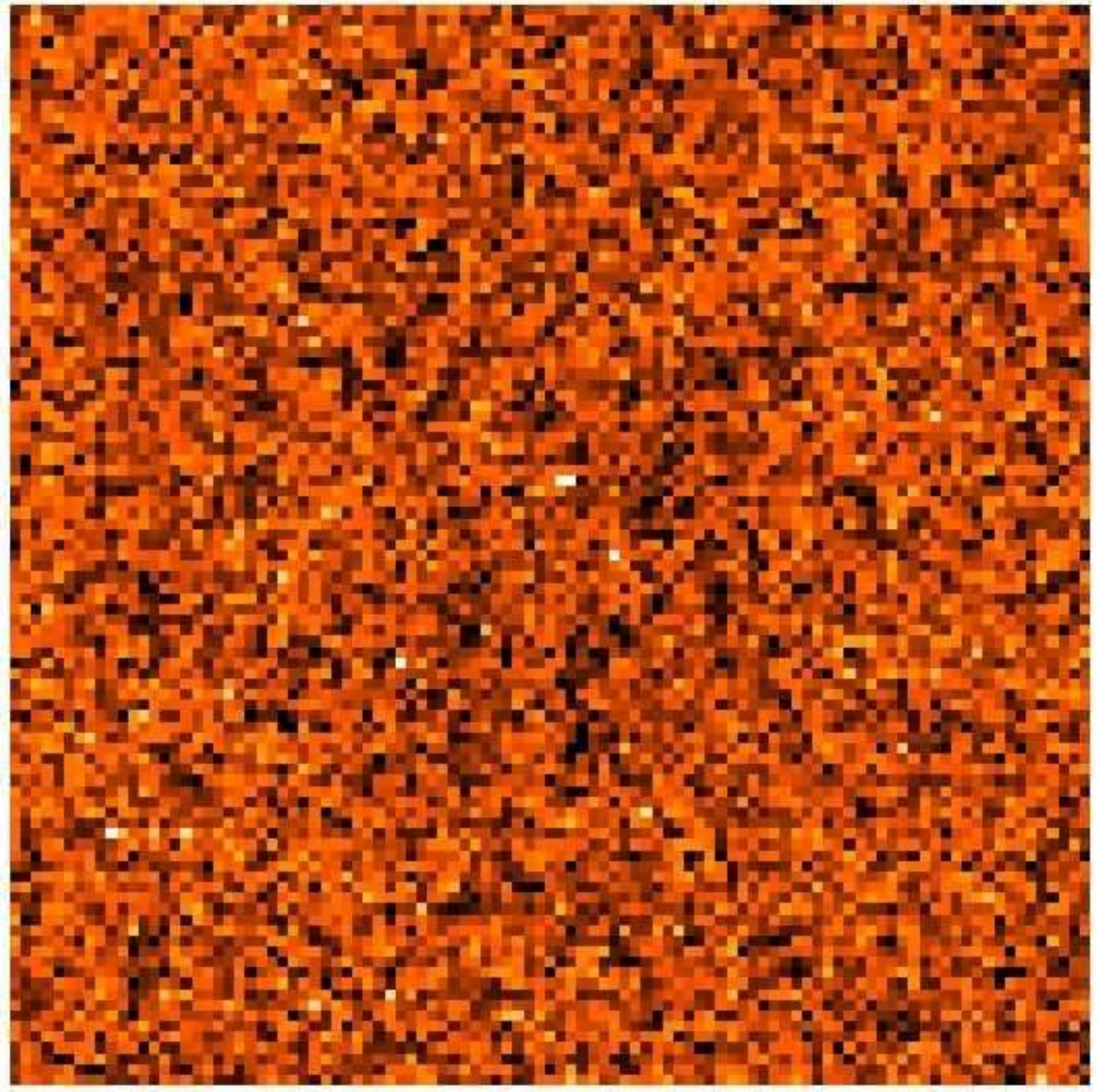} & \includegraphics[width=0.23\textwidth]{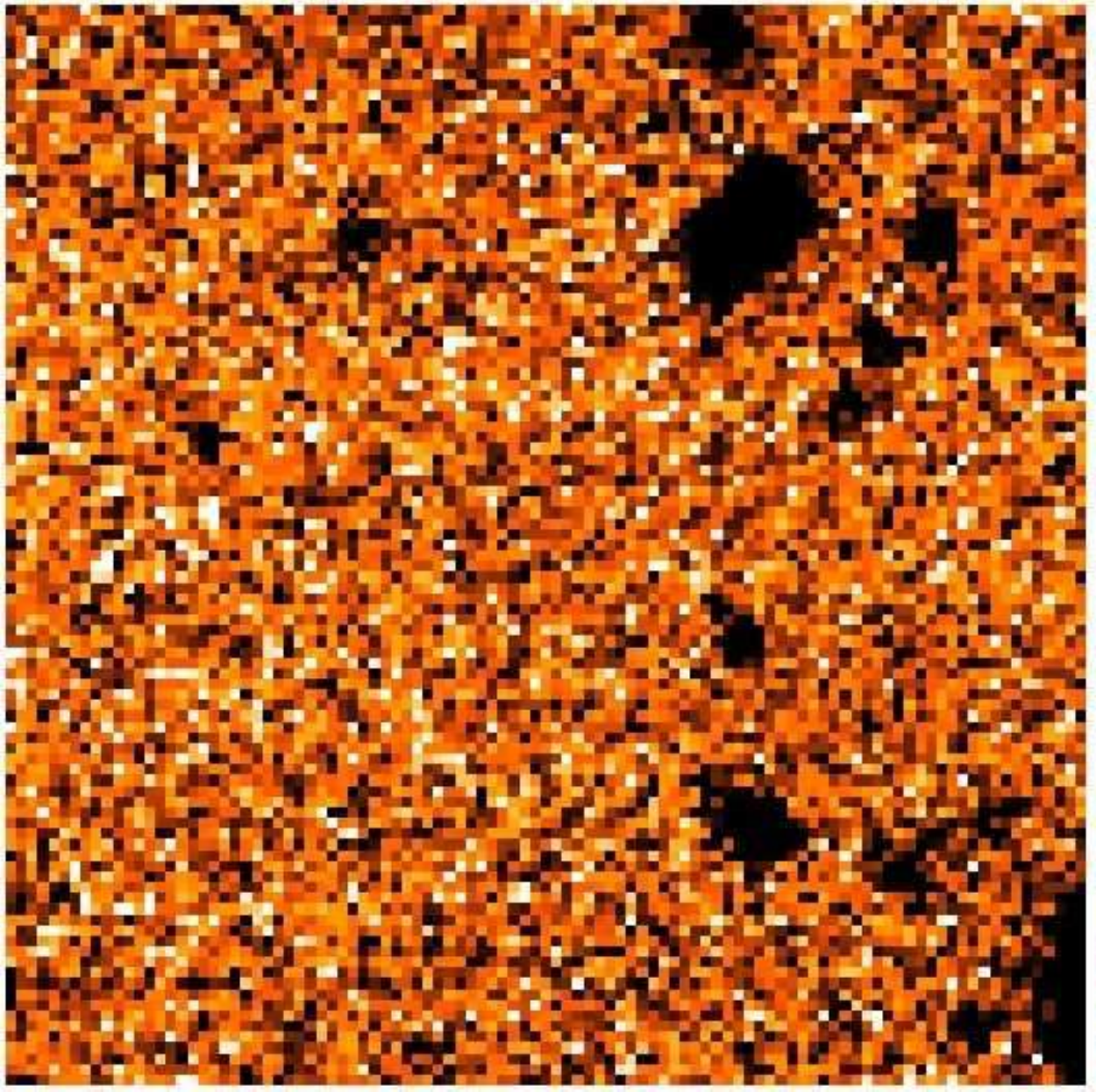} &\includegraphics[width=0.23\textwidth]{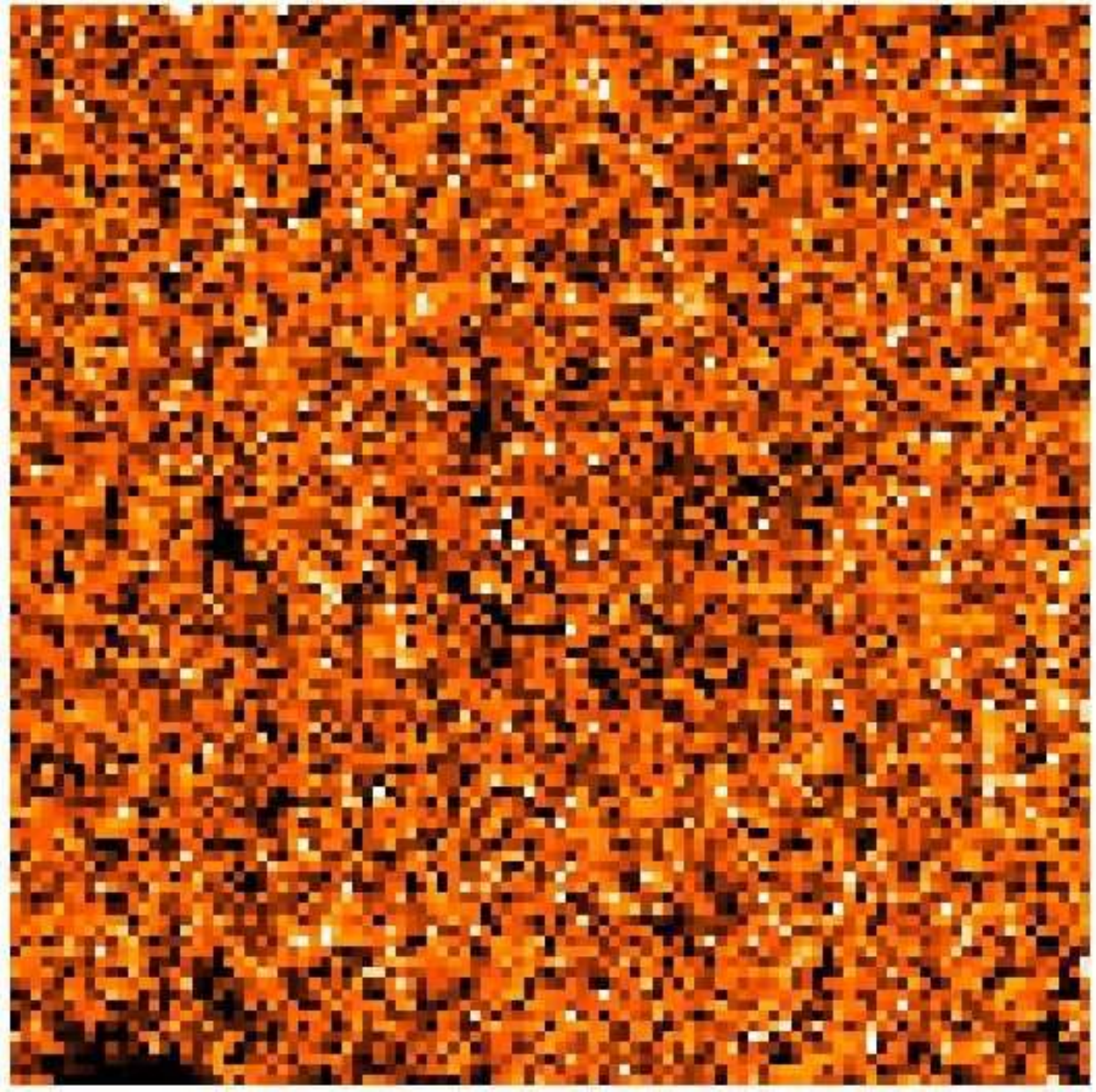}  \\
	\bf{0202$+$149}										 &\bf{0306$+$102}  									  & \bf{1150$+$497} 								 	  	&  	\bf{1156$+$295} 									 \\    										
	\includegraphics[width=0.23\textwidth]{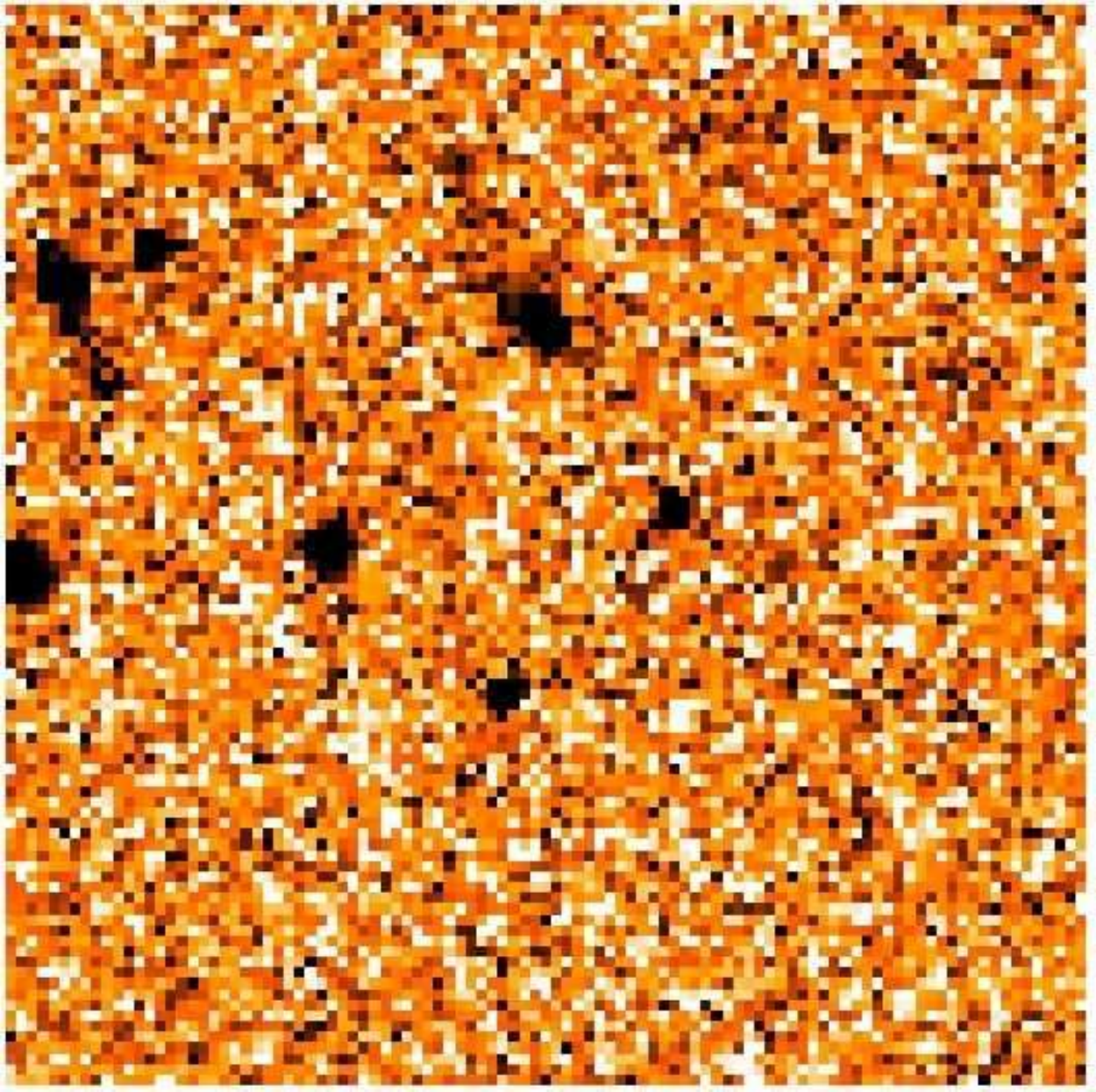}   &  \includegraphics[width=0.23\textwidth]{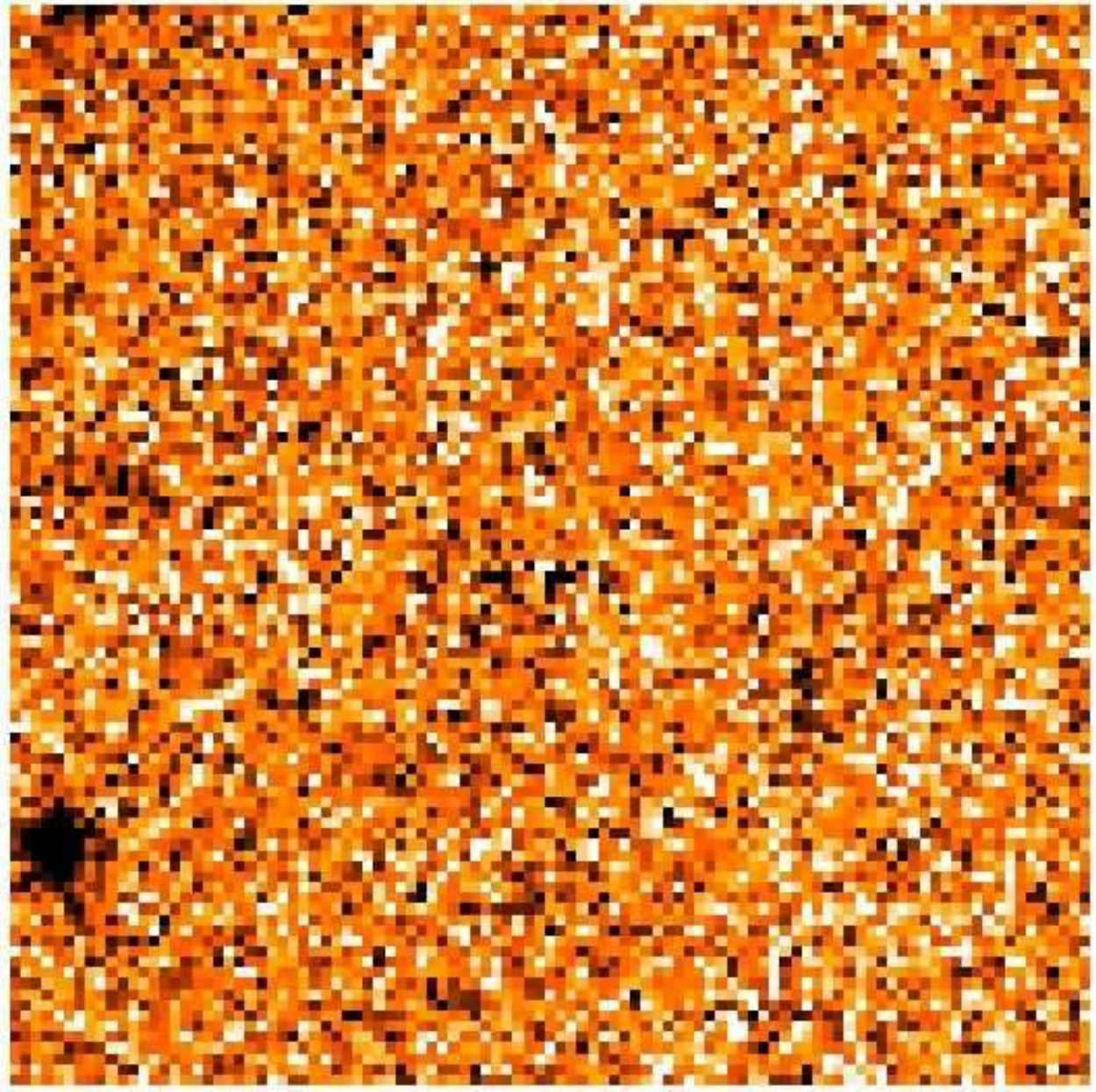}  &     \includegraphics[width=0.23\textwidth]{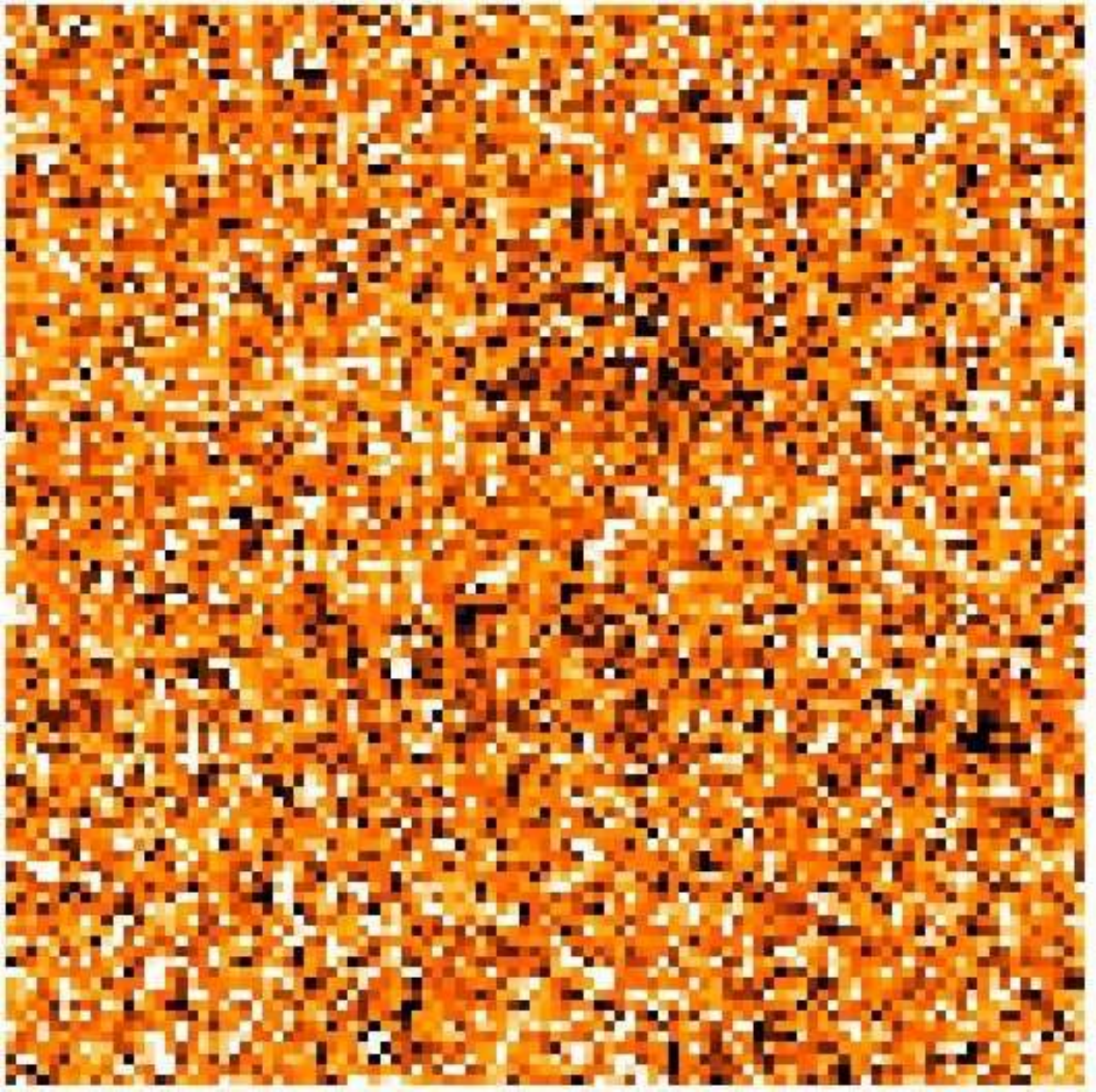} & \includegraphics[width=0.23\textwidth]{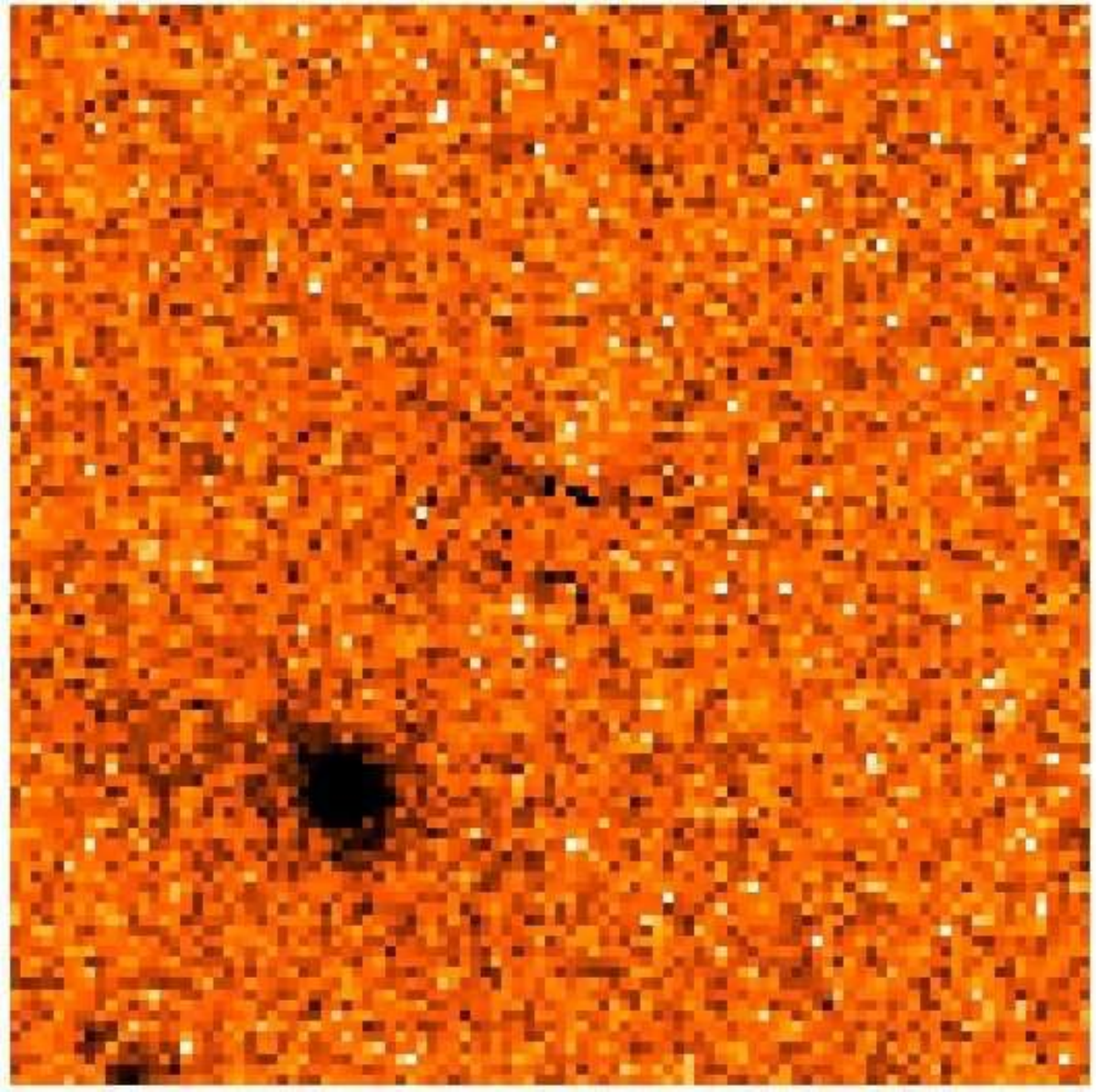}  \\
	\bf{1219$+$044}									   	 &	\bf{1308$+$326}  										&\bf{1510$-$089}  										 & \bf{1546$+$027} 								 	   \\
	\includegraphics[width=0.23\textwidth]{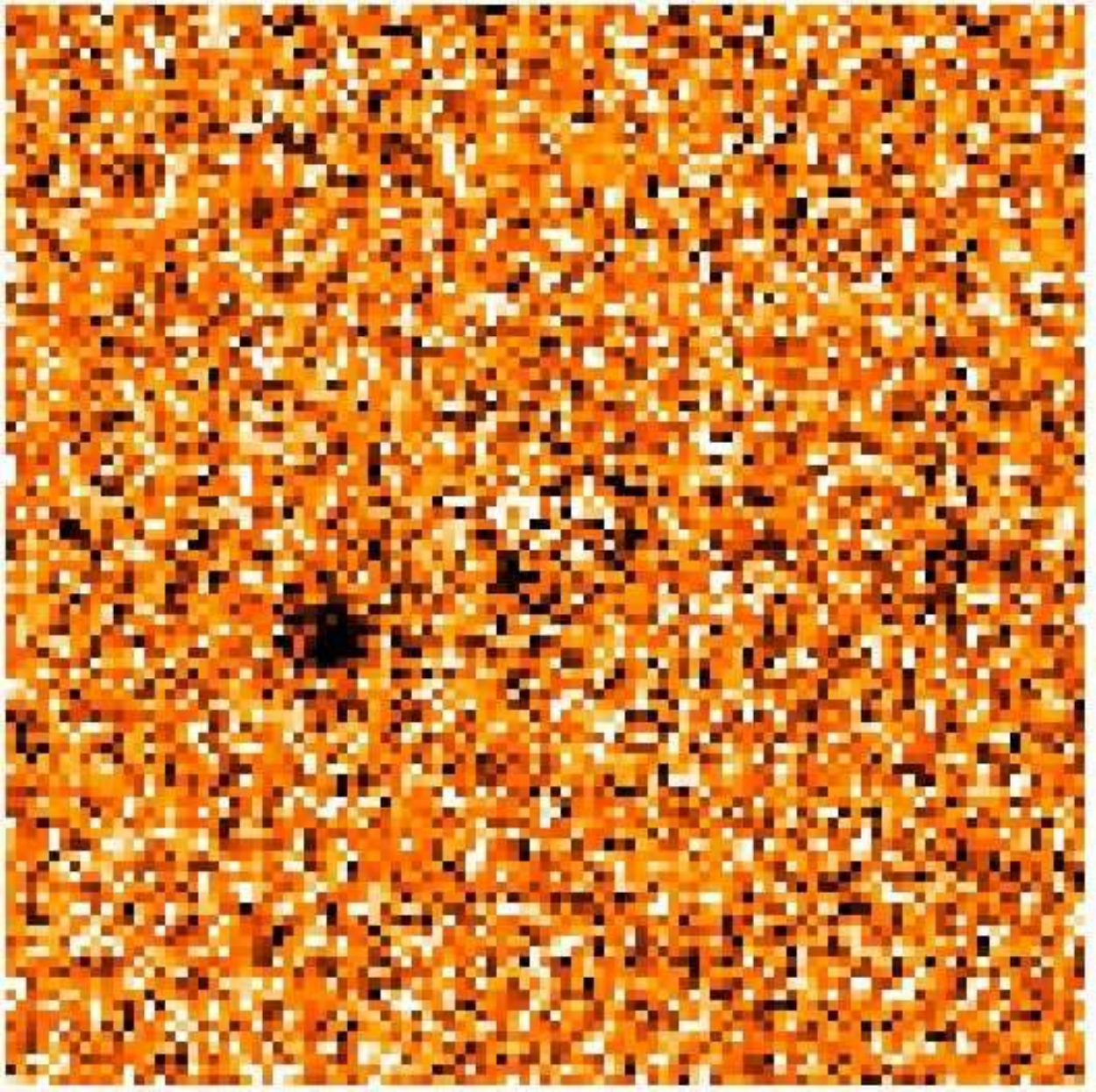}   &	\includegraphics[width=0.23\textwidth]{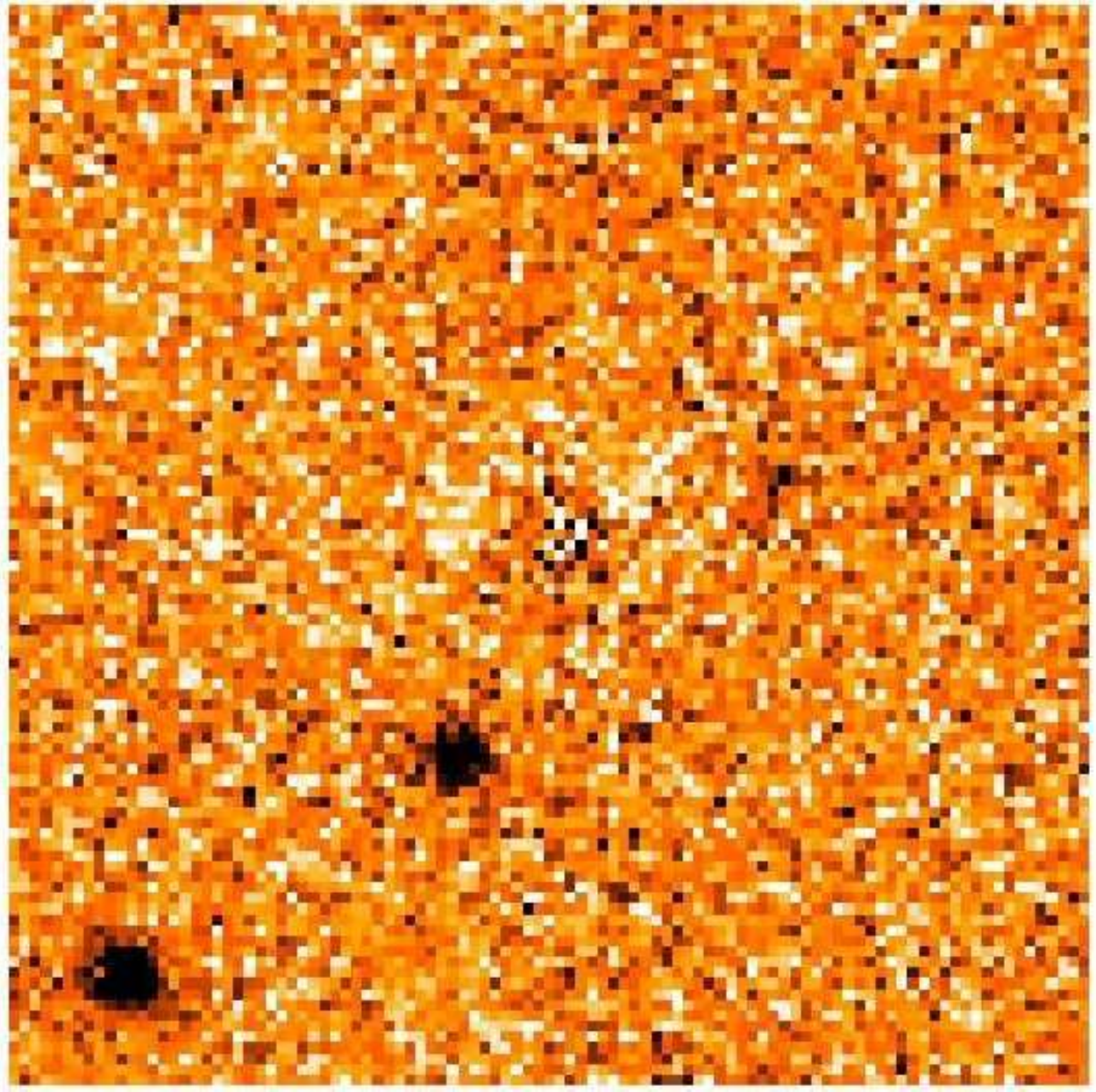}  & \includegraphics[width=0.23\textwidth]{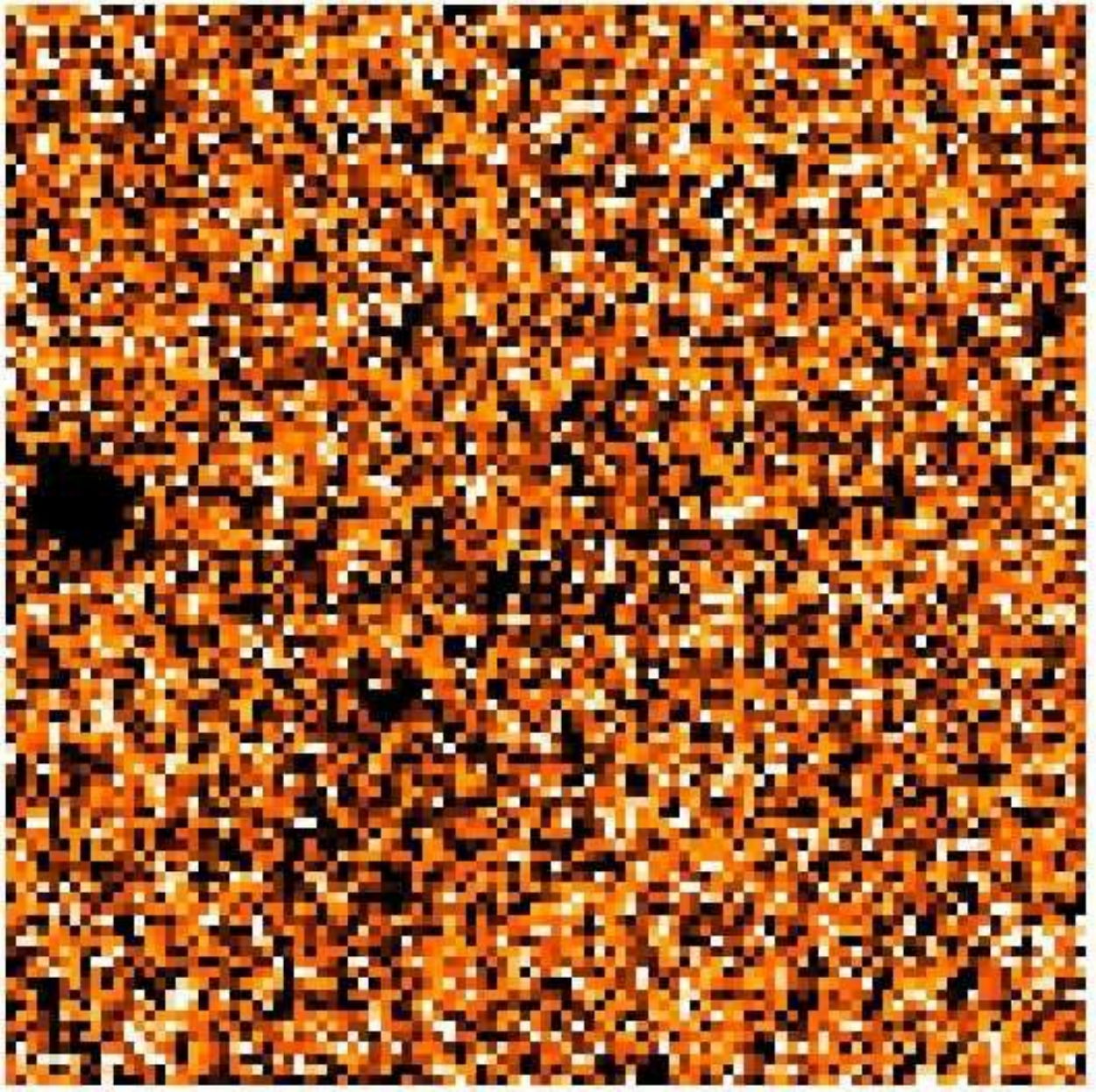}  &     \includegraphics[width=0.23\textwidth]{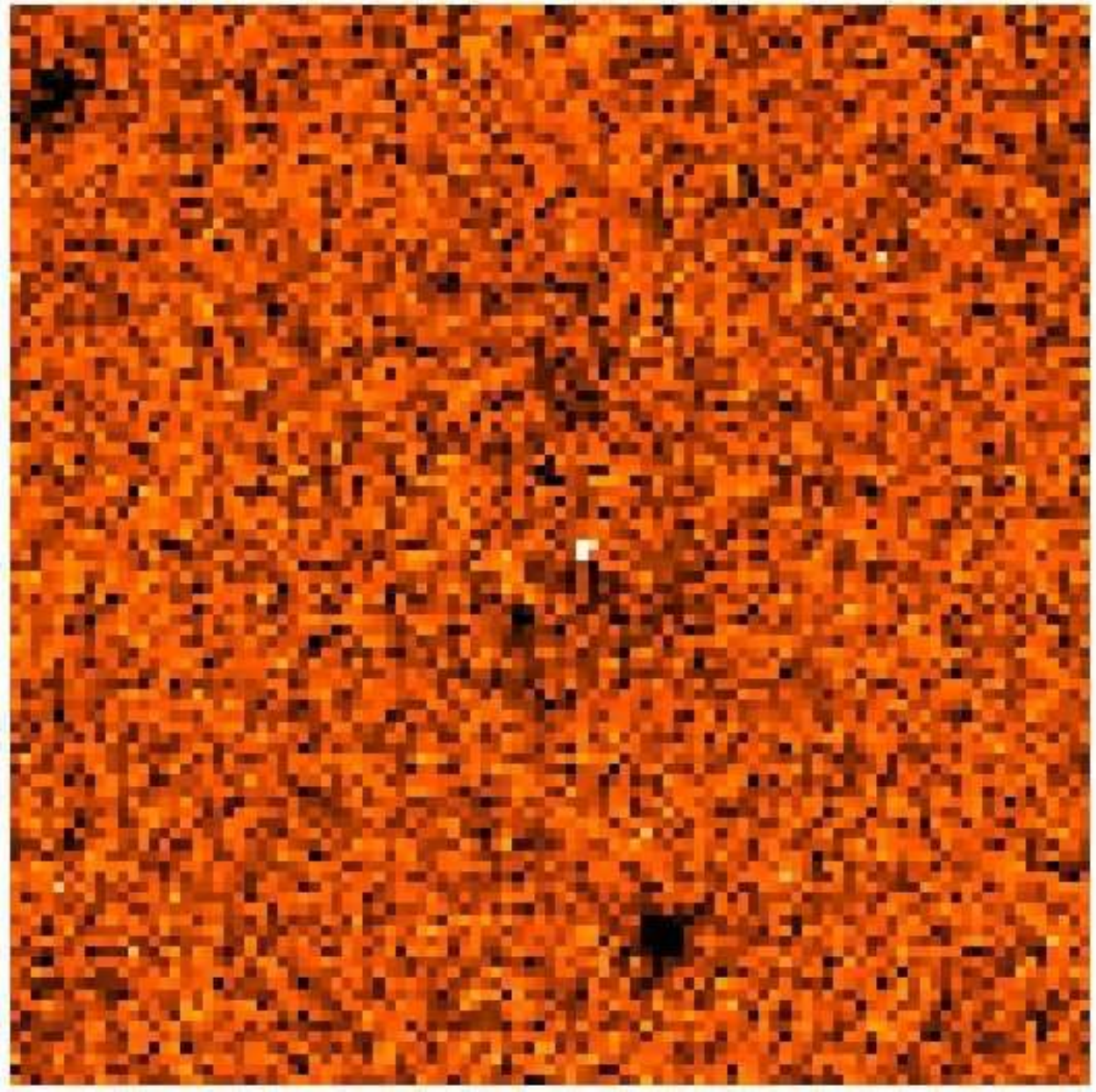} \\
		\bf{1641$+$399 Interacting?} 									&	\bf{1642$+$690}      								&	\bf{1828$+$487 Interacting?}									& \bf{1849$-$670} 										\\    
	 \includegraphics[width=0.23\textwidth]{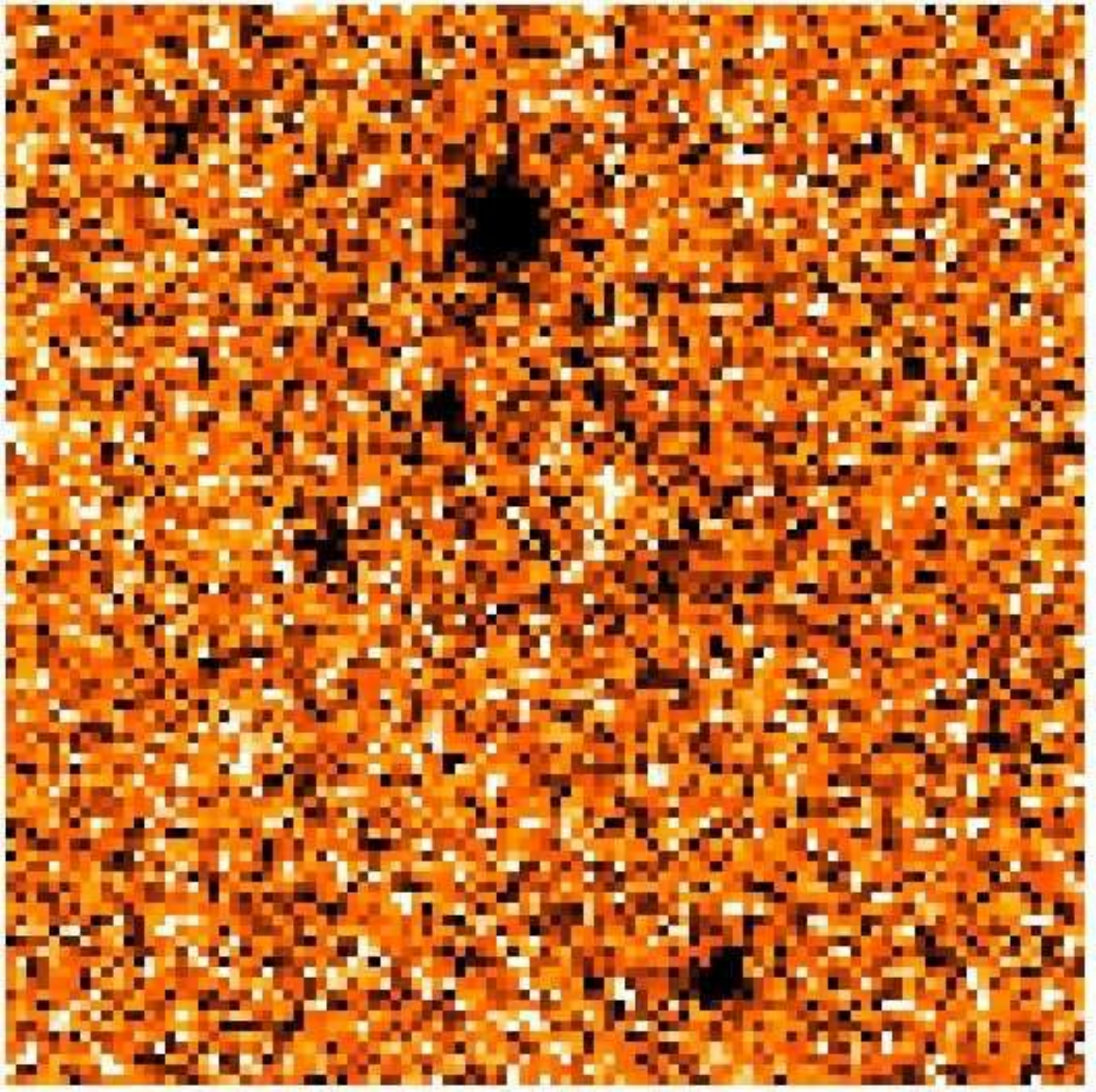} &\includegraphics[width=0.23\textwidth]{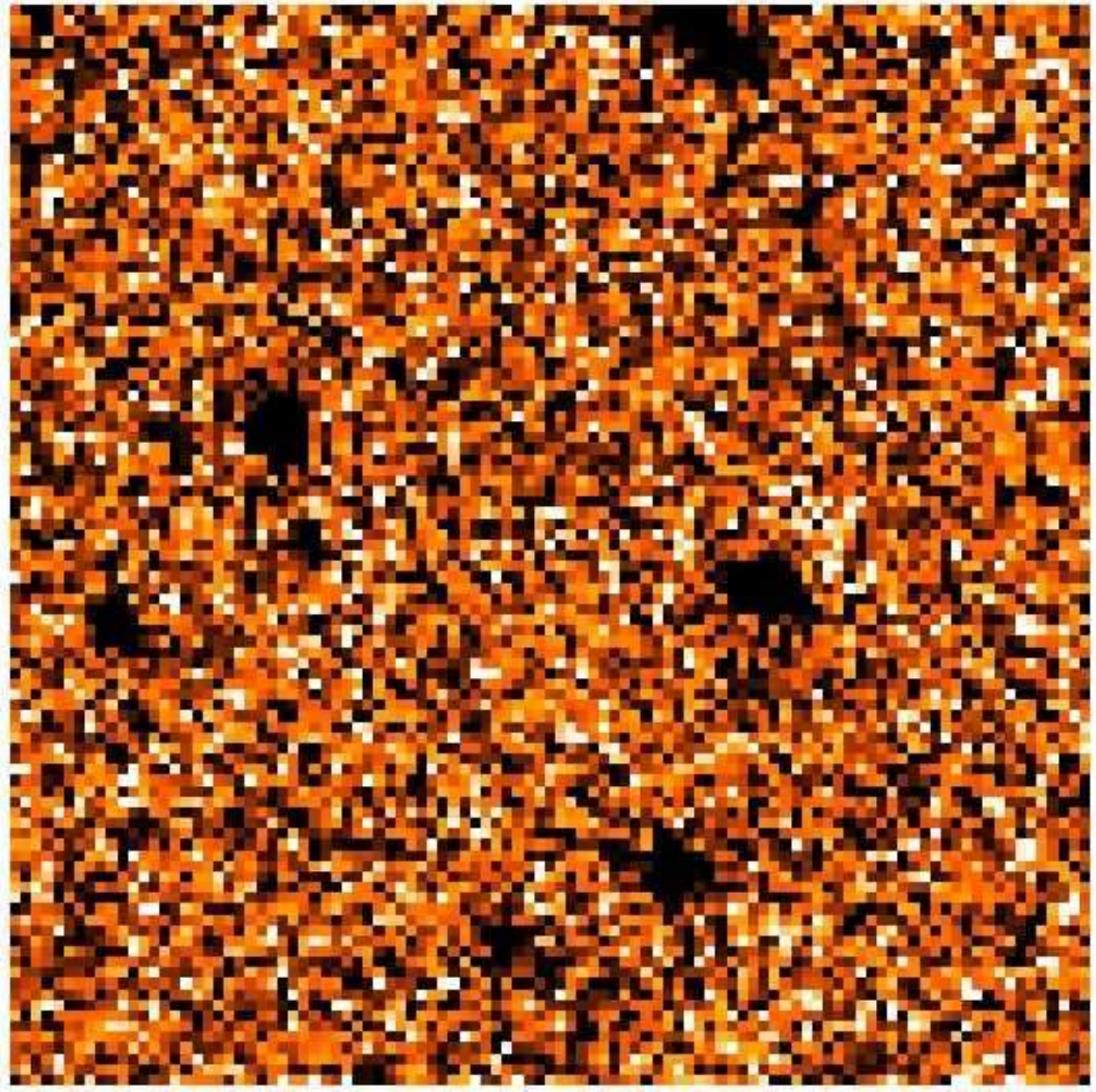} &	\includegraphics[width=0.23\textwidth]{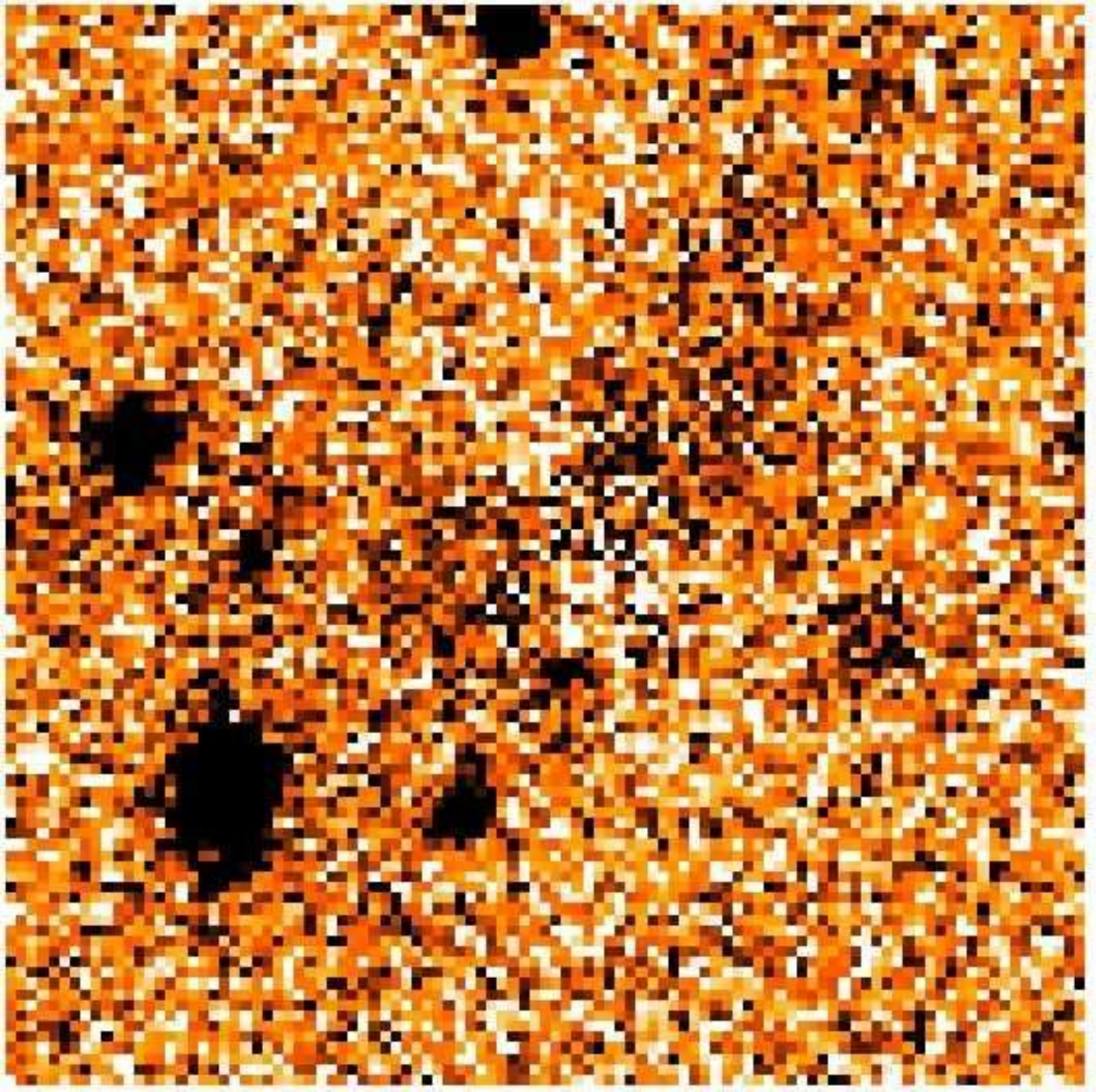} & \includegraphics[width=0.23\textwidth]{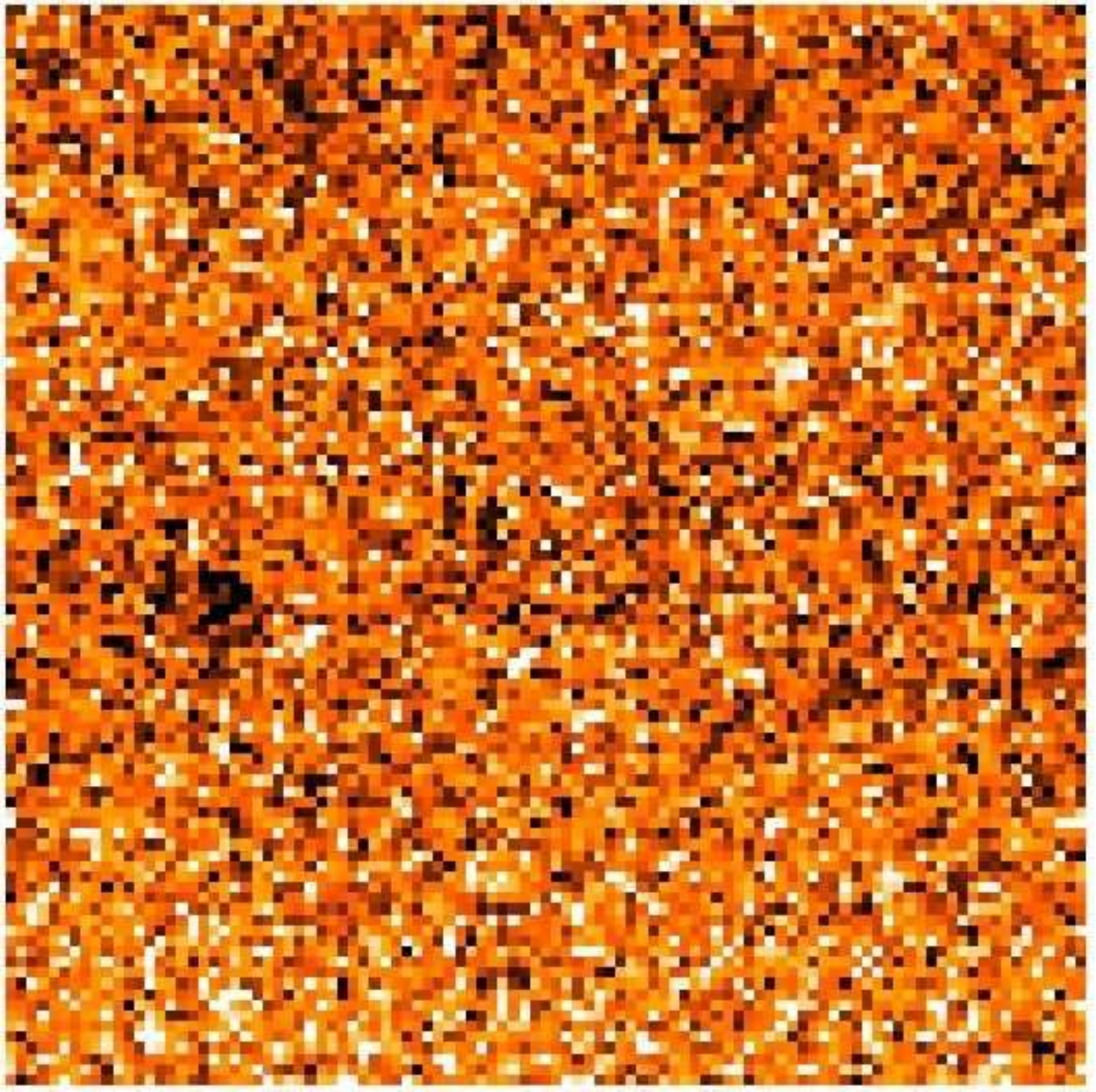}  \\
\bf{1928$+$738 Interacting?}										   &  	\bf{2216$-$038 Interacting?}							     	   	     &	\bf{2234$+$282 Interacting?}                					&  \\
     \includegraphics[width=0.23\textwidth]{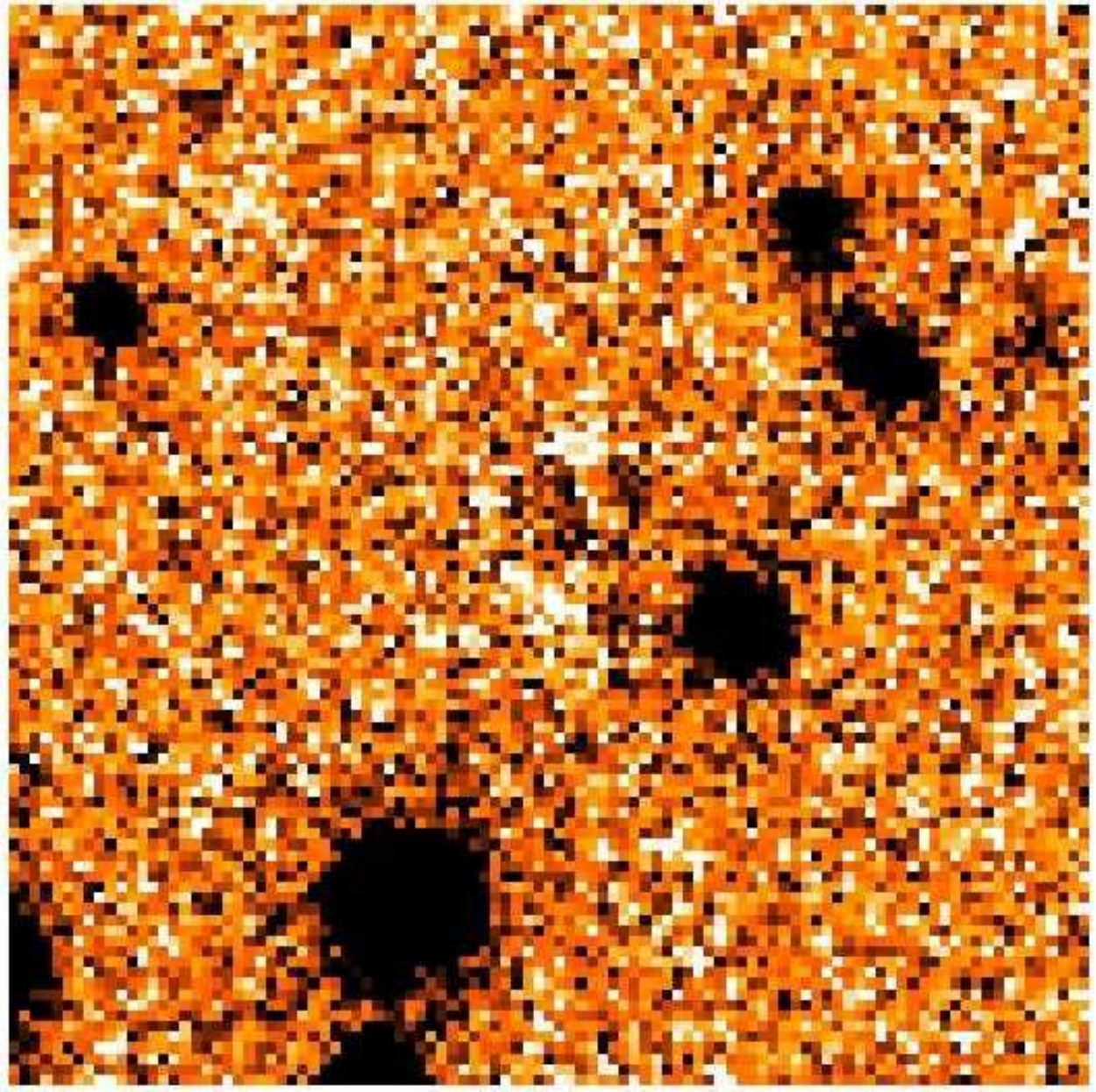}& \includegraphics[width=0.23\textwidth]{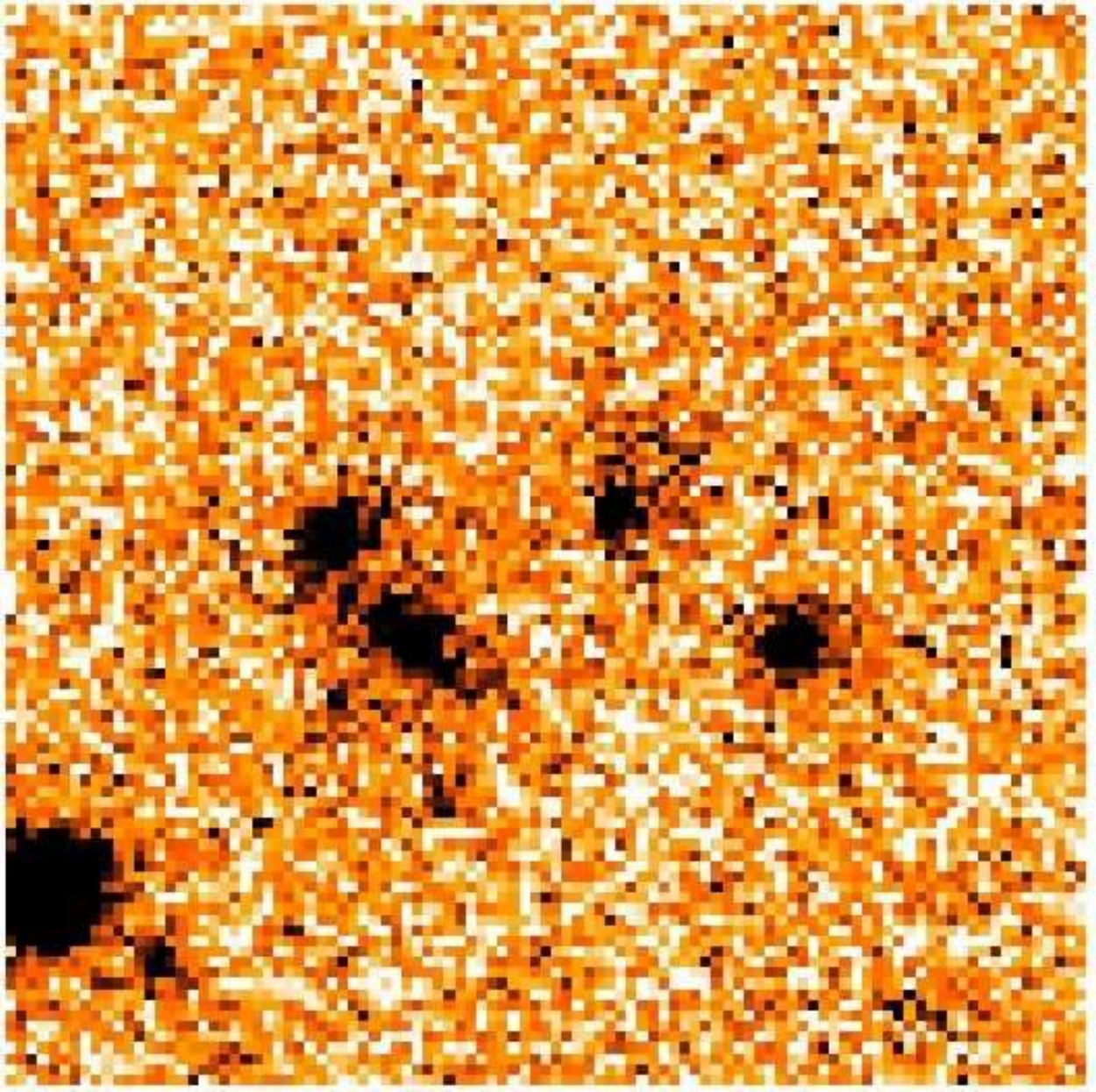}  &\includegraphics[width=0.23\textwidth]{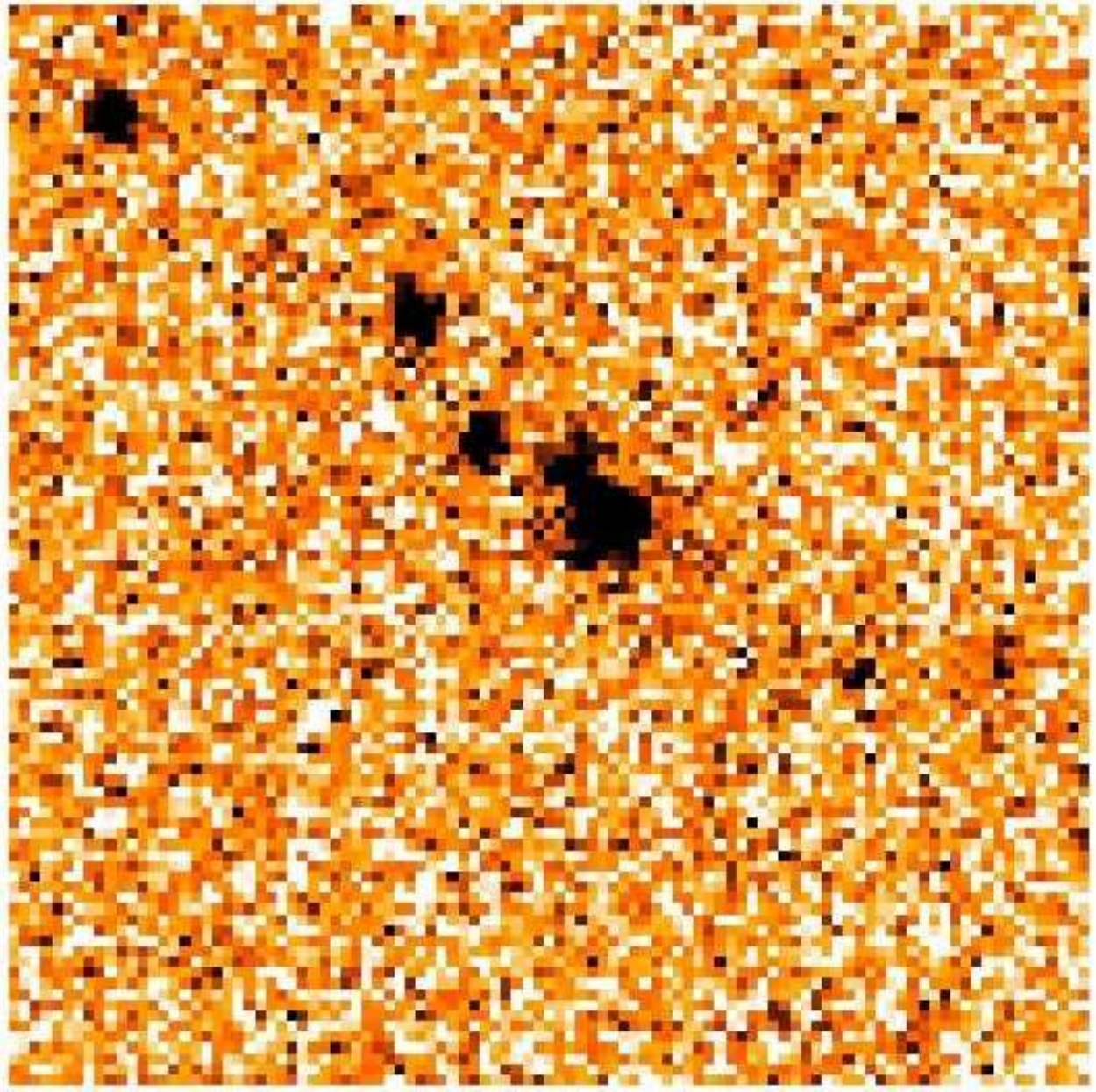} &	 \\
   \end{tabular}
  \caption{Residual images (best--fit model subtracted from the galaxy image) of the galaxies in this work. The galaxies location is at the center of the image. Some galaxies show suggestive evidence of recent interaction. All images are $23"\times23"$ (north is up and east is left).}
  \label{fig:interaction}
 \end{minipage}
\end{figure*}

%%%%%%%%%%%%
%%%%%%%%%%%%

\end{document}